\documentclass[sigconf,authorversion,nonacm]{acmart}

\usepackage{graphicx} 
\usepackage{hyperref}
\usepackage{amsfonts}
\usepackage{multirow}
\usepackage{enumitem}
\usepackage{subfig}
\usepackage{graphicx}
\usepackage{balance} 
\usepackage{comment}
\usepackage{epsfig}
\usepackage{epstopdf}
\usepackage{graphics}
\usepackage{amsmath}
\usepackage{booktabs} 
\usepackage{bookmark}
\usepackage{MnSymbol}%
\usepackage{xspace}
\usepackage{tikz}
\usepackage{amsthm}
\usepackage{mathtools}

\usepackage[ruled,linesnumbered]{algorithm2e}

\usepackage{filecontents}

\newtheorem{theorem}{Theorem}

\newtheorem{definition}{Definition}
\newcommand{\ignore}[1]{}
\newcommand{\revise}[1]{{\color{black}#1}}
\newcommand{\red}[1]{{\color{red}#1}}
\newcommand{\needcheck}[1]{{\color{purple}#1}}
\newcommand{\ourtech}{$\text{RiseFL}$}
\newcommand{\eiffel}{$\text{EIFFeL}$}
\newcommand{\rofl}{$\text{RoFL}$}

\usepackage{stfloats} 

\AtBeginDocument{%
  }

\ignore{
\setcopyright{acmcopyright}
\copyrightyear{2024}
\acmYear{2024}
\acmDOI{XXXXXXX.XXXXXXX}

\acmConference[SIGMOD '24]{Make sure to enter the correct
  conference title from your rights confirmation email}{June 11--16,
  2024}{Santiago, Chile}
\acmPrice{15.00}
\acmISBN{978-1-4503-XXXX-X/18/06}
}

\begin{document}

\title{Secure and Verifiable Data Collaboration with Low-Cost Zero-Knowledge Proofs}
\titlenote{This is an updated version of \cite{risefl-old} with additional experimental evaluations on tabular and medical image datasets.}


\author{Yizheng Zhu}
\affiliation{%
\institution{National University of Singapore}
\country{Singapore}
}
\email{yzhu@nus.edu.sg}

\author{Yuncheng Wu}
\affiliation{%
\institution{National University of Singapore}
\country{Singapore}
}
\email{wuyc@comp.nus.edu.sg}

\author{Zhaojing Luo}
\affiliation{%
\institution{National University of Singapore}
\country{Singapore}
}
\email{zhaojing@comp.nus.edu.sg}

\author{Beng Chin Ooi}
\affiliation{%
\institution{National University of Singapore}
\country{Singapore}
}
\email{ooibc@comp.nus.edu.sg}

\author{Xiaokui Xiao}
\affiliation{%
\institution{National University of Singapore}
\country{Singapore}
}
\email{xkxiao@nus.edu.sg}


\begin{abstract}
  Organizations are increasingly recognizing the value of data collaboration for data analytics purposes. Yet, stringent data protection laws prohibit the direct exchange of raw data.
To facilitate data collaboration, federated Learning (FL) emerges as a viable solution, which enables multiple clients to collaboratively train a machine learning (ML) model under the supervision of a central server while ensuring the confidentiality of their raw data. 
However, existing studies have unveiled two main risks: (i) the potential for the server to infer sensitive information from the client's uploaded updates (i.e., model gradients), compromising client input privacy, and (ii) the risk of malicious clients uploading malformed updates to poison the FL model, compromising input integrity. 
Recent works utilize secure aggregation with zero-knowledge proofs (ZKP) to guarantee input privacy and integrity in FL. Nevertheless, they suffer from extremely low efficiency and, thus, are impractical for real deployment.
%
In this paper, we propose a novel and highly efficient solution \ourtech{} for secure and verifiable data collaboration, ensuring input privacy and integrity simultaneously.
Firstly, we devise a probabilistic integrity check method that significantly reduces the cost of ZKP generation and verification. 
Secondly, we design a hybrid commitment scheme to satisfy Byzantine robustness with improved performance. 
%
Thirdly, we theoretically prove the security guarantee of the proposed solution.
Extensive experiments on synthetic and real-world datasets suggest that our solution is effective and is highly efficient in both client computation and communication. 
For instance, \ourtech{} is up to 28x, 53x and 164x faster than three state-of-the-art baselines ACORN, \rofl{} and \eiffel{} for the client computation.
\end{abstract}

\ignore{

}


\maketitle


\section{Introduction} \label{sec:intro}

\begin{figure}[t]
    \centering
    \includegraphics[width=0.48\textwidth]{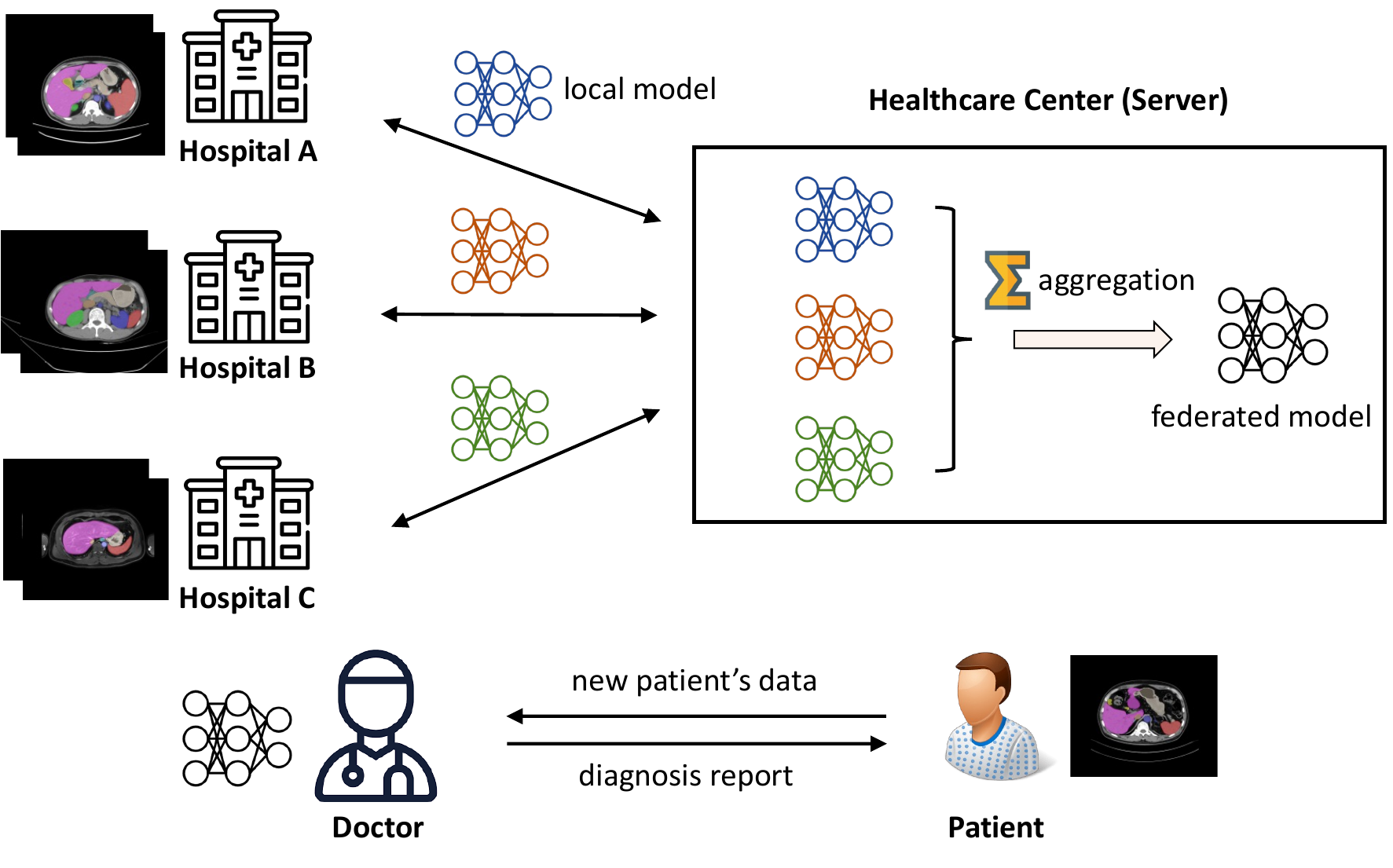}
    \caption{An example for healthcare data collaboration.}
    \label{fig:fl-example}
\end{figure}

Organizations and companies are progressively embracing digitization and digitalization as pathways to transformation, targeting enhancements in profitability, efficiency, or sustainability.
Data plays a crucial role in such transformations. 
For example, financial analysts may use users' historical data to adjudicate on credit card applications, and clinicians may use patients' electronic health records for disease diagnosis. 
While many organizations may lack expansive or pertinent datasets, there is a growing inclination towards data collaboration~\cite{WangWC0O23, 10.14778/3603581.3603588, Baunsgaard0CDGG21} for analytics purposes. 
Nevertheless, the ascent of rigorous data protection legislation, such as GDPR~\cite{GDPR2016}, deters direct raw data exchange.
\revise{For example, in a healthcare data collaboration scenario shown in Figure~\ref{fig:fl-example}, three hospitals aim to share their respective brain image databases to build a more accurate machine learning (ML) model. However, due to the highly sensitive nature of patients' data, directly sharing those brain images among hospitals is not permissible. 
}

In this landscape, federated learning~\cite{mcmahan2017communication, YangLCT19, li2020federated, LiuLXLM21, NguyenRCYMFMMMZ22, WuCXCO20, LiuWFWW21, BaoZXYOTA22, FuXCT022, FuSYJXT021, LiDCH22, FanFZPFLZ22, FuWX023, XieWGCYKLDZ23, WangTZZPF023, Zeng0FCPCWG23} emerges as a viable solution to facilitate data collaboration.
It enables multiple data owners (i.e., clients) to collaboratively train an ML model without necessitating the direct sharing of raw data, thereby complying with data protection acts. 
Typically, there is a centralized server that coordinates the FL training process as follows. 
For example, in Figure~\ref{fig:fl-example}, the server (i.e., the healthcare center) first initializes the model parameter and broadcasts it to all the hospitals. 
Then, in each iteration, each hospital computes a local model update (i.e., model gradients) on its own patients' data and uploads it to the server. 
The server aggregates all hospitals' updates to generate a global update and sends it back to the hospitals for iterative training~\cite{mcmahan2017communication}. Finally, the hospitals obtain a federated model that is trained on the three hospitals' brain datasets. After that, the doctor in each hospital can use it to assist in diagnosing new patients.  

Despite the fact that FL could facilitate data collaboration among multiple clients,
two main risks remain. 
The first is the client's \textit{input privacy}. 
Even without disclosing the client's raw data
to the server, recent studies~\cite{MelisSCS19, NasrSH19} have shown that the server can recover the client's sensitive data through the uploaded update with a high probability. 
The second is the client's \textit{input integrity}. In FL, there may exist a set of malicious clients that aim to poison the collaboratively trained model via Byzantine attacks, such as  imposing backdoors so that the model is susceptible to specific input data~\cite{BagdasaryanVHES20, XieHCL20}, contaminating the training process with malformed updates to degrade model accuracy~\cite{HayesO18, abs-1808-04866, BhagojiCMC19}, and so on. 

A number of solutions~\cite{BellBGL020, ZhengLLYYW23, KairouzL021, YinCRB18, YinCRB19, PanZWXJY20, CaoF0G21, MaSWLCD22, XuJZZJS22} have been proposed to protect input privacy and ensure input integrity in FL. 
On the one hand, instead of uploading the plaintext local updates to the server, the clients can utilize secure aggregation techniques~\cite{bonawitz2017practical, BellBGL020, KairouzL021}, such as secret sharing~\cite{shamir1979share, Keller20} and homomorphic encryption~\cite{Damg01, Elgamal85}, to mask or encrypt the local updates so that the server can aggregate the clients' updates correctly without knowing each update. 
In this way, the client's input privacy is preserved. However, these solutions do not ensure input integrity because it is difficult to distinguish a malicious encrypted update from benign ones.
On the other hand, \cite{YinCRB18, YinCRB19, PanZWXJY20, CaoF0G21, MaSWLCD22, XuJZZJS22} present various Byzantine-robust aggregation algorithms, allowing the server to identify malformed updates and eliminate them from being aggregated into the global update. Nevertheless, these algorithms require the clients to send plaintext updates to the server for the integrity verification, compromising the client's input privacy. 
%
%

In order to ensure input integrity while satisfying input privacy, \cite{roy2022eiffel, burkhalter2021rofl, bell2023acorn}
use secure aggregation to protect each client's update and allow the server to check the encrypted update's integrity using zero-knowledge proof (ZKP)~\cite{LiWXWR23} protocols. The general idea is to let each client compute a commitment of its local update and generate a proof that the update satisfies a publicly-known predicate, for example, the $L_2$-norm is within a specific range; then, the server can verify the correctness of the proofs based on the commitments without the need of knowing the plaintext values and securely aggregate the valid updates. 
Unfortunately, these solutions suffer from extremely low efficiency in proof generation and verification,
making them impractical for real deployments.
%
\revise{For example, under the experiment settings in Section~\ref{sec:experiment}, \eiffel{}~\cite{roy2022eiffel} takes 161 seconds to generate and verify proofs on 10K model parameters, compared to 7.6 seconds for client local training.
} 

{To 
introduce a practical FL system which ensures
both input privacy and input integrity,
%
}
we propose a secure and verifiable federated learning approach, called \ourtech{},
with high efficiency. In this paper, we focus on the $L_2$-norm integrity check, i.e., the $L_2$-norm of a client's local update is less than a threshold, which is widely adopted in existing works~\cite{sun2019can, burkhalter2021rofl, corrigan2017prio, roy2022eiffel}.
Our key observation of the low-efficiency in~\cite{burkhalter2021rofl, roy2022eiffel} is that their proof generation and verification costs for each client are linearly dependent on the number of parameters $d$ in the FL model. Therefore, we aim to reduce the proof cost, which particularly makes sense in the FL scenario because the clients often have limited computation and communication resources.
%
Specifically, our approach has the following novelties. 
%
%
First, we propose a probabilistic $L_2$-norm check method that decreases the cryptographic operation cost of proof generation and verification from $O(d)$ to $O(d/\log{d})$. The intuitive idea is to sample a set of public vectors and let each client generate proofs with respect to the inner product between its update and each public vector, instead of checking the $L_2$-norm of the update directly. This allows us to reduce the proof generation and verification time significantly. 
Second, we devise a hybrid commitment scheme based on Pedersen commitment~\cite{pedersen2001non} and verifiable Shamir secret sharing commitment~\cite{shamir1979share, feldman1987practical}, ensuring Byzantine-robustness while achieving further performance improvement on the client computation.
Third, although we consider the $L_2$-norm check, 
the proposed approach can be easily extended to various Byzantine-robust integrity checks~\cite{CaoF0G21, YinCRB18, SteinhardtKL17} based on different $L_2$-norm variants, such as cosine similarity~\cite{BagdasaryanVHES20, CaoF0G21}, sphere defense~\cite{SteinhardtKL17}, Zeno++~\cite{xie2019zeno++}, and so on. 

%

\ignore{
In \cite{roy2022eiffel}, a Byzantine-robust SAVI protocol EIFFel is introduced. It makes robust secure training possible when the number of Byzantine malicious clients is less than a third of the total. The robustness part is realized by a Zero-Knowledge Proof (ZKP) protocol that checks the integrity of clients' model updates. There are other systems that has various degrees of robustness and privacy \cite{burkhalter2021rofl,corrigan2017prio}. The main problem with leveraging ZKP in machine learning is the high cost of ZKP combined with the high dimensionality of ML. Each client has to produce a proof of the integrity of \emph{each of} the coordinate of its update. In \cite{roy2022eiffel}, with 126-bit encryption, on a ``large client", creating proofs on 1k coordinates takes about 0.5 second on average. (To expand: the probabilistic $L_\infty$ bound check is susceptible to attacks, so we ignore them here, and we need experiments to attack them). \cite{roy2022eiffel} has only 28-bit encryption \footnote{the solution to the discrete log problem in verifiable secret shares in Section 11.1 is only 56-bit, so it can be found in about $28 \cdot 2^{28}$ attempts.}, and the cost of creating proofs is even higher and grows linearly in both the total number of clients and the maximum number of malicious clients (which means quadratically in the total number of clients if the ratio of malicious clients is fixed). This makes training with large models very slow.

In this paper, we present a new FL system that significantly reduces the ZKP cost in robust secure FL under $l_2$-norm based checks. Experiments show that our system is over 100x faster than previous work \cite{roy2022eiffel,burkhalter2021rofl} when weight dimension is at least 100K. The key idea is a sampling method that reduces the number of multiplications and bound proofs from $O(d)$ to $O(1)$. A detailed analysis of cost comparison between our system and existing works will be in Section \ref{sec:analysis}.
}

\ignore{
\begin{figure}[t]
    \centering
    \includegraphics[width=0.48\textwidth]{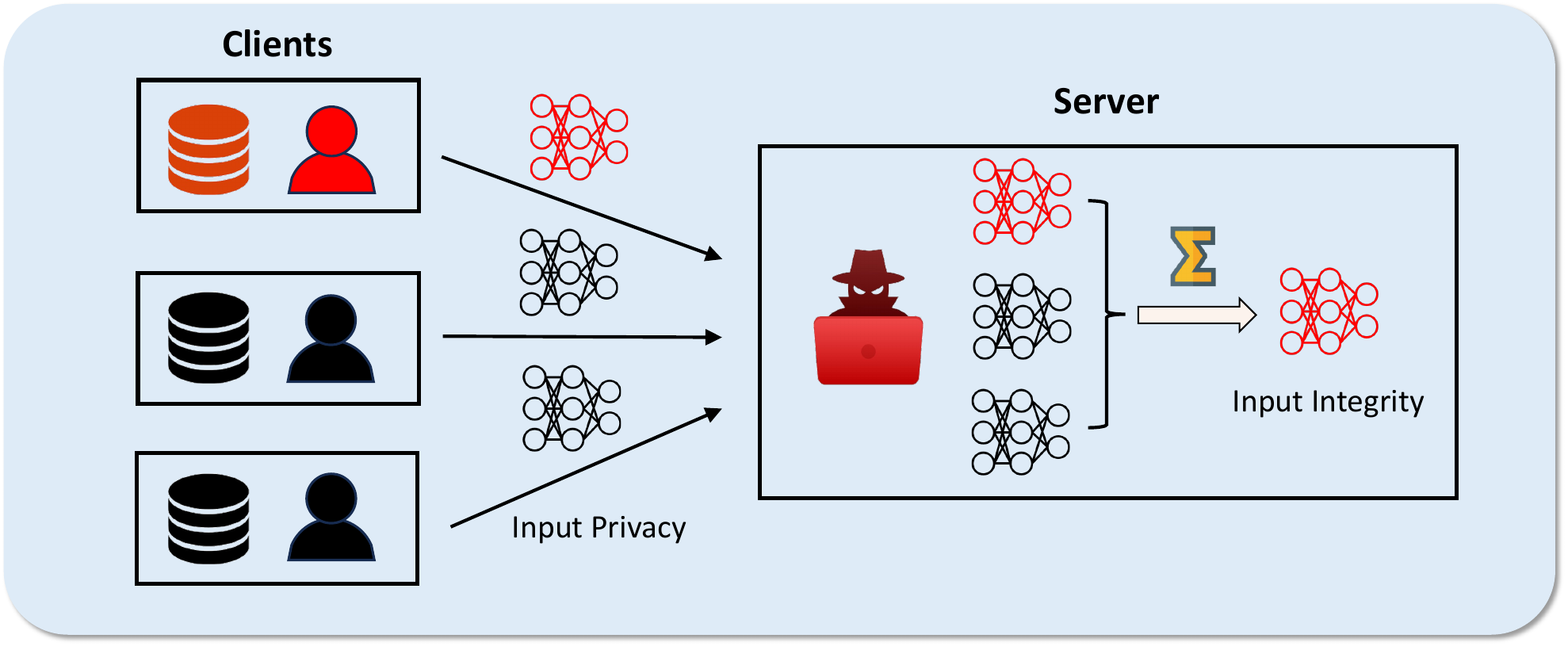}
    \caption{The input privacy and input integrity risks in FL.}
    \label{fig:fl-risks}
\end{figure}
}

In summary, we make the following contributions.
\begin{itemize}[leftmargin=20pt]
    \item We propose a novel and highly efficient solution \ourtech{} for FL-based data collaboration, which simultaneously ensures each client's input privacy and input integrity. 
    \item We present a probabilistic $L_2$-norm integrity check method and a hybrid commitment scheme, which significantly reduces the ZKP generation and verification costs.  
    \item We provide a formal security analysis of \ourtech{} and theoretically compare its computational and communication costs with three state-of-the-art solutions. 
    \item We implement \ourtech{} and evaluate its performance with a set of micro-benchmark experiments as well as FL tasks on two real-world datasets. The results demonstrate that \ourtech{} is effective in detecting various attacks and is up to 28x, 53x and 164x faster than ACORN~\cite{bell2023acorn}, \rofl{}~\cite{burkhalter2021rofl} and \eiffel{}~\cite{roy2022eiffel} for the client computation, respectively.
\end{itemize}

The rest of the paper is organized as follows. Section~\ref{sec:preliminary} introduces preliminaries and Section~\ref{sec:rsfl-overview} presents an overview of our solution. We detail the system design in Section~\ref{sec:rsfl-design} and analyze the security and cost in Section~\ref{sec:analysis}.
The evaluation is provided in Section~\ref{sec:experiment}. We review the related works in Section~\ref{sec:related} and conclude the paper in Section~\ref{sec:conclusion}.
\section{Preliminaries} \label{sec:preliminary}

We first introduce the notations used in this paper. 
Let $\mathbb{G}$ denote a cyclic group with prime order $p$, where the discrete logarithm problem~\cite{pomerance1990cryptology} is hard. 
Let $\mathbb{Z}_p$ denote the set of integers modulo the prime $p$.
%
We use $x$, $\textbf{x}$, and $\textbf{X}$ to denote a scalar, a vector, and a matrix, respectively. We use $\mathsf{Enc}_{\mathsf{K}}(x)$ to denote an encrypted value of $x$ under an encryption key $\mathsf{K}$, and $\mathsf{Dec}_{\mathsf{K}}(y)$ to denote a decrypted value of $y$ under the same key $\mathsf{K}$.
Since the data used in ML are often in the floating-point representation, we use fixed-point integer representation to encode floating-point values. 
In the following, we introduce the cryptographic building blocks used in this paper.

\vspace{1mm}
\noindent
\textbf{Pedersen Commitment.} 
A commitment scheme is a cryptographic primitive that allows one to commit a chosen value without revealing the value to others while still allowing the ability to disclose it later~\cite{Goldreich2001}. 
Commitment schemes are widely used in various zero-knowledge proofs. 
In this paper, we use the Pedersen commitment~\cite{pedersen2001non} for a party to commit its secret values. 
Given the independent group elements $(g,h)$, the Pedersen commitment encrypts a value $x \in \mathbb{Z}_p$ to $C(x,r) = g^x h^r$, where $r \in \mathbb{Z}_p$ is a random number. 
An important property of the Pedersen commitment is that it is additively homomorphic. 
Given two values $x_1, x_2$ and two random numbers $r_1, r_2$, the commitment follows $C(x_1,r_1) \cdot C(x_2,r_2) = C(x_1+x_2, r_1+r_2)$. 

\vspace{1mm}
\noindent
\textbf{Verifiable Shamir's Secret Sharing Scheme.}
Shamir's $t$-out-of-$n$ secret sharing (SSS) scheme~\cite{shamir1979share} allows a party to distribute a secret among a group of $n$ parties via shares so that the secret can be reconstructed given any $t$ shares but cannot be revealed given less than $t$ shares. 
The SSS scheme is verifiable (aka. VSSS) if auxiliary information is provided to verify the validity of the secret shares. 
We use the VSSS scheme~\cite{shamir1979share,feldman1987practical} to share a number $r \in \mathbb{Z}_p$. 
Specifically, the VSSS scheme consists of three algorithms, namely, $\mathsf{SS.Share}$, $\mathsf{SS.Verify}$, and $\mathsf{SS.Recover}$. 
\begin{itemize}
    \item $((1, r_1), \dots, (n, r_n), \Psi) \leftarrow \mathsf{SS.Share}(r, n, t, g)$. Given a secret $r \in \mathbb{Z}_p$, $g \in \mathbb{G}$, and $0 < t \leq n$, this algorithm outputs a set of $n$ shares $(i, r_i)$ for $i \in [n]$ and a check string $\Psi$ as the auxiliary information to verify the shares. Specifically, it generates a random polynomial $f$ in $\mathbb{Z}_p$ of degree at most $(t-1)$ whose constant term is $r$. 
    We set $r_i = f(i)$ and $\Psi = (g^r, g^{f_1}, \dots, g^{f_{t-1}})$ where $f_i$ is the $i$-th coefficient.
    \item $r \leftarrow \mathsf{SS.Recover}(\{(i, r_i)\}:{i \in A})$. 
    For any subset $A \subset [n]$ with size at least $t$, this algorithm recovers the secret $r$.
    \item $\mathsf{True}/\mathsf{False} \leftarrow \mathsf{SS.Verify}(\Psi, i, r_i, n, t, g)$. 
    Given a share $(i, r_i)$ and the check string $\Psi$, it verifies the validity of this share so that it outputs $\mathsf{True}$ if $(i, r_i)$ was indeed generated by $\mathsf{SS.Share}(r, n, t, g)$ and $\mathsf{False}$ otherwise.
\end{itemize}
This scheme is additively homomorphic in both the shares and the check string. 
If $((1, r_1), \dots, (n, r_n), \Psi_r) \leftarrow \mathsf{SS.Share}(r, n, t, g)$ and $((1, s_1), \dots, (n, s_n), \Psi_s) \leftarrow \mathsf{SS.Share}(s, n, t, g)$, then:
\begin{itemize}
    \item $r + s\leftarrow \mathsf{SS.Recover} (\{(i, r_i + s_i)\}:{i \in A})$ for any subset $A \subset [n]$ with size at least $t$, 
    \item $\mathsf{True} \leftarrow \mathsf{SS.Verify}(\Psi_r\cdot\Psi_s, i, r_i + s_i, n, t, g)$.
\end{itemize}

\ignore{
Given $0 < t \leq n$ and $g \in \mathbb{G}$, the algorithm $\mathsf{SS.Share}(r, n, t, g)$ outputs $(r_1, \dots, r_n, \Psi)$ such that:
\begin{enumerate}
    \item For any $0 < i_1 < \dots < i_t \leq n$, $\mathsf{SS.Recover}(i_1, r_{i1}, \dots, i_t, r_{it})$ outputs $r$.
    \item For any $i$, $\mathsf{SS.Verify}(\Psi, r_i, i, n, t, g)$ outputs $\mathsf{True}$.
    \item It is computationally infeasible to create $(r_1, \dots, r_n, \Psi)$ that satisfies Condition 2 but does not satisfy Condition 1.
    \item For any subset $A \subset [n]$ with size less than $t$, it is impossible to recover any information about $r$ from $\{(i, r_i) : i \in A\}$. 
\end{enumerate}
}

\vspace{1mm}
\noindent 
\textbf{Zero-Knowledge Proofs.} 
A zero-knowledge proof (ZKP) allows a prover to prove to a verifier that a given statement is true, such as a value is within a range, without disclosing any additional information to the verifier~\cite{BlumFM88}. 
We utilize two additively homomorphic ZKP protocols based on the Pedersen commitment as building blocks.
Note that the zkSNARK protocols~\cite{ben2013snarks,gennaro2013quadratic,parno2016pinocchio} are not additively homomorphic, thus they cannot support the secure aggregation required in federated learning.



The first is the $\Sigma$-protocol~\cite{camenisch1997proof} for proof of square and proof of relation. 
For proof of square $((x, r_{1}, r_{2}),(y_{1}, y_{2}))$, 
denote $(g, h)$ as the independent group elements, and let $y_{1} = C(x, r_1)$ and $y_{2} = C(x^2, r_2)$ be the Pedersen commitments, where $x,r_{1},r_{2} \in \mathbb{Z}_p$ are the secrets. 
The function $\mathsf{GenPrfSq}((x, r_{1}, r_{2}), (y_{1}, y_{2}))$ generates a proof $\pi$ that the secret value in $y_2$ is the square of the secret value in $y_1$. Accordingly, the function $\mathsf{VerPrfSq}(\pi, y_{1}, y_{2})$ verifies the correctness of this proof.
Similarly, for proof of a relation $((r, v, s), (z, e, o))$, denote $(g, q, h)$ as the independent group elements, and let $z = g^r, e = g^v h^r, o = g^v q^s$ be the Pedersen commitments, where $r,v,s \in \mathbb{Z}_p$ are the secrets. 
The function $\mathsf{GenPrfWf}((r, v, s), (z, e, o))$ generates a proof $\pi$ that the secrets in $e$ and $o$ are equal, and that the secret in $z$ is equal to the blind in $e$. 
The function $\mathsf{VerPrfWf}(\pi, z, e, o)$ verifies the proof.
The second ZKP protocol used is the Bulletproofs protocol~\cite{bunz2018bulletproofs} for checking the bound of a value $x$ with its Pedersen commitment $y = C(x,r)$. 
We denote $\mathsf{GenPrfBd}(x, y, r, b)$ as the function that generates a proof $\pi$ that $x \in [0, 2^b)$, where $2^b$ is the bound to be ensured. 
The corresponding function $\mathsf{VerPrfBd}(\pi, y, b)$ verifies the proof. 
We refer the interested readers to~\cite{bunz2018bulletproofs} for more details.

These building blocks can be easily extended to the vector form such that the proof generation and verification execute on a batch of values. Unless noted otherwise, we use the vector form in the rest of this paper.

\ignore{
The first ZKP protocol is the $\Sigma$-protocol~\cite{camenisch1997proof} for proof of square and proof of relation.
For proof of square $((x, r_{1}, r_{2}),(y_{1}, y_{2}))$,
denote $g, h \in \mathbb{G}$ the independent group elements, and let $y_{1} = g^{x} h^{r_{1}}$ and $y_{2} = g^{x^2} h^{r_{2}}$ be the commitments, where $x,r_{1},r_{2} \in \mathbb{Z}_p$ are the secrets. To handle the square in power, we rewrite $y_{2} = y_{1}^x h^{r_{2} - r_{1} x}$. 
The function $\mathsf{GenPrfSq}()$\footnote{We detail the $\mathsf{GenPrfSq}()$, $\mathsf{VerPrfSq}()$, $\mathsf{GenPrfWf}()$, $\mathsf{VerPrfWf}()$ functions with batch forms in Algorithms~\ref{alg:gen_prf_sq}-\ref{alg:ver_prf_wf} in Appendix~\ref{appendix:pre}, respectively.} generates a proof $\pi$ that $(y_{1}, y_{2})$ is of the form $(g^{x} h^{r_{1}}, g^{x^2} h^{r_{2}})$ for $x, r_{1}, r_{2} \in \mathbb{Z}_p$. Accordingly, the function $\mathsf{VerPrfSq}()$ verifies this proof based on $(y_{1}, y_{2})$. %
%
For proof of a relation $((r, v, s), (z,e, o))$, 
denote $g, q, h \in \mathbb{G}$ the independent group elements, and we let $z=g^r$, $e = g^v h^r$, $o = g^v q^s$ be the commitments, where $r,v,s \in \mathbb{Z}_p$ are the secrets. 
Then, the function $\mathsf{GenPrfWf}()$ generates a proof $\pi$ that $(z, e,o)$ is of the form $(g^r, g^{v} h^r, g^v q^s)$ given $x,r$ and $s$, and the function $\mathsf{VerPrfWf}()$ verifies the proof $\pi$ according to $(z,e,o)$.

The second ZKP protocol used is the Bulletproofs protocol~\cite{bunz2018bulletproofs} for checking the bound of $\textbf{x} = (x_1, \dots, x_k)$ in the vector of Pedersen commitments $\textbf{y} = (g^{x_1}h^{r_1}, \dots, g^{x_k}h^{r_k})$. 
To generate and verify a proof that $x_j \in [0, 2^b)$ for every $j = \{ 1, \dots, k \}$, 
%
we express $x_j$ in binary: $x_j = \sum_{i=0}^{b-1} x_{ji} 2^i$, each $x_{ji} \in \{0, 1\}$. Then, we use group elements $\textbf{f} \in \mathbb{G}^{2bk}$ to commit $x_{ji}$ and $x_{ji} - 1$ for each $j = \{ 1, \dots, k \}$ and each $i \in \{0, \cdots, b-1\}$, and use the algorithm in ~\cite{bunz2018bulletproofs} to prove and check that $x_{ji}(x_{ji}-1) = 0$. 
We denote $\mathsf{GenPrfBd}(g, h, \textbf{f}, b, \textbf{y}, \textbf{x}, \textbf{r})$ and $\mathsf{VerPrfBd}(g, h, \textbf{f}, b, \textbf{y}, \pi)$ the proof generation and verification functions for a statement that $x_j \in [0, 2^b)$ for every $j$, and refer the interested readers to~\cite{bunz2018bulletproofs} for more details. 
}


\ignore{
We can use independent group elements $g, h \in \mathbb{G}$, $f \in \mathbb{G}^{2b}$. 
The following functions are introduced: 
\begin{itemize}
    \item $\mathsf{GenPrfBd}(g, h, f, b, x, r)$, which generates a proof $\pi$ that $x \in [0, 2^b)$. 
    \item $\mathsf{VerPrfBd}(g, h, f, b, y, \pi)$, which verifies that $\pi$ is a proof that $y = g^x h^r$ and $x \in [0, 2^b)$.
\end{itemize}
Express $x$ in binary: $x = \sum_{i=0}^{b-1} x_i 2^i$, each $x_i \in \{0, 1\}$. The group elements $f \in \mathbb{G}^{2b}$ is used to commit $x_i$ and $x_i - 1$ for each $i$. The heavy-lifting part of Bulletproofs is to prove and check that $x_i(x_i-1) = 0$.
}
\section{System Overview} \label{sec:rsfl-overview}

In this section, we present the system model and the threat model, 
and then we give an overview of the proposed \ourtech{} system which ensures input privacy and input integrity.

\subsection{System Model} \label{subsec:sys-model}

There are $n$ clients $\{\mathcal{C}_1, \cdots, \mathcal{C}_n\}$ and a centralized server in the system. 
Each client $\mathcal{C}_i (i \in [1,n])$ holds a private dataset $\mathcal{D}_i$ to participate in data collaboration for training a federated learning (FL) model $\mathcal{M}$. 
Let $d$ be the number of parameters in $\mathcal{M}$. 
In each iteration, the training process consists of three steps. 
Firstly, the server broadcasts the current model parameters to all the clients. 
Secondly, each client $\mathcal{C}_i$ locally computes a model update (i.e., gradients) $\textbf{u}_i$ given the model parameters and its dataset $\mathcal{D}_i$, and submits $\textbf{u}_i$ to the server. Thirdly, the server aggregates the clients' gradients to a global update $\mathcal{U} = \sum\nolimits_{i\in[1,n]} \textbf{u}_i$ and updates the model parameters of $\mathcal{M}$ for the next round of training until convergence.

\ignore{
\begin{figure}[t]
    \centering
    \includegraphics[width=0.49\textwidth]{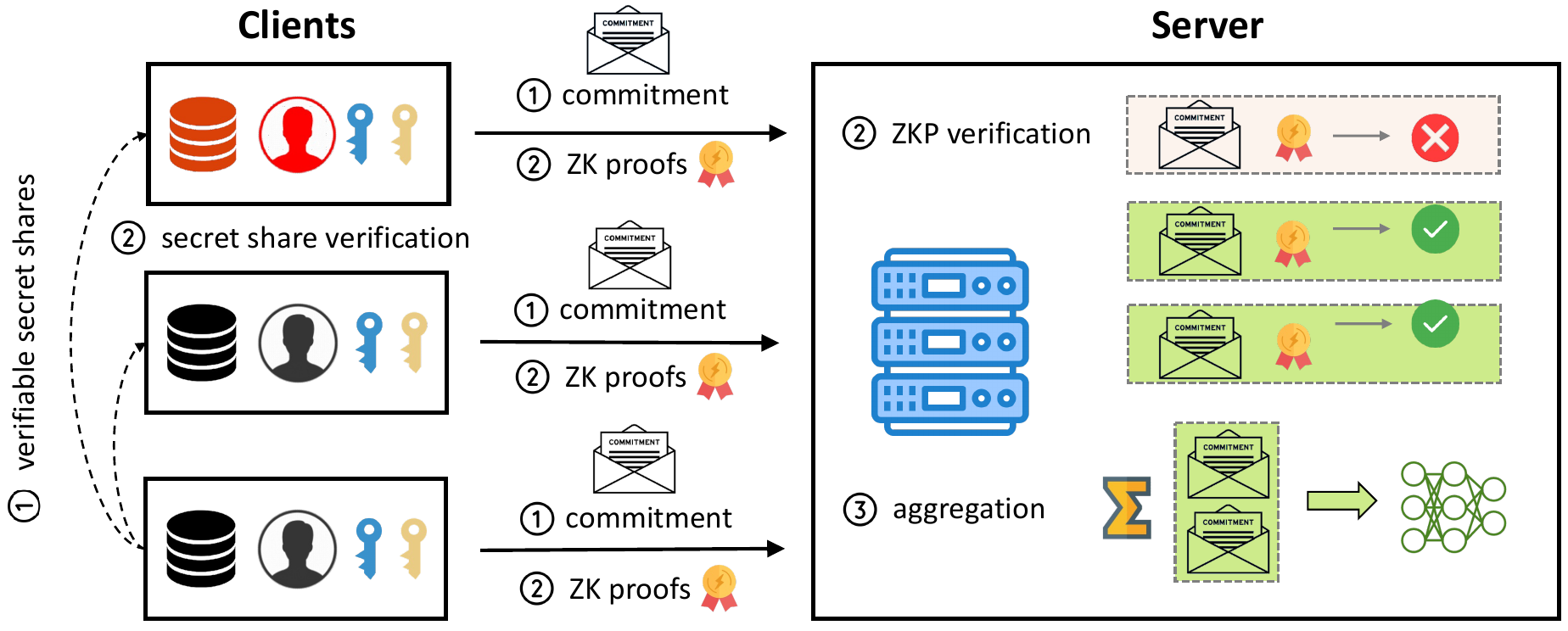}
    \caption{An overview of the proposed \ourtech{} system. \red{expand this figure to illustrate the three rounds}}
    \label{fig:rsfl-overview}
\end{figure}
}

\begin{figure*}[t]
    \centering
    \subfloat[System initialization]{{\includegraphics[width=0.235\textwidth]{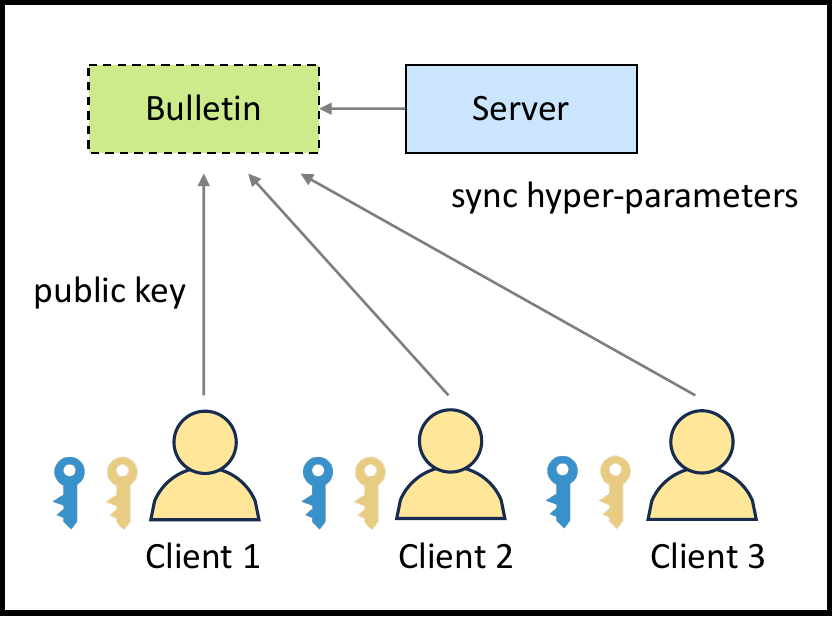} } \label{subfig:system-init}}%
    \subfloat[Commitment generation]{{\includegraphics[width=0.235\textwidth]{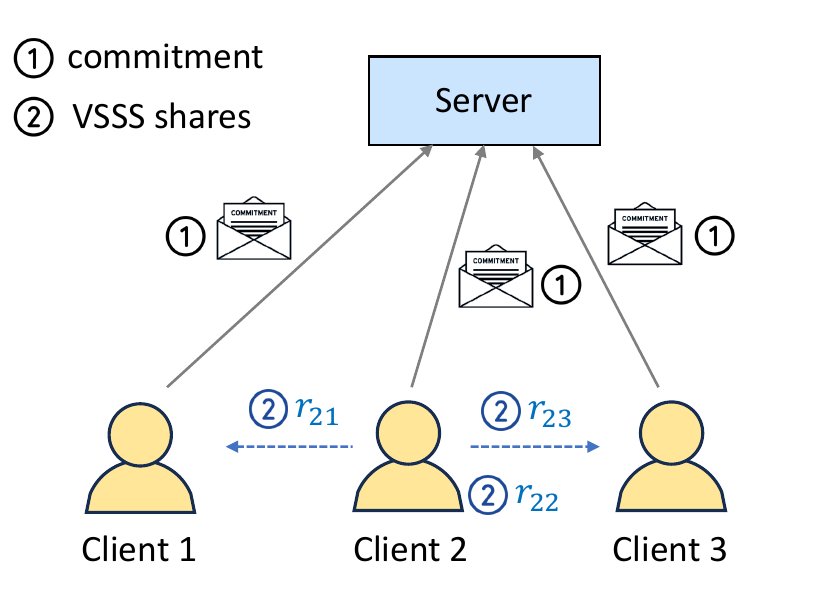} \label{subfig:commitment} }}%
~
    \subfloat[Proof generation and verification]{{\includegraphics[width=0.235\textwidth]{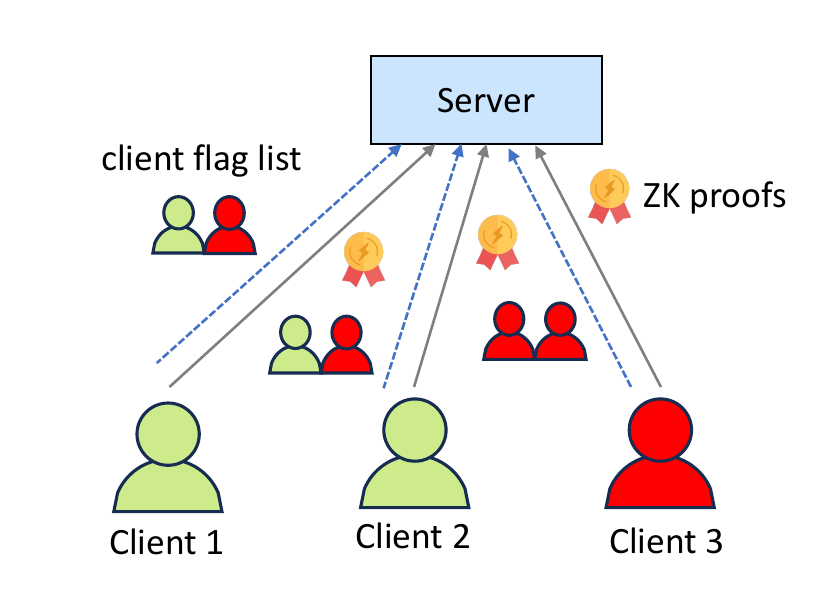} } \label{subfig:prof-gen-ver}}%
    \subfloat[Secure aggregation]{{\includegraphics[width=0.235\textwidth]{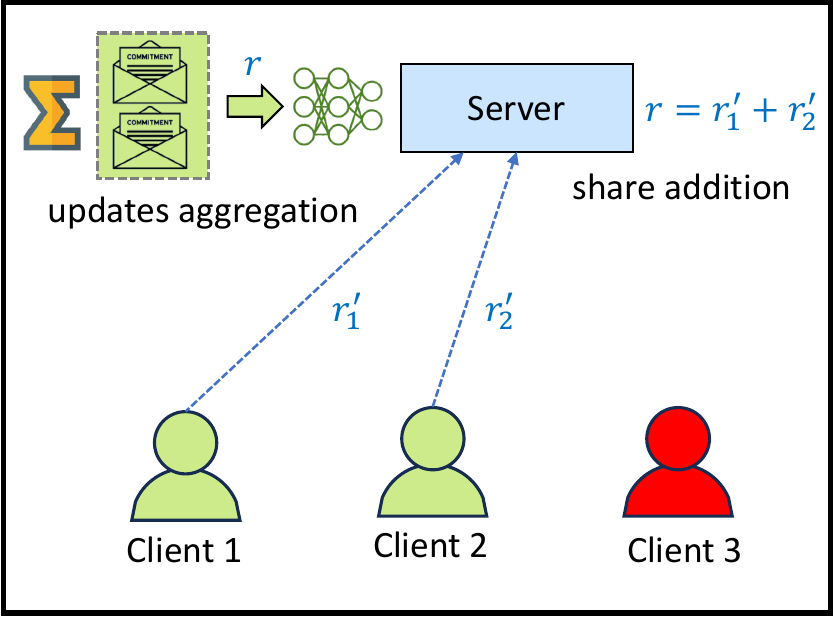} \label{subfig:secagg} }}%
    \caption{An overview of the proposed \ourtech{} system.}
    \label{fig:rsfl-overview}
\end{figure*}

\subsection{Threat Model} \label{subsec:threat-model}

%
We consider a malicious threat model in two aspects. 
First, regarding input privacy, we consider a malicious server (i.e., the adversary) that can deviate arbitrarily from the specified protocol to infer each client's uploaded model update. 
Also, the server may collude with some of the malicious clients to compromise the honest clients' privacy. 
Similar to~\cite{roy2022eiffel}, we do not consider the scenario that the server is malicious against the input integrity because its primary goal is to ensure the well-formedness of each client's uploaded update. 
Second, regarding input integrity, we assume that there are at most $m$ malicious clients in the system, where $m < n / 2$.
The malicious clients can also deviate from the specified protocol arbitrarily, such as sending malformed updates to the server to poison the aggregation of the global update, or intentionally marking an honest client as malicious to interfere with the server's decision on the list of malicious clients.


\ignore{
\begin{figure*}[h]
    \centering
    \subfloat[system initialization]{{\includegraphics[width=0.24\textwidth]{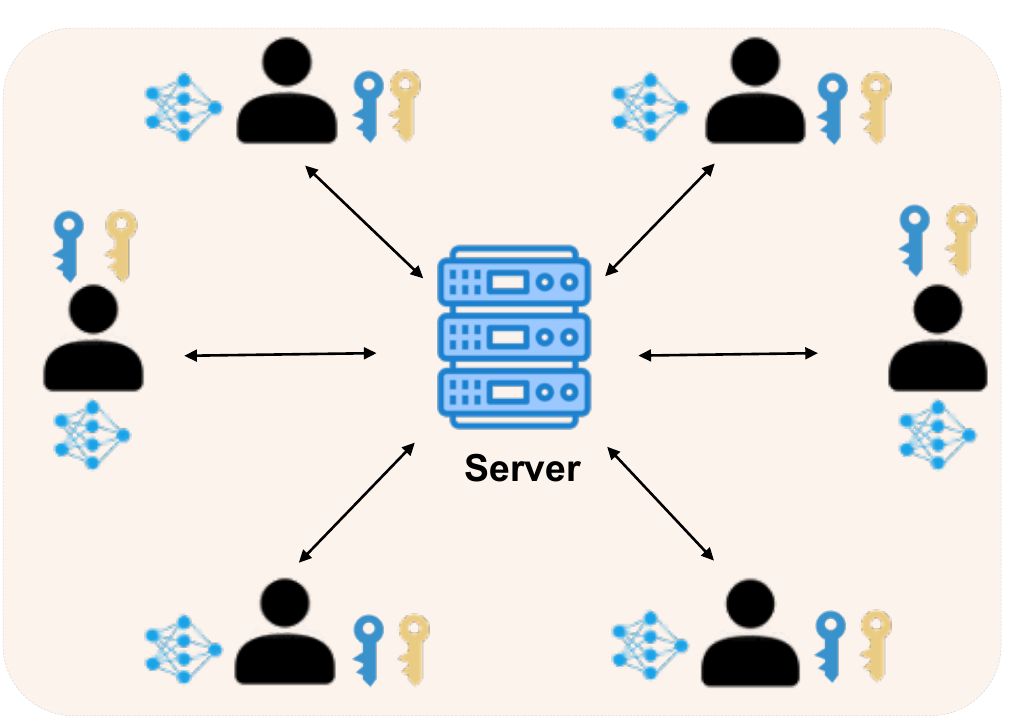} }}%
    \subfloat[commitment generation]{{\includegraphics[width=0.24\textwidth]{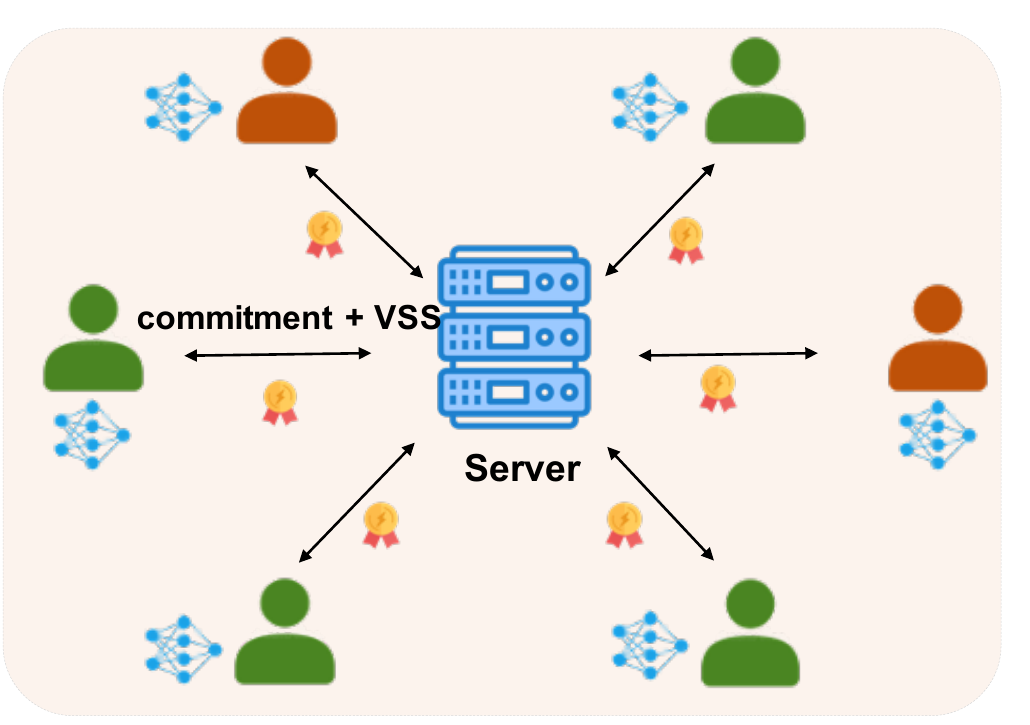} }}%
    \subfloat[proof generation and verification]{{\includegraphics[width=0.24\textwidth]{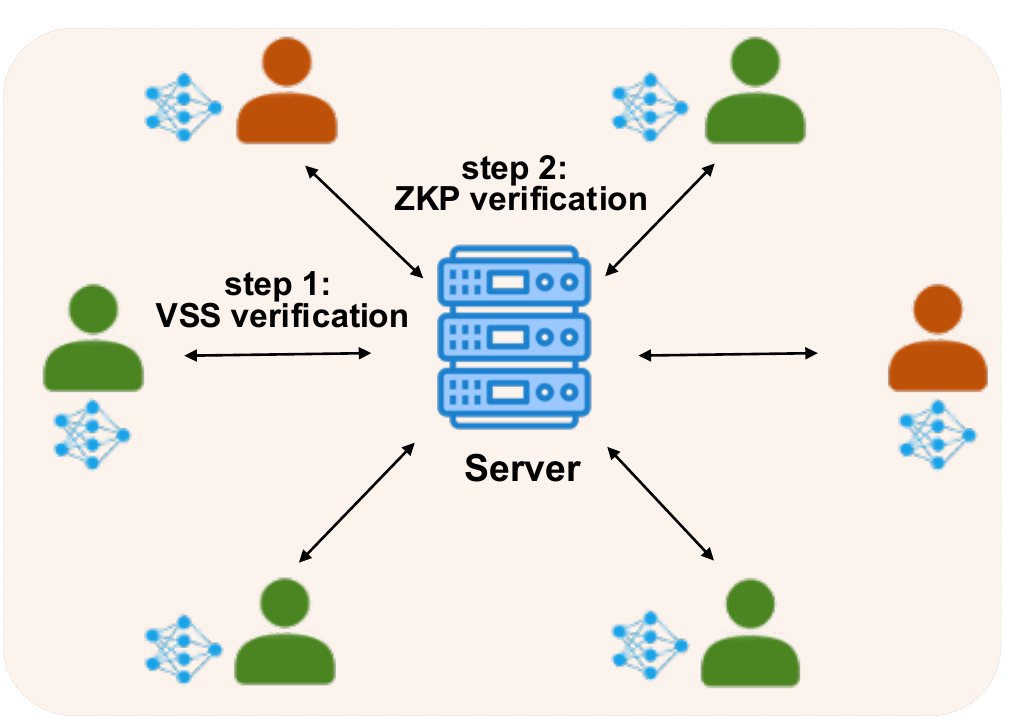} }}%
    \subfloat[aggregation of updates]{{\includegraphics[width=0.24\textwidth]{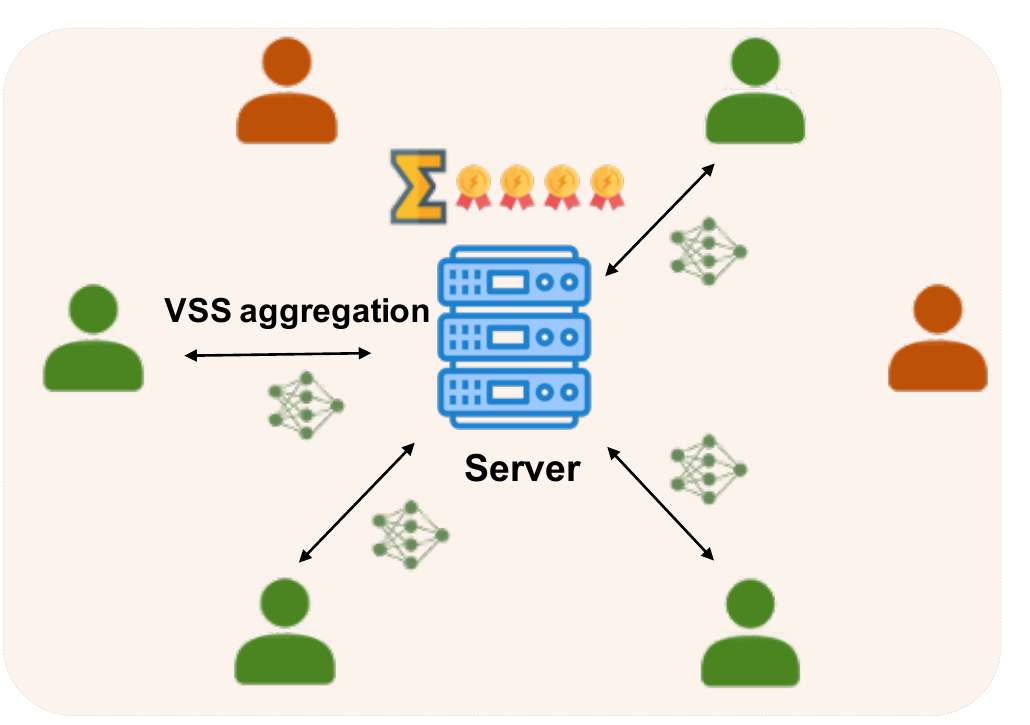} }}%
    \caption{An overview of the proposed \ourtech{} system. \needcheck{(To replace it with a smaller figure).}}
    \label{fig:rsfl-overview}
\end{figure*}
}

\subsection{Problem Formulation} \label{subsec:problem}

We aim to ensure both input privacy (for the clients) and input integrity (for the server) under the threat model described in Section~\ref{subsec:threat-model}. 
We formulate our problem as a relaxed variant of the secure aggregation with verified inputs (SAVI) problem in~\cite{roy2022eiffel}, 
namely $(D,F)$-relaxed SAVI, as defined in Definition~\ref{def:problem}. It achieves the same level of input privacy while relaxing the input integrity for significant efficiency improvement.
\ignore{
but relaxing the input integrity check for efficiency.
Definition~\ref{def:problem} formulates our problem, namely 
$( D,F)$-relaxed SAVI. 
}

\begin{definition} \label{def:problem}
Given a security parameter $\kappa$, a function $D : \mathbb{R}^d \to \mathbb{R}$ satisfying that $\textbf{u}$ is malicious if and only if $D(\textbf{u}) > 1$,
a function $F: (1, +\infty) \to [0,1]$, 
a set of inputs $\{\textbf{u}_1, \cdots, \textbf{u}_n\}$ from clients $\mathcal{C} = \{\mathcal{C}_1, \dots, \mathcal{C}_n\}$ respectively, 
and a list of honest clients $\mathcal{C}_H$, a protocol $\Pi$ is a 
$( D,F)$-relaxed SAVI protocol for $\mathcal{C}_H$ if:
\begin{itemize}[leftmargin=*]
    \item Input Privacy. 
    The protocol $\Pi$ realizes the ideal functionality $\mathcal{F}$ such that for an adversary $\mathcal{A}$ that consists of the malicious server and the malicious clients $\mathcal{C}_M = \mathcal{C} \setminus \mathcal{C}_H$ attacking the real interaction, there exist a simulator $\mathcal{S}$ attacking the ideal interaction, and
    \begin{align*} 
        |\mathrm{Pr}[\mathrm{Real}_{\Pi, \mathcal{A}} (\{\textbf{u}_{\mathcal{C}_H}\}) = 1] - \mathrm{Pr}[\mathrm{Ideal}_{\mathcal{F}, \mathcal{S}} (\mathcal{U}_H) = 1]| \leq \mathrm{negl}(\kappa),
    \end{align*}
    where $\mathcal{U}_H = \sum\nolimits_{\mathcal{C}_i \in \mathcal{C}_H} \textbf{u}_i$.
    \item Input Integrity. The protocol $\Pi$ outputs $\sum_{\mathcal{C}_i \in \mathcal{C}_{\mathrm{Valid}}} \textbf{u}_i$ with probability of at least $1 - \mathrm{negl}(\kappa)$, where $\mathcal{C}_H \subseteq \mathcal{C}_{\mathrm{Valid}}$. 
    For any malformed input $\textbf{u}_j$ from a malicious client $\mathcal{C}_j$, the probability that it passes the integrity check satisfies:
    \begin{align*}
        \mathrm{Pr}[ \mathcal{C}_j \in \mathcal{C}_\mathrm{Valid}] \leq F(D(\textbf{u}_j)).
    \end{align*}
\end{itemize}
\end{definition}

\ignore{
\begin{definition} \label{def:problem}
Given a security parameter $\kappa$, a function $D : \mathbb{R}^d \to \mathbb{R}$ satisfying that $\textbf{u}$ is malicious if and only if $f(\textbf{u}) > 1$, a function $F: (1, +\infty) \to [0,1]$, a set of inputs $\{\textbf{u}_1, \cdots, \textbf{u}_n\}$ from clients $\mathcal{C} = \{\mathcal{C}_1, \dots, \mathcal{C}_n\}$ respectively, and a list of honest clients $\mathcal{C}_H$, a protocol $\Pi$ is a $(\epsilon, D,F)$-relaxed secure aggregation with verified inputs, i.e., $(\epsilon, D,F)$-relaxed SAVI protocol, for $\mathcal{C}_H$ if:
\begin{itemize}[leftmargin=20pt]
    \item $\epsilon$-Input Privacy. For any probabilistic polynomial-time (PPT) algorithm $h$ that involves an execution of protocol $\Pi$ in which the malicious server $\mathcal{S}$ and the malicious clients $\mathcal{C}_M = \mathcal{C} \setminus \mathcal{C}_H$ collude, there is a PPT algorithm $f$ such that for any $x$, 
    \begin{align} 
        |\mathrm{Pr}[h(\{\textbf{u}_i\}_{i \in C_H}) = x] - \mathrm{Pr}[f(\sum_{i \in C_H} \textbf{u}_i) = x]| \leq \mathrm{negl}(\kappa) + \epsilon.
    \end{align}
    \item \needcheck{$(D,F)$-Relaxed Input Integrity. $\Pi$ outputs $\sum_{i \in \mathcal{C}_{\mathrm{Valid}}} \textbf{u}_i$ with probability at least $1 - \mathrm{negl}(\kappa)$, where $H \subseteq {\mathrm{Valid}}$. For any malformed input $\textbf{u}_i$ from a malicious client $\mathcal{C}_i$, the probability that it passes the integrity check satisfies:
    \begin{align}
        \mathrm{Pr}[ i \in \mathrm{Valid}] \leq F(D(\textbf{u}_i)).
    \end{align}
    }
\end{itemize}
\end{definition}
}
For input privacy in Definition~\ref{def:problem}, it ensures that the server can only learn the aggregation of honest clients' updates.
%
For input integrity, 
Definition~\ref{def:problem} relaxes the integrity verification by introducing a malicious pass rate function $F$. $F$ is a function that maps the degree of maliciousness of an input to the pass rate of the input, and the degree of maliciousness is measured by the function $D$. 
For instance, with an $L_2$-norm bound $B$, a natural choice is $D(\textbf{u}) = ||\textbf{u}||_2 / B$. Intuitively, the higher the degree of maliciousness, the lower the pass rate. 
So $F$ is usually decreasing. 
Our system also satisfies $\limsup_{x \to +\infty} F(x) \leq \mathrm{negl}(\kappa)$.
It means that any malicious client's malformed update whose degree of maliciousness passes a threshold can be detected with an overwhelming probability.
When $F \equiv \mathrm{negl}(\kappa)$, a protocol that satisfies $(D,F)$-relaxed SAVI will also satisfy SAVI. 

\subsection{Solution Overview} \label{subsec:overview}

To solve the problem in Definition~\ref{def:problem}, we propose a secure and verifiable federated learning system \ourtech{} with high efficiency. 
%
%
It tolerates $m < n / 2$ malicious clients for input integrity, which means that the server can securely aggregate the clients' inputs as long as a majority of the clients are honest. 
Figure~\ref{fig:rsfl-overview} gives an overview of \ourtech{}, which is composed of a system initialization stage and three iterative rounds: commitment generation, proof generation and verification, and secure aggregation. 
In the initialization stage (Figure~\ref{subfig:system-init}), all the parties agree on some hyper-parameters, such as the number of clients $n$, the maximum number of malicious clients $m$, the security parameters, e.g., key size, and so on. 
%

In each iteration of the FL training process, each client $\mathcal{C}_i (i \in [n])$ commits its model update $\textbf{u}_i$ using the hybrid commitment scheme based on Pedersen commitment and verifiable Shamir's secret sharing (VSSS) in Section~\ref{subsec:commit}, and then sends the commitment to the server and the secret shares to the corresponding clients (see Figure~\ref{subfig:commitment}). 
In the proof generation and verification round (Figure~\ref{subfig:prof-gen-ver}), there are two steps. 
In the first step, 
each client verifies the authenticity of other clients' secret shares. 
For the secret shares that are verified to be invalid, the client marks the respective clients as malicious.
With the marks from all clients, the server can then identify a subset of malicious clients. 
In the second step, the server uses a probabilistic integrity check method presented in Section~\ref{subsec:proof-gen-ver} to check each client's update $\textbf{u}_i$. 
Next, the server filters out the malicious client list $\mathcal{C}^*$ and broadcasts it to all the clients. 
In the secure aggregation round (Figure~\ref{subfig:secagg}), each client aggregates the secret shares from clients $\mathcal{C}_j (j \notin \mathcal{C}^*)$ and sends the result to the server. 
The server reconstructs the sum of secret shares and securely aggregates the updates $\textbf{u}_j (j \notin \mathcal{C}^*)$ based on the Pedersen commitments.
We summarize the process of each iteration in Algorithm~\ref{alg:risefl}, and we shall detail the steps in the following section.

\begin{algorithm}[t]
  \small
  \SetAlgoLined
  \SetKwFunction{CommitmentGeneration}{CommitmentGeneration}
  \SetKwFunction{ProofGenVer}{ProofGenerationAndVerification}
  \SetKwFunction{SecAgg}{SecureAggregation}
  \caption{RiseFL (One Iteration)}
  \label{alg:risefl}

  \KwIn{the number of clients $n$, the maximum number of malicious clients $m$, the number of random vectors $k$ }
  \KwOut{the aggregated model update $\textbf{u}$ of honest clients}
  \SetKwProg{Fn}{Procedure}{}{}
  \Fn{\CommitmentGeneration}{
    \textbf{Client $i$:} \\
    $\quad$ generate a random secret $r_i$ \\
    $\quad$ compute commitment $\textbf{y}_i = C(\textbf{u}_i, r_i)$ in Eqn~\ref{eq:commitment} \\
    $\quad$ share $r_i$ via VSSS and encrypt shares $r_ij, \forall j \in [1,n]$ \\
    $\quad$ send $\textbf{y}_i$ and $\mathsf{Enc}(r_{ij}), \forall j \in [1,n]$ with check string to server \\
    \textbf{Server:} \\
    $\quad$ send encrypted shares and check strings to clients
  }
  \SetKwProg{Fn}{Procedure}{}{}
  \Fn{\ProofGenVer}{
    \textit{Step 1: Verify the authenticity of secret shares (Section~\ref{subsubsec:vsss})} \\
    \textbf{Client $i$:} \\
    $\quad$ initialize local malicious client set $\mathcal{C}_i^* = \emptyset$ \\
    $\quad$ decrypt to get $r_{ji}, \forall j \in [1,n]$ and check their authenticity \\
    $\quad$ put clients do not pass the check to $\mathcal{C}_i^*$ and send it to server \\
    \textit{Step 2: Verify the integrity of each client's update (Section~\ref{subsubsec:integrity})} \\
    \textbf{Server:} \\
    $\quad$ initialize malicious client set $\mathcal{C}^* = \emptyset$ \\
    $\quad$ generate a random $s$ and broadcast it to clients \\
    $\quad$ sample $k+1$ random vectors $\textbf{A}=(\textbf{a}_0, \cdots, \textbf{a}_k)$ \\
    $\quad$ compute $h_t, \forall t \in [0,k]$ and broadcast it to clients \\ 
    \textbf{Client $i$:} \\
    $\quad$ compute $\textbf{A}$ according to $s$ \\
    $\quad$ compute proof $\pi$ based on and send it to server \\
    \textbf{Server:} \\
    $\quad$ verify $\mathcal{C}_i^*, \forall i \in [1,n]$ and the proofs \\
    $\quad$ update $\mathcal{C}^*$ and broadcast 
  }
  \SetKwProg{Fn}{Procedure}{}{}
  \Fn{\SecAgg}{
    \textbf{Client $i$:} \\
    $\quad$ sends aggregated shares $r_i'$ from honest clients to server \\
    \textbf{Server:} \\
    $\quad$ aggregate shares $r_i', \forall i \in [1,n]$ \\
    $\quad$ compute aggregated model update $\textbf{u}$ in Eqn~\ref{eq:derivation-agg} \\
  }
\end{algorithm}

\section{RiseFL Design} \label{sec:rsfl-design}

\ignore{
\begin{figure*}[thb!]
  \centering
  \begin{tikzpicture}
    \draw[semithick] (0,0) rectangle (17.5,3.6); 
    \node[anchor=north west] at (0,3.6) {
        \begin{minipage}{17.2cm}
                \begin{enumerate}[topsep=0pt,itemsep=0ex]
                    \item The client and server agree on independent $g, q \in \mathbb{G}$, $\textbf{w} \in \mathbb{G}^d$, $\textbf{f} \in \mathbb{G}^{2 k b_{\mathsf{max}}}$, the bound $B_0$ of sum of squares of inner products, the maximum number of bits $b_{\mathsf{max}}$ of $B_0$, the number of bits of each inner product $b_{\mathsf{ip}} < b_{\mathsf{max}}$. 
                    \item The client sends $z_i = g^{r_i} \in \mathbb{G}$ and $\textbf{y}_i = C(\textbf{u}_i, r_i) \in \mathbb{G}^d$ to the server.
                    \item The server randomly samples $\textbf{a}_0 \in \mathbb{Z}_p^d$ from the uniform distribution, $\textbf{a}_1, \dots, \textbf{a}_k \in \mathbb{Z}_p^d$ from the discrete normal distribution and sends $\textbf{A}$ to the client, where $\textbf{A}$ is the $(k+1) \times d$ matrix whose rows are $\textbf{a}_{0}, \dots, \textbf{a}_k$.
    \item The server computes $h_t = \prod_l w_l^{a_{tl}}$ for $t \in [0, k]$ and sends $\textbf{h} = (h_0, \dots, h_k)$ to the client. 
    \item The client 
    generates a proof {$\pi \leftarrow \mathsf{GenPrf}(g, q, \textbf{w}, \textbf{f}, b_{\mathsf{ip}}, b_{\mathsf{max}}, B_0, \textbf{A}, \textbf{h}, z_i, r_i, \textbf{u}_i)$} by Algorithm~\ref{alg:gen_prf} 
    and sends $\pi$ to the server. 
    \item The server checks $\pi$ using $\mathsf{VerPrf}(g, q, \textbf{w}, \textbf{f}, b_{\mathsf{ip}}, b_{\mathsf{max}}, B_0, \textbf{A}, \textbf{h}, z_i, \textbf{y}_i, \pi)$ in Algorithm~\ref{alg:ver_prf}.
    \end{enumerate}
        \end{minipage}
    };
  \end{tikzpicture}
  \caption{Probabilistic input integrity check 
  between the server and one client. \red{to be updated}}
  \label{fig:wf-protocol}
\end{figure*}
}

In this section, we introduce our system design. 
We first present the rationale of the protocol in Section~\ref{subsec:rationale}. 
Then, we introduce the initialization stage in Section~\ref{subsec:init} and the three steps in each iteration of the training stage in Sections~\ref{subsec:commit}-\ref{subsec:secagg}, respectively. 
Finally, we discuss the extension of our system in Section~\ref{subsec:extension}.

\ignore{
Then, we introduce the hybrid commitment scheme and probabilistic integrity check method in Sections~\ref{subsec:hybrid-scheme} and \ref{subsec:sampling-check}, respectively. Finally, we detail the protocol in Section~\ref{subsec:protocol}.
}

\subsection{Rationale} \label{subsec:rationale}
The most relevant work to our problem is \eiffel{}~\cite{roy2022eiffel}, which also ensures input privacy and integrity in FL training. 
However, its efficiency is low, and thus is impractical to be deployed in real-world systems. 
For example, under the experiment settings in Section~\ref{sec:experiment}, given 100 clients and 1K model parameters, \eiffel{} takes around 15.3 seconds for proof generation and verification on each client. 
More severely, the cost is increased to 152 seconds when the number of model parameters $d$ is 10K. 
%
The underlying reason is that the complexity of its proof generation and verification is linearly dependent on $d$, making \eiffel{} inefficient and less scalable. 
We shall detail the cost analysis of \eiffel{} in Section~\ref{subsec:cost}. 
\ignore{
Specifically, in \eiffel{}, each client $\mathcal{C}_i (i \in [n])$ uses VSSS to share each coordinate of its update $\textbf{u}_i$, i.e., $O(d)$, with all the clients as commitments. To ensure the secret shares can tolerate $m$ malicious clients, each client needs to perform $O(m)$ cryptographic group exponentiations to calculate the check strings for each coordinate. Thus, the complexity for commitment generation is $O(dm)$.
%
Also, \eiffel{} adopts the secret-shared non-interactive proof (SNIP)~\cite{Corrigan-GibbsB17} for ZKP generation and verification, which requires $O(bmnd)$ field multiplications, where $b$ is the bit length of weight updates. \needcheck{Should say details are in Section~\ref{sec:analysis}?}
%
These make \eiffel{} inefficient and less scalable. 
}

\begin{figure}[t]
    \centering
    \subfloat[Sample random vectors and \\ compute inner products]{{\includegraphics[width=0.21\textwidth]{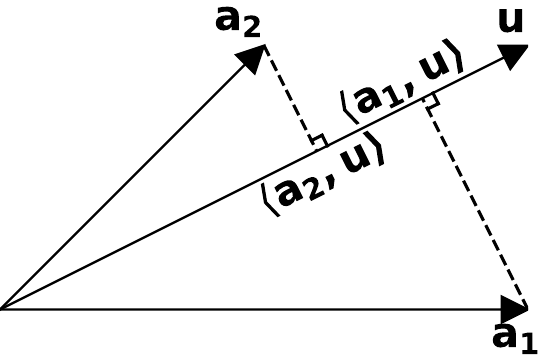} } \label{subfig:prob_1}}%
    \subfloat[Check the bound of the sum of \\ inner products ]{\includegraphics[width=0.23\textwidth]{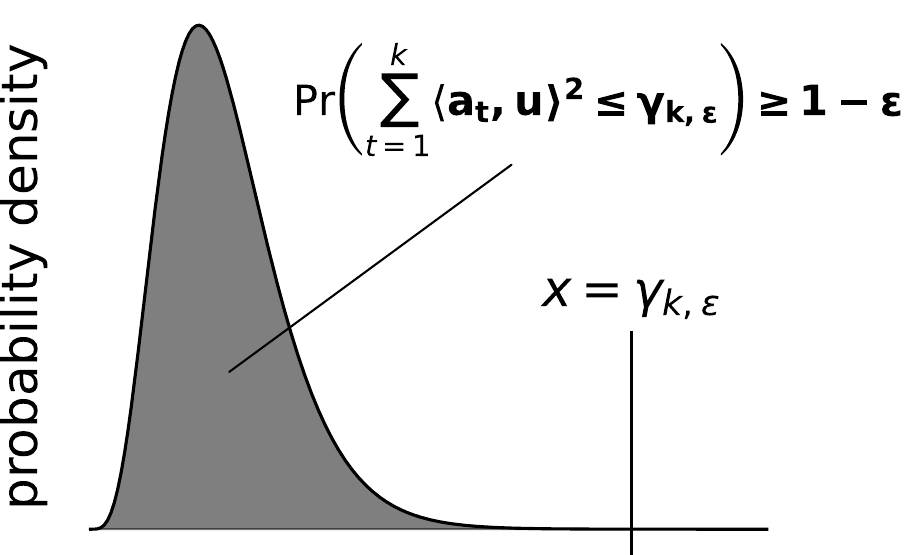} \label{subfig:prob_2} }%
    \caption{The probabilistic check when $||\textbf{u}||_2 = 1$.}
    \label{fig:prob}
\end{figure}

The rationale behind our idea is to reduce the complexity of expensive group exponentiations in ZKP generation and verification. 
To do so, we design a probabilistic $L_2$-norm integrity check method, as shown in Algorithm~\ref{alg:prob_check}. 
%
%
The intuition is that, instead of generating and verifying proofs for the $L_2$-norm of an update ${|| \textbf{u} ||}_2$, where $\textbf{u} \in \mathbb{R}^d$, we randomly sample $k$ points $\textbf{a}_1,\dots, \textbf{a}_k$ from the normal distribution $\mathcal{N}(\mathbf{0}, \mathbf{I}_d)$. 
Then random variable is:
%
\begin{align} 
   \frac{1}{||\textbf{u}||_2^2} \sum\nolimits_{i=1}^k  \langle \textbf{a}_i, \textbf{u} \rangle ^2, \label{eq:prob-check}
\end{align}
which follows a chi-square distribution $\chi_k^2$ with $k$ degrees of freedom.
In Algorithm \ref{alg:prob_check}, if $||\textbf{u}||_2 \leq B$, then the probability that $\textbf{u}$ passes the check is at least $1 - \epsilon$.
%
where $\epsilon$ is chosen to be cryptographically small, e.g., $2^{-128}$.
In this way, the probability that the client fails the check is of the same order as the probability that the client's encryption is broken.
%
\ignore{
We shall present the detailed constructions of proof generation and verification based on this method in Section~\ref{subsec:sampling-check}. 
Moreover, we propose a hybrid commitment scheme to efficiently support this probabilistic check method, as will be presented in Section~\ref{subsec:hybrid-scheme}. As a consequence, we can significantly reduce the cryptographic operation costs from $O(d)$ to $O(d/\log d)$.}
Figure~\ref{fig:prob} gives an illustration of this method.
Assume that $||u||_2=1$. We sample two random normal samples $\textbf{a}_1, \textbf{a}_2$; then, the inner products $\langle \textbf{a}_1, \textbf{u} \rangle, \langle \textbf{a}_2, \textbf{u} \rangle$ are the projections of $\textbf{a}_1, \textbf{a}_2$ onto $\textbf{u}$. Figure~\ref{subfig:prob_2} shows the probability density function of $\chi_k^2$. With a specific bound $\gamma_{k, \epsilon}$ (we shall introduce this bound in Section~\ref{subsec:security}), there is an overwhelming probability such that $\sum_{t=1}^k \langle \textbf{a}_t, \textbf{u} \rangle^2\leq \gamma_{k, \epsilon}$.

Based on this method, we further devise an optimization technique, as will be introduced in Section~\ref{subsec:proof-gen-ver}, which can 
significantly reduce the cryptographic operation costs from $O(d)$ to $O(d/\log d)$. 
We shall present the details in the following subsections.

%

\ignore{
Second, in order to reduce the commitment generation cost, we propose a hybrid commitment scheme, which uses Pedersen commitment for each client to commit each coordinate in its update and uses VSSS to commit a single secret for Byzantine-tolerant aggregation. This scheme reduces the commitment generation complexity to $O(d)$, while supporting the probabilistic $L_2$-norm bound check method efficiently. We shall detail this scheme in Section~\ref{subsec:hybrid-scheme}.
}

\ignore{
\begin{algorithm}[t]
  \caption{Probabilistic $L_2$-norm bound check}\label{alg:prob_check}
  \begin{algorithmic}[1]
    \STATE \textbf{Input:} $\textbf{u} \in \mathbb{R}^d$, $B$, $k$, $\epsilon$
    \STATE \textbf{Output:} ``Pass'' or ``Fail'' 
    \STATE Sample $\textbf{a}_1, \dots, \textbf{a}_k \in \mathbb{R}^d$ i.i.d.\ from $\mathcal{N}(\mathbf{0}, \mathbf{I}_d)$ 
    \STATE Compute $\gamma_{k,\epsilon}$ which satisfies that $\mathrm{Pr}_{t \sim \chi_k^2}[t < \gamma_{k,\epsilon}] = 1 - \epsilon$.
    \IF {$\sum_{i=1}^k  \langle \textbf{a}_i, \textbf{u} \rangle ^2 \leq B^2 \gamma_{k,\epsilon}$}
      \STATE Return ``Pass''
    \ELSE
      \STATE Return ``Fail''
     \ENDIF
\end{algorithmic}
\end{algorithm}
}

\begin{algorithm}[t]
  \small
  \SetAlgoLined
  \caption{Probabilistic $L_2$-norm bound check}
  \label{alg:prob_check}

  \KwIn{$\textbf{u} \in \mathbb{R}^d$, $B$, $k$, $\epsilon$}
  \KwOut{``Pass'' or ``Fail''}
  Sample $\textbf{a}_1, \dots, \textbf{a}_k \in \mathbb{R}^d$ i.i.d.\ from $\mathcal{N}(\mathbf{0}, \mathbf{I}_d)$ \\
  Compute $\gamma_{k,\epsilon}$ which satisfies that $\mathrm{Pr}_{t \sim \chi_k^2}[t < \gamma_{k,\epsilon}] = 1 - \epsilon$ \\
  \eIf{$\sum_{i=1}^k  \langle \textbf{a}_i, \textbf{u} \rangle ^2 \leq B^2 \gamma_{k,\epsilon}$}{
  \KwRet{``Pass''}
  } {
  \KwRet{``Fail''}
  }
\end{algorithm}

\subsection{System Initialization} \label{subsec:init}

In this stage, all parties (the server and the clients) are given the system parameters, including the number of clients $n$, the maximum number of malicious clients $m$, the bound on the number of bits $b_{\text{ip}}$ of each inner product,
the maximum number of bits $b_{\text{max}} > b_{\text{ip}}$ of the sum of squares of inner product,  the bound of the sum of inner products $B_0 < 2^{b_{\text{max}}}$, the number of samples $k$ for the probabilistic check, a set of independent group elements $g,q \in \mathbb{G}, \textbf{w} = (w_1,\cdots,w_d) \in \mathbb{G}^d$, the factor $M>0$ used in discretizing the normal distribution samples, and a cryptographic hash function $H(\cdot)$. 
Note that $b_{\text{ip}}$ and $b_{\text{max}}$ define the maximum values of $\langle \textbf{a}_i, \textbf{u} \rangle$ and $\sum\nolimits_{i=1}^k  \langle \textbf{a}_i, \textbf{u} \rangle ^2$ in Equation~\ref{eq:prob-check}, respectively. 

Since there is no direct channel between any two clients in the FL setting considered in this paper, we let the server forward some of the messages. To prevent the server from accessing the secret information, each client $\mathcal{C}_i (i \in [n])$ generates a public/private key pair $(pk_i, sk_i)$ and sends the public key $pk_i$ to a public bulletin. 
Subsequently, each client fetches the other clients' public keys such that each pair of clients can establish a secure channel via the Diffie-Hellman protocol~\cite{Merkle78} for exchanging messages securely.

\subsection{Commitment Generation} \label{subsec:commit}

Recall that in each iteration of the FL training process, each client $\mathcal{C}_i (i \in [n])$ obtains a model update $\textbf{u}_i$ via local training on its dataset $\mathcal{D}_i$.
In order to prove to the server that the $L_2$-norm of $\textbf{u}_i$ is within a bound $B_0$, the client $\mathcal{C}_i$ needs to commit its update $\textbf{u}_i$ before generating the proofs. 
In \ourtech{}, the server is expected to not only identify malicious clients, but also aggregate well-formed model updates. 
%
Therefore, we propose a novel hybrid commitment scheme based on Pedersen commitment and VSSS, where VSSS is used to protect the random secret in the Pedersen commitment for secure aggregation.  

\vspace{1mm}
\noindent
\textbf{Hybrid commitment scheme.} 
Note that the clients and server agree on independent group elements $g, w_1, \dots, w_d \in \mathbb{G}$, where $w_j (j \in [d])$ is used for committing the $j$-th coordinate in $\textbf{u}_i$. 
Then, $\mathcal{C}_i$ generates a random secret $r_i \in \mathbb{Z}_p$ and encrypts $\textbf{u}_i$ with Pedersen commitment as follows:
\begin{align} \label{eq:commitment}
    \textbf{y}_i = C(\textbf{u}_i, r_i) & = (C({u_{i1}}, r_i), \dots, C({u_{id}}, r_i)) \nonumber \\ 
    & =  (g^{u_{i1}} w_1^{r_i}, \dots, g^{u_{id}} w_d^{r_i}),
\end{align}
where $u_{ij}$ is the $j$-th coordinate in $\textbf{u}_i$. Each client $\mathcal{C}_i$ sends $\textbf{y}_i = C(\textbf{u}_i, r_i)$ and $z_i = g^{r_i}$ to the server as commitments. Given that $r_i$ is held by each client $\mathcal{C}_i$, the server knows nothing regarding each update $\textbf{u}_i$. 

To facilitate the server to aggregate well-formed updates, we also require each client $\mathcal{C}_i$ to share its secret $r_i$ with other clients using VSSS. Specifically, $\mathcal{C}_i$ computes $((1, r_{i1}), \dots, (n, r_{in}), \Psi_{r_i}) \leftarrow \mathsf{SS.Share}(r_i, n, m+1, g)$ and sends $((j, r_{ij}), \Psi_{r_i})$ to $\mathcal{C}_j$, where $g^{r_i} = \Psi_{r_i}(0)$.
Since the clients do not have communication channels, $\mathcal{C}_i$ encrypts each share $r_{ij}$ with the corresponding symmetric key established by $(pk_j, sk_i)$ for each client $\mathcal{C}_j (j \in [1,n])$, obtaining $\mathsf{Enc}(r_{ij})$. 
The encrypted $\mathsf{Enc}(r_{ij})$ ensures that only $\mathcal{C}_j$ can decrypt it. 
Finally, $\mathcal{C}_i$ sends the commitment $\textbf{y}_i$, the encrypted share $\mathsf{Enc}(r_{ij}) (j \in [1,n])$, and the check string $\Psi_{r_i}$ to the server.
For example, suppose in Figure~\ref{subfig:commitment}, Client 2 has a model update $\textbf{u}_2 = (5,1)$ and its generates a random secret $r_2 = 10$. Let $w_1 = g^3, w_2 = g^4$ (ignoring the independence requirement of $g, w_1, w_2$). Client 2 can commit $\textbf{y}_2 = (g^5 \cdot w_1^{10}, g^1 \cdot w_2^{10}) = (g^{35}, g^{41})$. Then, it can generate secret shares of $r_2 = (r_{21}, r_{22}, r_{23}) = (3, -4, -11)$ using the polynomial $-7x+10$, where $r_{21} = 3$ and $r_{23} = -11$ will be encrypted using keys shared with Client 1 and Client 3 respectively, before they are sent to the server together with the check string. 

After receiving the messages from all clients, the server forwards the encrypted shares $\mathsf{Enc}(r_{ij}) (i \in [1,n])$ to client $\mathcal{C}_j$, and broadcasts the check strings $\Psi_{r_i} (i \in [1,n])$ to all clients.
We will introduce how the server aggregates well-formed updates from honest clients based on these commitments in Section~\ref{subsec:secagg}.


\subsection{Proof Generation and Verification} \label{subsec:proof-gen-ver}

In this round, the clients and the server jointly flag the malicious clients. 
Specifically, the server initializes a list $\mathcal{C}^* = \emptyset$ to record the malicious clients that will be identified in the current iteration. 
This round consists of two steps: verifying the authenticity of secret shares and verifying the integrity of each client's update. 

\subsubsection{Verify the authenticity of secret shares.}\label{subsubsec:vsss}
After receiving the encrypted shares $\mathsf{Enc}(r_{ji})$ and check strings $\Psi_{r_j}$ for $j \in [1,n]$, each client $\mathcal{C}_i$ decrypts $r_{ji}$ and checks against $\Psi_{r_j}$. 
If a check fails, $\mathcal{C}_i$ marks the corresponding client $\mathcal{C}_j$ as malicious because the share is not authenticated. 
Subsequently, $\mathcal{C}_i$ sends a list of candidate malicious clients that do not pass the check to the server.

The server follows two rules to flag the malicious clients~\cite{roy2022eiffel}. 
First, if a client $\mathcal{C}_i$ flags more than $m$ clients as malicious or is flagged as malicious by more than $m$ clients, the server puts $\mathcal{C}_i$ to $\mathcal{C}^*$. 
This is because there are at most $m$ malicious clients in the system. 
An honest client cannot mark more than $m$ clients as malicious clients and will not be marked as malicious by more than $m$ clients. For example, in Figure~\ref{subfig:prof-gen-ver}, Client 3 marks the other two clients as malicious, then the server can ensure that this client is malicious.

Second, if a client $\mathcal{C}_i$ is flagged as malicious by $[1,m]$ clients, the server requests the shares $r_{ij}$ in the clear for all $\mathcal{C}_j$ that flags $\mathcal{C}_i$ and checks against $\Psi_{r_i}$, if the clear $r_{ij}$ fails the check, the server puts $\mathcal{C}_i$ to $\mathcal{C}^*$. 
Note that if client $\mathcal{C}_i$ receives more than $m$ requests of the clear-text shares $r_{ij}$ from the server, it marks the server as malicious and quits the protocol because this should not happen. 
If $\mathcal{C}_i$ receives $m'\in[1,m]$ requests of clear-text $r_{ij}$, it sends them to the server. 
Under the assumption of at most $m$ malicious clients, there are at most $m-m'$ malicious clients left. Even if all of the malicious clients collude with the server, at most $m$ values of $r_{ij}$ would be revealed, so $r_i$ remains secure. 
%

\subsubsection{Verify the integrity of each client's update.}\label{subsubsec:integrity}
Next, we present how the server and clients jointly execute the probabilistic integrity check described in Section~\ref{subsec:rationale}.
Without loss of generality, we describe the check for one client $\mathcal{C}_i$. 

\vspace{1mm}
\noindent \textbf{Sampling $k$ random vectors.}
At the beginning, the server and clients need to agree on $k$ random samples $\textbf{a}_1, \cdots, \textbf{a}_k$ with dimensionality $d$. 
Intuitively, we can let the server sample the $k$ vectors and broadcast them to all clients. 
However, there are two main issues. 
First, the communication cost can be very high. 
As will be discussed in Section~\ref{subsec:security}, the choice of $k$ typically requires to be several thousand. 
Therefore, it is communication inefficient for the server to broadcast $k$ vectors (each vector contains $d$ elements).
Second, the server may select special vectors that are not random, i.e., the vectors may allow the server to infer the client's update easily. 
For example, if the server selects $\textbf{a}_t = (1, 0, \dots, 0)$ for all $t\in [1,k]$, it may infer the range of the first element in a client's update.
%
To solve these issues, we let the server select a random value $s$ and broadcast it to all clients. 
Based on $s$, the server and each client $\mathcal{C}_i (i \in [n])$ can compute a seed $H(s, \{pk_i\}_{1 \leq i \leq n})$ using $s$ and all clients' public keys. 
Hence, the clients and the server can generate the same set of random samples because the seed is the same. 
%
Note it is computationally infeasible to find a seed that produces specific vectors since $H(\cdot)$ is a cryptographic hash function.
%

\vspace{1mm}
\noindent \textbf{Optimization of cryptographic operations.}
Given the random vectors $\textbf{a}_1, \cdots, \textbf{a}_k$, a naive way of realizing the probabilistic bound check in Algorithm~\ref{alg:prob_check} is to let client $\mathcal{C}_i$ commit to $\langle \textbf{a}_t, \textbf{u}_i\rangle$ with $z_t = g^{\langle \textbf{a}_t, \textbf{u}_i \rangle} w_{d+t} ^{r_i}$ for $t \in [1,k]$ based on additional independent group elements $w_{d+1}, \dots, w_{d+k}$. It can then generate a $\Sigma$-protocol~\cite{camenisch1997proof} proof that the secret in $z_t$ is indeed the inner product of the secrets in $\textbf{y}_i$ and $\textbf{a}_t$ for $t \in [1,k]$ and a proof that the sum of the squares of the secrets in $z_t$ is bounded. 
However, its computational cost will be $O(d)$ in terms of group exponentiations. 
We thus propose an optimization to reduce the cost of proof generation to $O(d/\log d)$ group exponentiations. 

We observe that the following term $e_{t} (t \in [1,k])$ is used in the inner product proof generation and verification:
\begin{align} \label{eq:commit-inner-prod}
    e_{t} := y_{i1}^{a_{t1}} \dots y_{id}^{a_{td}} & = (g^{u_{i1}} w_1^{r_i})^{a_{t1}} \dots (g^{u_{id}} w_d^{r_i})^{a_{td}} \\
    & = g^{\langle \textbf{a}_t, \textbf{u}_i \rangle} (w_1^{a_{t1}} \dots w_d^{a_{td}})^{r_i}.
\end{align}
Let $h_t = w_1^{a_{t1}} \dots w_d^{a_{td}}$, which is the same for all the clients as the random vector $\textbf{a}_t$ and the group elements $w_1, \cdots, w_d$ are shared by the server and clients. 
Thus, we let the server precompute $h_t$ for $t \in [1,k]$ and broadcast them to all clients before they generate proofs.
Given $h_t$, the cost of computing the values $e_t$ in Equation~\ref{eq:commit-inner-prod} for $t \in [1, k]$ is $O(k)$ group exponentiations 
and $O(kd)$ finite field operations. 
Still, the client cannot blindly compute the value in Equation~\ref{eq:commit-inner-prod} by $e_t = g^{\langle \textbf{a}_t, \textbf{u}_i \rangle} h_t^{r_i}$ as $h_t$ is received from the server, which could be malicious. 
Therefore, we let the client use batch verification\footnote{To check whether $x_1=\dots = x_n = 1$, it is enough to randomly sample $\alpha_1, \dots, 
\alpha_n$ from the uniform distribution on $\mathbb{Z}_p$ and check whether $x_1^{\alpha_1} \dots x_n^{\alpha_n} = 1$.} to check whether $h_t =  w_1^{a_{t1}} \dots w_d^{a_{td}}$ for $t \in [1,k]$. 
The cost of batch verification is $O(d/\log d)$ group exponentiations and $O(kd)$ finite field operations. 
%
Note that the group exponentiations are the major cost as they are much more complex than finite field operations. 
The reduction of the number of group exponentiations from $O(d)$ to $O(d/\log d)$ significantly reduces client cost, while the additional cost of $O(kd)$ finite field operations is not large.\footnote{The corresponding micro-benchmark results will be presented in Section~\ref{subsec:exp-micro}.}

To make sure that $e_t$ is indeed the commitment of the inner product of $\textbf{a}_t$ and the secret in $\textbf{y}_i$, the server uses batch verification to check that the values $e_t$, $t \in [1,k]$, submitted by client $\mathcal{C}_i$ satisfy Equation~\ref{eq:commit-inner-prod}: $e_t  \stackrel{?}{=}  y_{i1}^{a_{t1}} \dots y_{id}^{a_{td}}$. 
The only issue at this step is that even if all of the equations $e_t = y_{i1}^{a_{t1}} \dots y_{id}^{a_{td}}$ are satisfied, the server is still not sure that client $\mathcal{C}_i$ possesses a value $\textbf{u}_i$ that is used to produce $\textbf{y}_i$.
To address this issue, the server additionally samples $\textbf{a}_0 \in \mathbb{Z}_p^d$ from the uniform distribution on $\mathbb{Z}_p$ with cryptographically secure pseudo-random number generator (PRNG), computes the corresponding $h_0 = \prod_l w_l^{a_{il}}$ and broadcasts it together with $h_1, \dots, h_k$. 
The client needs to additionally verify the correctness of $h_0$ together with $h_1, \dots, h_k$ by batch verification, which is detailed in Algorithm~\ref{alg:ver_crt}. 
The server then uses batch verification in Algorithm~\ref{alg:ver_crt} to check the correctness of $e_t  \stackrel{?}{=}  y_{i1}^{a_{t1}} \dots y_{id}^{a_{td}}$ for $t \in [0,k]$. With the additional commitment $e_0$, the server can acquire an additional $\Sigma$-protocol proof that client $\mathcal{C}_i$ possesses a value $\gamma$ that satisfies $e_0 = g^{\gamma} h_0^{r_i}$. Once this additional proof is satisfied, the server is sure that client $\mathcal{C}_i$ possesses $\textbf{u}_i$ that satisfies $y_{il} = g^{u_{il}} w_l^{r_i}$, and that $\gamma = \langle \textbf{a}_0, \textbf{u}_i \rangle$. 
Therefore, the secrets in $e_t$ are indeed the inner product between $\textbf{a}_t$ and $\textbf{u}_i$, $t \in [0,k]$.

Finally, the commitment $e_t$ uses group element $h_t$, which is different for different $t$. In order to verify the bound of the sum of the squares of the secrets in $e_t$, we need to fix another independent group element $q$ and convert $e_t$ to another commitment $o_t=g^{\langle \textbf{a}_t, \textbf{u}_i \rangle} q^{s_t}$, where $s_t$ is another blind chosen by client $\mathcal{C}_i$. $o_t$, $t \in [1,k]$ are all based on $q$. We then proceed to check the bound of the squares of the secrets in $o_t$.

\ignore{
\begin{algorithm}[t]
  \caption{{$\mathsf{VerCrt}(\textbf{w}, \textbf{h}, \textbf{A})$}}
  \label{alg:ver_crt}
  \begin{algorithmic}[1]
\STATE \textbf{Input:} $\textbf{w} = (w_1, \dots, w_d) \in \mathbb{G}^d$, $\textbf{h}  = (h_0, \dots, h_k) \in \mathbb{Z}_p^{(k+1)}$, $\textbf{A} \in \mathbb{M}_{(k+1) \times d}(\mathbb{Z}_p)$. 
\STATE Randomly Sample $\textbf{b} = (b_0, \dots, b_k) \in \mathbb{Z}_p^{k+1}$.
\STATE Compute $\textbf{c} = (c_1, \dots, c_d) =  \textbf{b} \cdot \textbf{A} \in \mathbb{Z}_p^d$. 
    \RETURN $h_0^{b_0} \dots h_k^{b_k} == w_1^{c_1} \dots w_d^{c_d}$.
\end{algorithmic}
\end{algorithm}
}

\begin{algorithm}[t]
  \small
  \SetAlgoLined
  \caption{{$\mathsf{VerCrt}(\textbf{w}, \textbf{h}, \textbf{A})$}}
  \label{alg:ver_crt}
  
  \KwIn{$\textbf{w} = (w_1, \dots, w_d) \in \mathbb{G}^d$, $\textbf{h}  = (h_0, \dots, h_k) \in \mathbb{Z}_p^{(k+1)}$, $\textbf{A} \in \mathbb{M}_{(k+1) \times d}(\mathbb{Z}_p)$.}
  Randomly Sample $\textbf{b} = (b_0, \dots, b_k) \in \mathbb{Z}_p^{k+1}$. \\
  Compute $\textbf{c} = (c_1, \dots, c_d) =  \textbf{b} \cdot \textbf{A} \in \mathbb{Z}_p^d$. \\
  \KwRet{$h_0^{b_0} \dots h_k^{b_k} == w_1^{c_1} \dots w_d^{c_d}$.}
\end{algorithm}

\begin{figure}[t]
    \centering
    \includegraphics[width=0.49\textwidth]{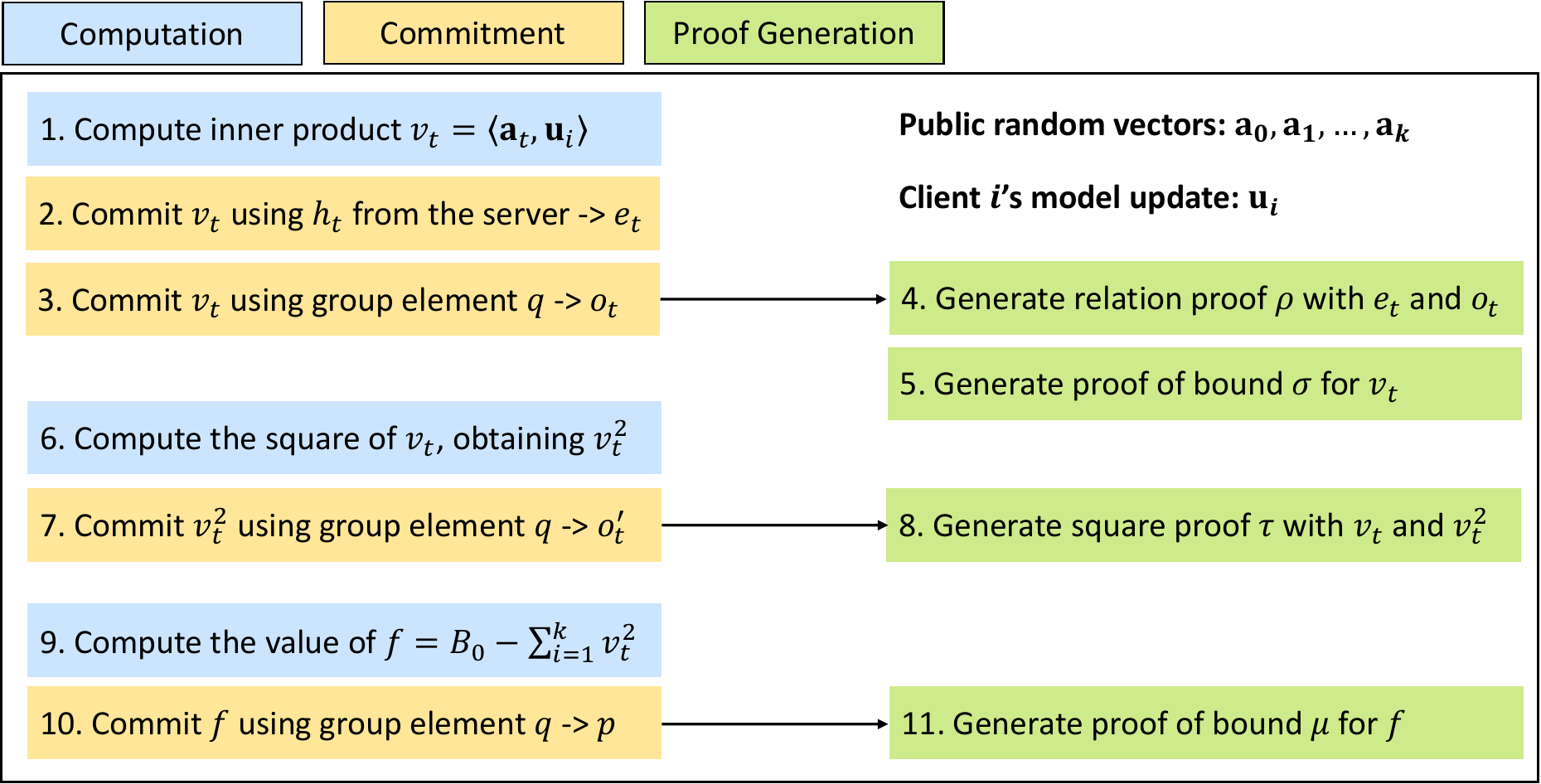}
    \caption{An illustration of the ZKP proof generation on the $i$-th client.}
    \label{fig:zkp}
\end{figure}

\vspace{1mm}
\noindent \textbf{Probabilistic input integrity verification.}
Now we detail the probabilistic input integrity verification in Algorithm~\ref{alg:prob_check}.
%
Assume the server and clients generate $k+1$ random samples $\textbf{A} = (\textbf{a}_0, \textbf{a}_1, \cdots, \textbf{a}_k) \in \mathbb{Z}_p^d$ using the aforementioned techniques.
\ignore{
$\textbf{a}_0$ is sampled from the uniform distribution on $\mathbb{Z}_p$ with cryptographically secure pseudo-random number generator (PRNG) for checking the integrity of Pedersen commitments $\textbf{y}_i$. 
$\textbf{a}_1, \dots, \textbf{a}_k$ are sampled from the discrete normal distribution with insecure PRNG for fast execution of probabilistic check in Algorithm~\ref{alg:prob_check}. 
%
}
After that, the server computes $h_t = \prod_l w_l^{a_{tl}}$ for $t \in [0,k]$. 
%
%
Let $\textbf{h} = (h_0, h_1, \cdots, h_k)$. 
The server sends $\textbf{h}$ to the client. 
Upon receiving the information, the client first verifies the correctness of $\textbf{h}$ using $\mathsf{VerCrt}(\textbf{w}, \textbf{h}, \textbf{A})$ in Algorithm~\ref{alg:ver_crt}. 
%
%
If it is correct, the client computes the following items for generating the proof that Eqn~\ref{eq:prob-check} is less than the bound $B_0$. An illustration of the client's proof generation is given in Figure~\ref{fig:zkp}.
%
\begin{itemize}[leftmargin=*]
    \item The client computes the inner products between $\textbf{u}_i$ and each row of $\textbf{A}$, obtaining $\textbf{v}^* = (v_0, v_1, \cdots, v_k)$, where $v_t = \langle \textbf{a}_t, \textbf{u}_i \rangle$ for $t \in [0, k]$. The client commits $e_t = g^{v_t} h_t^{r_i}$ using its secret $r_i$ for $t \in [0, k]$. Let $\textbf{e}^* = (e_0, e_1, \cdots, e_k)$ and $\textbf{e} = (e_1, \cdots, e_k)$. The commitment $e_0$ is used for integrity check of $\textbf{y}_i$. The commitments $\textbf{e}$ are used for bound check of $\textbf{v} = (v_1, \cdots, v_k)$.
    \item The client commits $v_t$ using $o_t = g^{v_t} q^{s_t}$ for $t \in [1, k]$, where $s_t$ is a random number. Let $\textbf{o} = (o_1, \cdots, o_k)$ be the resulted commitment. 
    Note that $e_t$ and $o_t$ commit to the same secret $v_t$ using different group elements $h_t$ and $q$. 
    \item The client generates a proof $\rho$ to prove that $(z, \textbf{e}^*, \textbf{o})$ is well-formed, which means that the secret in $z$ is used as the blind in $e_t$, $t \in [0,k]$, and that the secrets in $e_t$ and $o_t$ are equal, $t \in [1, k]$.
    Note that $z = g^{r_i} = \Psi_{r_i}(0)$ is the $0$-th coordinate of the check string of Shamir's share of $r_i$. 
    \item The client generates a proof $\sigma$ that the secret in $o_t$ is in the interval $[-2^{b_{\mathsf{ip}}}, 2^{b_{\mathsf{ip}}})$ for $t \in [1, k]$. This ensures the inner product of $\textbf{a}_t$ and $\textbf{u}_i$ does not cause overflow when squared. 
    \item The client commits $o_t' = g^{v_t^2} q^{s_t'}$ for $t \in [1, k]$, where $s_t'$ is a random number. Let $\textbf{o}' = (o_1', \cdots, o_k')$ be the resulted commitment. This commitment will be used in the proof generation and verification for proof of square. 
    \item The client generates a proof $\tau$ to prove that the secret in $o_t'$ is the square of the secret in $o_t$ for $t \in [1, k]$, using the building block described in Section~\ref{sec:preliminary}.
    \item The client computes $B_0 - \sum_t v_t^2$ and further commits it with $p = g^{B_0} (\prod_t o_t')^{-1}$.
    \item The client generates a proof $\mu$ that $B_0 - \sum_t v_t^2$ is in the interval $[0, 2^{b_{\mathsf{max}}})$. This proof is to guarantee that Eqn~\ref{eq:prob-check} is less than the bound of the probabilistic check. 
\end{itemize}

Finally, the client sends the proof $\pi = (\textbf{e}^*, \textbf{o}, \textbf{o}', p, \rho, \tau, \sigma, \mu)$ to the server for verifcation.

After receiving the proof, the server verifies it accordingly, including checking the correctness of $\textbf{e}^*$ using Algorithm~\ref{alg:ver_crt}, checking the well-formedness proof $\rho$, checking the square proofs of $(\textbf{o}', \textbf{o})$, and checking the two bound proofs. 
If all the checks are passed, the server guarantees that the client's update passes the probabilistic check in Algorithm~\ref{alg:prob_check}.
%
%
%
Consequently, the server can verify the proof of each client $\mathcal{C}_i$ 
and put it to the malicious client list 
$\mathcal{C}^*$ if the verification fails. 
The list $\mathcal{C}^*$ is broadcast to all the clients.

\subsection{Secure Aggregation} 
\label{subsec:secagg}

Let $\mathcal{H} = \mathcal{C} \setminus \mathcal{C}^*$ be the set of honest clients. 
In this round, each client $\mathcal{C}_i \in \mathcal{H}$ selects the corresponding secret shares from the honest clients $\mathcal{C}_j \in \mathcal{H}$, aggregates the shares $r_i' = \sum_{\mathcal{C}_j \in \mathcal{H}} r_{ji}$, and sends $r_i'$ to the server. 
The server uses $\mathsf{SS.Verify}$ to verify the integrity of each $r_i'$ and uses $\mathsf{SS.Recover}$ to recover $r' = \sum_{\mathcal{C}_i \in \mathcal{H}} r_i'$. 
%
According to the homomorphic property of VSSS, $r' = r = \sum_{\mathcal{C}_i \in \mathcal{H}} r_i$ is the summation of the honest clients' secrets.
The summation will be used for calculating the aggregation of honest clients' model updates $\mathcal{U} = \sum\nolimits_{\mathcal{C}_i \in \mathcal{H}} \textbf{u}_i$.

%
Specifically, the server executes the following steps. 
For each dimension $l \in [d]$ in the updates, the server multiplies the commitments from honest clients $\mathcal{C}_j \in \mathcal{H}$ and obtains:
\begin{align} \label{eq:commitment-aggregation}
    C(\mathcal{U}, r) & = (\prod\nolimits_{\mathcal{C}_i \in \mathcal{H}} C({u_{i1}}, r_i), \dots, \prod\nolimits_{\mathcal{C}_i \in \mathcal{H}} C({u_{id}}, r_i) ).  
\end{align}
Without loss of generality, we consider the $l$-th dimension and derive the calculation as follows.
\begin{align}
    \prod\nolimits_{\mathcal{C}_i \in \mathcal{H}} C({u_{il}}, r_i) 
    & = g^{\sum_{\mathcal{C}_i \in \mathcal{H}} u_{il}} w_l^{\sum_{\mathcal{C}_i \in \mathcal{H}} r_i} \label{eq:derivation-prod} \\
    & = g^{u_l} w_l^{r}, \label{eq:derivation-agg}
\end{align}
where $u_l$ is the aggregation of the $l$-th elements in honest clients' updates, $w_l$ is the common group element used for the commitment in Eqn~\ref{eq:commitment}, and $r$ is the summation of honest clients' secrets. 
Note that the derivation of Eqn~\ref{eq:derivation-prod} is due to clients using the same group elements $g$ and $w_l$.
Since the server has already calculated $r = r'$, it can thus compute $g^{u_l}$ for $l \in [d]$ and solve $u_l$ according to Eqn~\ref{eq:derivation-agg}.

\ignore{
\subsection{Protocol Description} \label{subsec:algorithm}
The protocol consists of four stages. The first stage is system initialization. At this stage, the server and the clients agrees on public parameters and the clients set up public/private key pairs as in Section~\ref{subsec:init}.

The second stage is commitment generation. At this stage, each client $\mathcal{C}_i$ commits its model update $\textbf{u}_i$ using a secret $r_i$, sends the commitment $\textbf{y}_i$ to the server, and send VSSS of $r_i$ to the server. The server forwards all the VSSS check strings $\Psi_{r_i}$ to all the clients. The details are in Section~\ref{subsec:commit}.

The third stage is proof generation and verification. At this stage, the server checks the proofs of well-formedness from all the clients using the probabilistic check, outlined in Section~\ref{subsec:rationale}. It consists of three substages. At the first substage, the server generates a random number $s$ that is used to generate $k$ random vectors, computes group elements $h_0, \dots, h_k$ as in Subsection~\ref{subsubsec:integrity} and broadcasts then to the clients. At the second substage, each client $\mathcal{C}_i$ uses $s$ to generate $k$ random vectors, and sends a proof of wellformedness of $\mathcal{C}_i$ to the server as described in Subsection~\ref{subsubsec:integrity}. $\mathcal{C}_i$ also checks the integrity of VSSS check strings $\Psi_{r_j}$ for all $j$ and sends a list of $r_j$ that fails the integrity check to the server. At the third substage, the server checks the well-formedness proofs as described in Subsection~\ref{subsubsec:integrity}, and requests the clear-text $r_{ij}$'s from $\mathcal{C}_i$ as described in Subsection~\ref{subsubsec:vsss}.

The fourth stage is aggregation. At this stage, the server broadcasts the list $\mathcal{H}$ of honest clients, receives the shares $r_i'$ from each $\mathcal{C}_i \in \mathcal{H}$, and computes $\sum_{\mathcal{C}_i \in \mathcal{H}} \textbf{u}_i$. The details are in Section~\ref{subsec:secagg}.

\red{may give a high-level algorithm for the protocol instead of the current figure}
}

\subsection{Extensions} \label{subsec:extension}

Although we focus on the $L_2$-norm bound check in this paper, our solution can be easily extended to support a wide range of defense methods.
For instance, we can support the sphere defense~\cite{SteinhardtKL17}, the cosine similarity defense~\cite{BagdasaryanVHES20, CaoF0G21} and the Zeno++ defense~\cite{xie2019zeno++}. 

For sphere defense, the server broadcasts a public vector $\textbf{v}$ and a bound $B$, and then checks whether the model update $\textbf{u}$ satisfies $||\textbf{u} - \textbf{v}||_2 \leq B$. 
We can change our protocol in which the $i$-th client commits to $\textbf{u}_i-\textbf{v}$ instead of $\textbf{u}_i$. 
The server then recovers $\sum_{i\in \mathcal{H}} (\textbf{u}_i-\textbf{v})$ and computes $\sum_{i\in \mathcal{H}} \textbf{u}_i = \sum_{i\in \mathcal{H}} (\textbf{u}_i-\textbf{v}) + \textbf{v} \cdot |\mathcal{H}|$, where $\mathcal{H}$ is the set of honest clients.
For cosine similarity defense, the server broadcasts $\textbf{v}, B$ and a public hyperparameter $\alpha$, and then checks whether the model update $\textbf{u}$
satisfies both $||\textbf{u}||_2 \leq B$ and $\langle \textbf{u}, \textbf{v} \rangle \geq \alpha ||\textbf{u}||_2 ||\textbf{v}||_2$. 
We can add a predicate to the protocol that checks $||\textbf{u}||_2 \leq \frac{\langle \textbf{u}, \textbf{v} \rangle}{\alpha ||\textbf{v}||_2}$ based on Algorithm~\ref{alg:prob_check}.
For the Zeno++ defense $\gamma\langle \textbf{v}, \textbf{u} \rangle - \rho ||\textbf{u}||_2^2 \geq \gamma \epsilon$, where $\rho, \gamma, \epsilon$ are public hyperparameters, 
it can be converted into sphere defense by $||\textbf{u} - \frac{\gamma}{2 \rho} \textbf{v}||_2 \leq \sqrt{\frac{\gamma}{\rho} \epsilon + \frac{\gamma^2}{4 \rho^2} ||\textbf{v}||_2^2}$ for the check. 

\ignore{
\subsection{Hybrid Commitment Scheme} \label{subsec:hybrid-scheme}

We first introduce our hybrid commitment scheme that will be used in the probabilistic check method. After each client $\mathcal{C}_i (i \in [n])$ computes the local update $\textbf{u}_i$, it first needs to commit $\textbf{u}_i$ to the server before generating the proofs. Assume that the clients and server agree on independent group elements $g, w_1, \dots, w_d \in \mathbb{G}$, where $w_j (j \in [d])$ is used for committing the $j$-th coordinate in $\textbf{u}_i$. 
Then, $\mathcal{C}_i$ generates a random secret $r_i \in \mathbb{Z}_p$ and encrypts $\textbf{u}_i$ with Pedersen commitment as follows:
%
\begin{align} \label{eq:commitment}
    C(\textbf{u}_i, r_i) & = (C({u_{i1}}, r_i), \dots, C({u_{id}}, r_i)) \nonumber \\ 
    & =  (g^{u_{i1}} w_1^{r_i}, \dots, g^{u_{id}} w_d^{r_i}),
\end{align}
where $u_{ij}$ is the $j$-th coordinate in $\textbf{u}_i$. Each client $\mathcal{C}_i$ sends $\textbf{y}_i = C(\textbf{u}_i, r_i)$ and $z_i = g^{r_i}$ to the server as commitments. Given that $r_i$ is held by each client $\mathcal{C}_i$, the server knows nothing regarding each update $\textbf{u}_i$. 

To facilitate the server to aggregate well-formed updates, we also require each client $\mathcal{C}_i$ to share its secret $r_i$ with other clients using VSSS. Specifically, $\mathcal{C}_i$ computes $((1, r_{i1}), \dots, (n, r_{in}), \Psi_{r_i}) \leftarrow \mathsf{SS.Share}(r_i, n, m+1, g)$ and sends $((j, r_{ij}), \Psi_{r_i})$ to $\mathcal{C}_j$. Note that $g^{r_i} = \Psi_{r_i}(0)$.
The secret $r_i$ allows the server to correctly aggregate the updates from honest clients. Let $\mathcal{C}_H^*$ be the set of honest clients identified by the server and clients (we will discuss how the server and clients collaboratively identify this set in Section~\ref{subsec:protocol}). The server can compute $\mathcal{U} = \sum\nolimits_{\mathcal{C}_i \in \mathcal{C}_H^*} \textbf{u}_i$ as follows. First, the server aggregates the commitments from $\mathcal{C}_H^*$ by:
\begin{align} \label{eq:commitment-aggregation}
    C(\mathcal{U}, r) & = (\prod\nolimits_{\mathcal{C}_i \in \mathcal{C}_H^*} C({u_{i1}}, r_i), \dots, \prod\nolimits_{\mathcal{C}_i \in \mathcal{C}_H^*} C({u_{id}}, r_i) )  \nonumber \\ 
    & = (g^{\sum_{\mathcal{C}_i \in \mathcal{C}_H^*} u_{i1}} w_1^{\sum_{\mathcal{C}_i \in \mathcal{C}_H^*} r_i}, \dots, g^{\sum_{\mathcal{C}_i \in \mathcal{C}_H^*} u_{id}} w_d^{\sum_{\mathcal{C}_i \in \mathcal{C}_H^*} r_i}) \nonumber \\
    & = (g^{u_1} w_1^{r}, \dots, g^{u_d} w_d^r),
\end{align}
where $r = \sum_{\mathcal{C}_i \in \mathcal{C}_H^*} r_i$. Note that $r$ can be computed by the clients in $\mathcal{C}_H^*$ using secure aggregation. That is, for each client $\mathcal{C}_i \in \mathcal{C}_H^*$, it sums the secret shares $r_i' = \sum_{\mathcal{C}_j \in \mathcal{C}_H^*} r_{ji}$ and sends it to the server. The server checks the integrity of each $r_i'$ against $\prod_{\mathcal{C}_j \in \mathcal{C}_H^*} \Psi_{r_j}$, and uses the ones that pass the check to recover $r'$.
According to the homomorphic property of VSSS, $r' = r$. 
%
%
Consequently, with the knowledge of $r$, the server computes $g^{u_l}$ for $l \in [d]$ and solves $u_l$ according to Eqn~\ref{eq:commitment-aggregation},
which is the aggregation of the $l$-coordinate in honest clients' updates. 
}

\ignore{
The benefits of this approach are that: 
\begin{enumerate}
\item Low commitment cost. The computation of the Pedersen commitment $C(x_i, r_i)$ takes $d$ cryptographic group exponentiations, ignoring the cost of the small exponentiation $g^{x_{il}}$. The check string of sharing $r_i$ takes $m+1$ group exponentiations, which is negligible compared to Pedersen commitment. On the other hand, if one uses VSS to share every coordinate $x_i$, the check string of every coordinate takes $2m+1$ group exponentiations\footnote{Because $x_i$ is small, Feldman's verifying scheme \cite{feldman1987practical} is not secure. We have to use the information-theoretic secure one \cite{pedersen2001non} which costs $2m+1$ group exponentiations.}. In total, the cost would be $d(2m+1)$ group exponentiations, which is $2m+1$ times more than ours. 
    \item Low cost of Pedersen commitments of inner products. The Pedersen commitment of an inner product can be computed from the individual commitments:
\begin{displaymath}
    g^{\langle a, x_i \rangle} (\prod_l P_l^{a_l})^{r_i} = \prod_l (g^{x_{il}}P_l^{r_i})^{a_l}.
\end{displaymath}
Client $i$ can compute this value in $O(1)$ time once it receives and trusts the server's computation of $\prod_l P_l^{a_l}$. The cost of verifying server's computation of the multi-exponential takes around $d / \log(d)$ group exponentiations using Pippenger-like algorithm \cite{pippenger1980evaluation}. Moreover, the client can verify multiple multi-exponentials at no extra cost of group operations using batch verification, as discussed in the next paragraph.
\end{enumerate}
}

\ignore{
\subsection{Probabilistic Input Integrity Check} \label{subsec:sampling-check}

Next, we present how the server and clients execute the probabilistic integrity check. Suppose the server and clients agree on some necessary parameters, including the number of samples $k$ in Eqn~\ref{eq:prob-check} for the probabilistic $L_2$-norm check. We will discuss the effect of the choice of $k$ in Section~\ref{subsec:security}. 

Without loss of generality, we describe the check for one client $\mathcal{C}_i$, as summarized in Figure~\ref{fig:wf-protocol}. 
The client first sends the commitments $z_i = g^{r_i}$ and $\textbf{y}_i = C(\textbf{u}_i, r_i)$ to the server using the hybrid commitment scheme.
Then, the server randomly generates $k+1$ random samples, say $\textbf{a}_0 ,\dots,  \textbf{a}_k\in \mathbb{Z}_p^d$.
$\textbf{a}_0$ is sampled from the uniform distribution on $\mathbb{Z}_p$ with cryptographically secure pseudo-random number generator (PRNG) for checking the integrity of Pedersen commitments $\textbf{y}_i$. 
$\textbf{a}_1, \dots, \textbf{a}_k$ are sampled from the discrete normal distribution with insecure PRNG for fast execution of probabilistic check in Algorithm~\ref{alg:prob_check}.
After that, the server computes  $h_t = \prod_l w_l^{a_{tl}}$ for $t \in [0,k]$.
%
%
Let $\textbf{A} = (\textbf{a}_0, \textbf{a}_1, \cdots, \textbf{a}_k)$ and $\textbf{h} = (h_0, h_1, \cdots, h_k)$. The server sends $\{ \textbf{A}, \textbf{h} \}$ to the client. 

\begin{algorithm}[t]
  \caption{{$\mathsf{VerCrt}(\textbf{w}, \textbf{h}, \textbf{A})$}}
  \label{alg:ver_crt}
  \begin{algorithmic}[1]
\STATE \textbf{Input:} $\textbf{w} = (w_1, \dots, w_d) \in \mathbb{G}^d$, $\textbf{h}  = (h_0, \dots, h_k) \in \mathbb{Z}_p^{(k+1)}$, $\textbf{A} \in \mathbb{M}_{(k+1) \times d}(\mathbb{Z}_p)$. 
\STATE Randomly Sample $\textbf{b} = (b_0, \dots, b_k) \in \mathbb{Z}_p^{k+1}$.
\STATE Compute $\textbf{c} = (c_1, \dots, c_d) =  \textbf{b} \cdot \textbf{A} \in \mathbb{Z}_p^d$. 
    \RETURN $h_0^{b_0} \dots h_k^{b_k} == w_1^{c_1} \dots w_d^{c_d}$.
\end{algorithmic}
\end{algorithm}

\begin{algorithm}[t]
  \caption{{$\mathsf{GenPrf}(g, q, \textbf{w}, \textbf{f}, b_{\mathsf{ip}}, b_{\mathsf{max}}, B_0, \textbf{A}, \textbf{h}, z, r, \textbf{u})$}}
  \label{alg:gen_prf}
  \begin{algorithmic}[1]
\STATE \textbf{Input:} $g,q,z \in \mathbb{G}$, 
$\textbf{w} \in \mathbb{G}^d$, 
$\textbf{f} \in \mathbb{G}^{2kb_{\mathsf{max}}}$, 
$b_{\mathsf{ip}} < b_{\mathsf{\max}}$, 
$B_0 \in (0, 2^{b_{\mathsf{max}}}]$,
$A \in \mathbb{M}_{(k+1) \times d}(\mathbb{Z}_p)$,
$\textbf{h} \in \mathbb{G}^{(k+1)}$, 
$r \in \mathbb{Z}_p$,
$\textbf{u} \in \mathbb{Z}_p^d$.
\IF {\NOT { $\mathsf{VerCrt}(\textbf{w}, \textbf{h}, \textbf{A})$ } }
      \STATE \textbf{Abort}
     \ENDIF
     \STATE Compute $\textbf{v}^* = \textbf{u} \cdot \textbf{A}^T \in \mathbb{Z}_p^{k+1}$. Denote $\textbf{v}^* = (v_0, \dots, v_k)$, and let $\textbf{v} = (v_1, \dots, v_k)$. 
     \STATE Compute $e_t = g^{v_{t}} h_{t}^r$ for $t \in [0, k]$, and let $\textbf{e} = (e_0, e_1, \dots, e_k)$. 
     \STATE Randomly sample $\textbf{s}, \textbf{s}' \in \mathbb{G}^k$.
     \STATE Compute $o_t = g^{v_{t}} q^{s_t}$, $o_t' =  g^{v_{t}^2} q^{s_t'}$ for $t \in [1,k]$. 
     \STATE Compute $\rho \leftarrow \mathsf{GenPrfWf}(g, q, \textbf{h}, z,  \textbf{e}, \textbf{o}, r, \textbf{v}^*,\textbf{s})$.
     \STATE Compute $\tau \leftarrow \mathsf{GenPrfSq}(g, q, \textbf{o}, \textbf{o}', \textbf{v}, \textbf{s}, \textbf{s}')$.
     \STATE Compute $\sigma \leftarrow \mathsf{GenPrfBd}(g, q, {\textbf{f}}, b_{\mathsf{ip}},  g^{2^{b_{\mathsf{ip}} - 1}} \cdot \textbf{o},  \textbf{v} + 2^{b_{\mathsf{ip}} - 1} \cdot \textbf{1}, \textbf{s})$. 
     \STATE Compute $\mu \leftarrow \mathsf{GenPrfBd}(g, q, {\textbf{f}}, b_{\max}, g^B (\prod_{t=1}^k o_t')^{-1}, B - \sum_{t=1}^k v_t^2, -\sum_{t=1}^k s_t')$. 
     \RETURN $\pi = (\textbf{e}, \textbf{o}, \textbf{o}', \rho, \tau, \sigma, \mu)$.
\end{algorithmic}
\end{algorithm}

\begin{algorithm}[t]
  \caption{{$\mathsf{VerPrf}(g, q, \textbf{w}, \textbf{f}, b_{\mathsf{ip}}, b_{\mathsf{max}}, B_0, \textbf{A}, \textbf{h}, z, \textbf{y}, \pi)$}}
  \label{alg:ver_prf}
  \begin{algorithmic}[1]
\STATE \textbf{Input:} $g \in \mathbb{G}$, 
$q \in \mathbb{G}$, 
$\textbf{w} \in \mathbb{G}^d$, 
$\textbf{f} \in \mathbb{G}^{2kb_{\mathsf{max}}}$, 
$b_{\mathsf{ip}} < b_{\mathsf{\max}}$, 
$B_0 \in (0, 2^{b_{\mathsf{max}}}]$,
$A \in \mathbb{M}_{(k+1) \times d}(\mathbb{Z}_p)$,
$\textbf{h} \in \mathbb{G}^k$, 
$\textbf{y} \in \mathbb{G}^d$,
$\pi$.
\STATE Unravel $\pi = (\textbf{e}, \textbf{o}, \textbf{o}', \rho, \tau, \sigma, \mu)$.
    \RETURN  $\mathsf{VerCrt}(\textbf{y}, \textbf{e}, \textbf{A})$  \AND 
    \STATE $\qquad\mathsf{VerPrfWf}(g, q, \textbf{h}, z, \textbf{e}, \textbf{o}, \rho)$ 
    \AND 
    \STATE  $\qquad\mathsf{VerPrfSq}(g, q, \textbf{o}, \textbf{o}', \tau)$ \AND 
    \STATE $\qquad\mathsf{VerPrfBd}(g, q, {\textbf{f}}, b_{\mathsf{ip}},  g^{2^{b_{\mathsf{ip}} - 1}} \cdot \textbf{o}, \sigma)$) \AND 
    \STATE $\qquad\mathsf{VerPrfBd}(g, q, {\textbf{f}}, b_{\max}, g^B (\prod_{t=1}^k o_t')^{-1}, \mu)$.
\end{algorithmic}
\end{algorithm}

Upon receiving the information, the client generates the proof $\pi$ using Algorithm~\ref{alg:gen_prf}. 
Specifically, the client first verifies the correctness of $\textbf{h}$ using $\mathsf{VerCrt}(\textbf{w}, \textbf{h}, \textbf{A})$ from Algorithm~\ref{alg:ver_crt}.
This is to ensure that the server does not steal information from the client by sending incorrect $\textbf{h}$.
Note that Algorithm~\ref{alg:ver_crt} uses batch verification to accelerate the verification of $h_t = \prod_l w_l^{A_{tl}}$ for $t \in [0, k]$.
%
If it is correct, the client computes the following items for generating the proof that Eqn~\ref{eq:prob-check} is less than a bound, based on Algorithm~\ref{alg:prob_check}. 
\begin{itemize}[leftmargin=*]
    \item The client computes the inner products between $\textbf{u}_i$ and each row of $\textbf{A}$, obtaining $\textbf{v}^* = (v_0, v_1, \cdots, v_k)$, where $v_t = \langle \textbf{a}_t, \textbf{u}_i \rangle$ for $t \in [0, k]$. The client commits $e_t = g^{v_t} h_t^{r_i}$ using its secret $r_i$ for $t \in [0, k]$. Let $\textbf{e}^* = (e_0, e_1, \cdots, e_k)$ and $\textbf{e} = (e_1, \cdots, e_k)$. The commitment $e_0$ is used for integrity check of $\textbf{y}_i$. The commitments $\textbf{e}$ are used for bound check of $\textbf{v}$.
    \item Let $\textbf{v} = (v_1, \cdots, v_k)$. The client commits $v_t$ using $o_t = g^{v_t} q^{s_t}$ for $t \in [1, k]$, where $s_t$ is a random number. Let $\textbf{o} = (o_1, \cdots, o_k)$ be the resulted commitment. 
     $e_t$ and $o_t$ commit to the same secret $v_t$ using different group elements $h_t$ and $q$. As a result, $o_1, \dots, o_k$ use the same group element $q$, ready for batch square checking and batch bound checking.
    \item The client further commits $o_t' = g^{v_t^2} q^{s_t'}$ for $t \in [1, k]$, where $s_t'$ is a random number. Let $\textbf{o}' = (o_1', \cdots, o_k')$ be the resulted commitment. This commitment will be used in the proof generation and verification for proof of square.
    \item The client generates a proof $\rho$ to prove that $(z, \textbf{e}^*, \textbf{o})$ is well-formed, which means that the secret in $z$ is used as the blind in $e_t$, $t \in [0,k]$, and that the secrets in $e_t$ and $o_t$ are equal, $t \in [1, k]$.
    Note that $z = g^{r_i} = \Psi_{r_i}(0)$ is the $0$-th coordinate of the check string of Shamir's share of $r_i$. 
    \item The client generates a proof $\tau$ to prove that the secret in $o_t'$ is the square of the secret in $o_t$ for $t \in [1, k]$, using the building block described in Section~\ref{sec:preliminary}.
    \item The client generates a proof $\sigma$ that the secret in $o_t$ is in the interval $[-2^{b_{\mathsf{ip}}}, 2^{b_{\mathsf{ip}}})$ for $t \in [1, k]$. This ensures the inner product of $\textbf{a}_t$ and $\textbf{u}_i$ does not cause overflow when squared. 
    \item The client generates a proof $\mu$ that $B - \sum_t v_t^2$ is in the interval $[0, 2^{b_{\mathsf{max}}})$ using the commitment $g^B (\prod_t o_t')^{-1}$. This proof is to guarantee that Eqn~\ref{eq:prob-check} is less than the bound of the probabilistic check. 
\end{itemize}

As a result, the client sends the proof $\pi = (\textbf{e}, \textbf{o}, \textbf{o}', \rho, \tau, \sigma, \mu)$ to the server.  
After receiving the proof, the server can verify it accordingly, including: checking the correctness of $e_t = \prod_{l} y_{il}^{a_{tl}}$, $t \in [0,k]$ using Algorithm~\ref{alg:ver_crt}, checking the well-formedness proof $\rho$ using Algorithm~\ref{alg:ver_prf_wf} in Appendix~\ref{appendix:pre}, checking the square proofs of $(\textbf{o}', \textbf{o})$, and checking the two bound proofs. If all the checks are passed, the server guarantees that the client's update passes the check in Algorithm~\ref{alg:prob_check}.
}

\ignore{
The server verifies the proof by checking the following items:
\begin{itemize}
    \item Check the correctness of $Y_t = \prod_l y_l^{a_{tl}}$ for $t \in [0, k]$ using $\mathsf{VerCrt}(\textbf{y}, \textbf{A}, \textbf{Y}^*)$ from Algorithm~\ref{alg:ver_crt}, where $Y^*_{t} = Y_{t - 1}$ for $t \in [1, k+1]$.
    \item Check the well-formedness proof of $(\Psi_{r_i}(0), Y_0, \textbf{Y}, \textbf{z})$.
    \item Check the square proof of $(\textbf{z}^*, \textbf{z})$.
    \item Check the bound proof of $\textbf{z}$.
    \item Check the bound proof of $g^B (\prod_t z_t^*)^{-1}$.
\end{itemize}
}

\ignore{
\subsection{Protocol Description} \label{subsec:protocol}

We present the full protocol of \ourtech{} in Figure~\ref{fig:protocol}, including a system initialization stage, followed by three iterative rounds. 

\vspace{1mm} 
\noindent
\textbf{System Initialization.} All parties are given the system parameters, including the number of clients $n$, the maximum number of malicious clients $m$,
the bound on the number of bits $b_{\text{ip}}$ of each inner product,
the maximum number of bits $b_{\text{max}} > b_{\text{ip}}$ of the sum of squares of inner product,  the bound of the sum of inner products $B_0 < 2^{b_{\text{max}}}$, the number of samples $k$ for the probabilistic check, a set of independent group elements $g \in \mathbb{G}, q \in \mathbb{G}, \textbf{w} \in \mathbb{G}^d, \textbf{f} \in \mathbb{G}^{2kb_{\text{max}}}$, the factor $M>0$ used in discretizing the normal distribution samples, and a cryptographic hash function $H(\cdot)$. 
Since there is no direct channel between any two clients, we let the server forward some of the messages. To prevent the server from accessing the secret information, each client $\mathcal{C}_i (i \in [n])$ generates a public/private key pair $(pk_i, sk_i)$ and sends the public key $pk_i$ to a public bulletin. Then, each client fetches the other clients' public keys such that each pair of clients can establish a secure channel via the Diffie-Hellman protocol~\cite{Merkle78} for exchanging messages.

\vspace{1mm} 
\noindent
\textbf{Round 1: Commitment Generation.} In every FL training iteration, each client $\mathcal{C}_i$ generates a random number $r_i$ as the secret. Then, $\mathcal{C}_i$ adopts the hybrid commitment scheme in Section~\ref{subsec:hybrid-scheme} to commit its update $\textbf{u}_i$ using Eqn~\ref{eq:commitment}, obtaining $\textbf{y}_i = C(\textbf{u}_i, r_i)$. 
Also, it generates the secret shares of $r_i$ using VSSS, obtaining $((1, r_{i1}), \dots, (n, r_{in}), \Psi_{r_i})$, and encrypts each share $r_{ij}$ ($\forall j \in [1,n] \wedge j \neq i$) using the encryption key based on $(pk_j, sk_i)$. 
Next, $\mathcal{C}_i$ sends the encrypted shares $\mathsf{Enc}(r_{ij})$ and the check string $\Psi_{r_i}$ to the server. Afterward, the server forwards the encrypted shares to respective clients and broadcasts the check strings to all the clients. 
In addition, the server initializes a list $\mathcal{C}^* = \emptyset$ for the current iteration to record the malicious clients that will be identified in the following round. 

\vspace{1mm} 
\noindent
\textbf{Round 2: Proof Generation and Verification.} In this round, the clients and the server jointly flag the malicious clients in two steps. 
The first step is to verify the authenticity of secret shares. After receiving the encrypted shares $\mathsf{Enc}(r_{ji})$ and check strings $\Psi_{r_j}$ for $j \in [1,n]$, each client $\mathcal{C}_i$ decrypts $r_{ji}$ and checks against $\Psi_{r_j}$. Then, $\mathcal{C}_i$ sends a list of candidate malicious clients that do not pass the check to the server. 
The server follows two rules to flag the malicious clients~\cite{roy2022eiffel}: (1) if a client $\mathcal{C}_i$ flags more than $m$ clients as malicious or is flagged as malicious by more than $m$ clients, the server puts $\mathcal{C}_i$ to $\mathcal{C}^*$; (2) if a client $\mathcal{C}_i$ is flagged as malicious by $[1,m]$ clients, the server requests the shares $r_{ij}$ in the clear for all $\mathcal{C}_j$ that flags $\mathcal{C}_i$ and checks against $\Psi_{r_i}$, if the clear $r_{ij}$ fails the check, the server puts $\mathcal{C}_i$ to $\mathcal{C}^*$. This ensures that if $\mathcal{C}_i$ is honest, then $\mathcal{C}_i$ passes the check and its clear $r_{ij}$ is sent only for malicious $\mathcal{C}_j$ and at most $m$ clear text shares are sent.


The second step is to verify the integrity of each client's update using the probabilistic method. Note that in Section~\ref{subsec:sampling-check}, we require the server to send $k+1$ random samples with dimensionality $d$ to the client. To reduce the communication cost, we let the server select a random value $s$ and broadcast it to all clients. Based on $s$, the server and each client $\mathcal{C}_i (i \in [n])$ first compute a seed $H(s, \{pk_i\}_{1 \leq i \leq n})$ using $s$ and all clients' public keys. Hence, the clients and the server can generate the same set of random samples $\textbf{A} = (\textbf{a}_0, \textbf{a}_1, \cdots, \textbf{a}_k)$ because the seed is the same. 
Then, the server computes $\textbf{h}$ (see Section~\ref{subsec:sampling-check}) and broadcasts it to all clients. 
Next, each client $\mathcal{C}_i (i \in [1,n])$ generates $\textbf{A}$, computes the proof $\pi_i$ according to Algorithm~\ref{alg:gen_prf}, and sends $\pi_i$ to the server. 
Consequently, the server verifies the proof of each client $\mathcal{C}_i$ and puts it to $\mathcal{C}^*$ if the verification fails. The list $\mathcal{C}^*$ is broadcast to all the clients.

\vspace{1mm} 
\noindent
\textbf{Round 3: Aggregation.} 
In this round, each client $\mathcal{C}_i$ selects the corresponding shares from $\mathcal{C}_j \notin \mathcal{C}^*$, aggregates them to get $r_i'$, and sends $r_i'$ to the server. The server uses $\mathsf{SS.Verify}$ to verify the integrity of each $r_i'$ and uses $\mathsf{SS.Recover}$ with the ones that pass the integrity check to recover $\sum_{\mathcal{C}_j \notin \mathcal{C}^*} r_j$. 
%
%
Therefore, the server can solve the equation in Eqn~\ref{eq:commitment-aggregation} to compute the aggregation of honest clients' updates. 
}

\ignore{
Setup

All parties are given the number of clients $n$, the bound on the malicious clients $m < n / 2$, the maximum bit number $b$ of bounds, public keys $h \in \mathbb{G}^d$ and $f \in \mathbb{G}^{2b}$, the bound $B$, the number of inner product samples $k$.

Round 1 (Announce Public Information)

Each client $\mathcal{C}_i$ generates a key pair $(pk_i, sk_i)$ with Diffie–Hellman and sends $pk_i$ to the server. (This has the property that the shared key between $\mathcal{C}_i$ and $\mathcal{C}_j$ can be computed from either $(pk_i, sk_j)$ or $(pk_j, sk_i)$)

Server initializes $Flag[i] = \emptyset$ for $i \in [n]$ and initializes $\mathcal{C}^* = \emptyset$. Server sends all the public keys $(pk_i)_i$ to all the clients.

Round 2 (Generate proof and Distribute shares of the secret)

Each client $\mathcal{C}_i$ commits its update $\mathbf{u}_i$ with $C(\mathbf{u}_i, r_i) = (g^{u_i^{(1)}} (h^{(1)})^{r_i}, \dots, g^{u_i^{(d)}} (h^{(d)})^{r_i})$ and sends $C(\mathbf{u}_i, r_i)$ to the server. $\mathcal{C}_i$ also makes shares $(r_{ij})_j$, encrypt $r_{ij}$ with $(pk_i, sk_j)$ and sends the encrypted $[r_{ij}]$ and the check string $\Psi_i$ to the server.

Round 3 (Verify Proof)

The server announces a random seed $s$ from which all parties compute: 

random $a_m \in \mathbb{F}^d$ for $m \in [k]$ that follows discretized $\mathcal{N}(0,1)$ using random seed $((pk_i)_i, s)$.

random $a_0 \in \mathbb{F}^d$ that follows the uniform distribution using random seed $((pk_i)_i, s)$.

The server also computes $h_m = \prod_l (h^{(l)})^{a_m^{(l)}}$ for $m = 0, \dots, k$ and sends them to all the clients. 

Each client $\mathcal{C}_i$ checks correctness of $h_m$, $m = 0, \dots, k$,  computes $v_{im} = \langle a_m, u_i \rangle$ for $m = 0, \ldots, k$, and sends $C(v_{im}, r_i)=g^{v_{im}} h_{m}^{r_i}$ to the server. Client $i$ also sends a proof that $\sum_{m=1}^k v_{im}^2$ is bounded using $f$.

Each client $\mathcal{C}_i$ downloads check strings $(\Psi_j)_j$, public keys $(pk_j)_j$ and encrypted $([r_{ji}])_j$, decrypts $r_{ji}$ using $(pk_j, sk_i)$ and checks $r_{ji}$ against $\Psi_j$, and sends the list of $j$'s that don't pass the check to the server. 

The server verifies well-formedness of $C(v_{im}, r_i)$ for all $i$ and checks the bound proofs. Put $i$ into $\mathcal{C}^*$ if client $i$ fails the proof.

The server puts $i$ into $\mathcal{C}^*$ (the list of malicious clients) if $i$ does not pass the check, puts $i$ into $\mathcal{C}^*$ if either $i$ flags more than $m$ clients or more than $m$ clients flag $i$. For all $i$ such that between $1$ and $m$ clients (inclusive) flag $i$, the server requests client $\mathcal{C}_i$ the share $r_{ij}$ (in clear) for all $j$ that flags $i$ and checks them against $\Psi_i$. 

The clients send the requested $r_{ij}$ (in clear). 

 If one of clear text $r_{ij}$ does not pass the check, the server puts $i$ into $\mathcal{C}^*$. If all pass the check, the server sends $r_{ij}$ (in clear) to $\mathcal{C}_j$.

Round 4 (Compute Aggregate)

Each client receives $\mathcal{C}^*$ and sends the aggregated share $R_i = \sum_{j \notin \mathcal{C}^*} r_{ji}$ to the server. The server checks every $R_i$ against $\prod_{j \notin \mathcal{C}^*} \Psi_j$, uses the ones that pass the check to reconstruct $R = \sum_{i \notin \mathcal{C}^*} r_i$. Then use $R$ and $\prod_{i \notin \mathcal{C}^*} C(\mathbf{u}_i, r_i)$ to compute $g^{\sum_{i \notin \mathcal{C}^*} u_i}$ and find $\sum_{i \notin \mathcal{C}^*} u_i$ using baby-step giant-step.
}
\section{Analysis} \label{sec:analysis}

\begin{figure}[t]
    \centering
    \subfloat[$k$ vs. $F_{k, \epsilon, d, M}$]{{\includegraphics[width=0.235\textwidth]{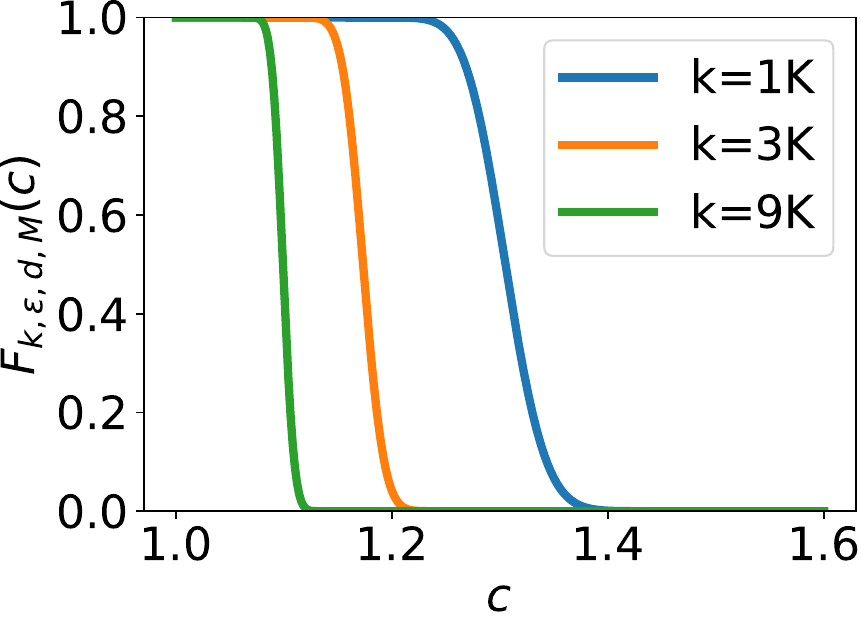} } \label{subfig:discussion-f}}%
    \subfloat[$k$ vs. maximum damage]{{\includegraphics[width=0.235\textwidth]{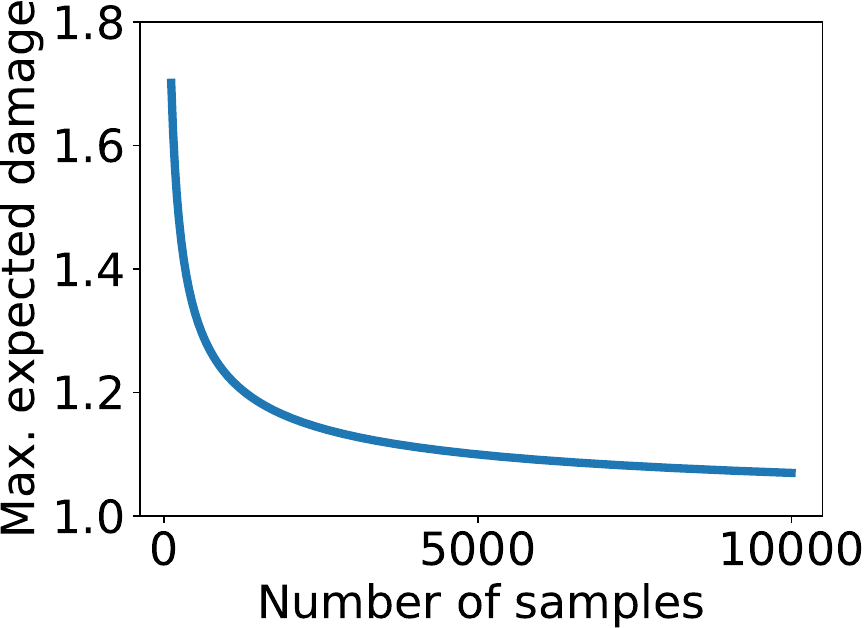} \label{subfig:discussion-dmage} }}%
    \caption{The effect of $k$ when $\epsilon = 2^{-128}$, $d = 10^6$, $M = 2^{24}$: (a) the trend of $F_{k,\epsilon, d, M}$ w.r.t. $k$; (b) the trend of the maximum damage w.r.t. $k$.}
    \label{fig:discussion}
\end{figure}

\subsection{Security Analysis} \label{subsec:security}

We state the formal security theorem and outline its proof. 

\begin{theorem} \label{theorem:security}
By choosing $\epsilon = \mathrm{negl}(\kappa)$ and $B_0 = B^2M^2( \sqrt{\gamma_{k, \epsilon}} +  \frac{\sqrt{kd}}{2M})^2$, for any list of honest clients $\mathcal{C}_H$ of size at least $n - m$, \ourtech{} satisfies $( D, F_{k, \epsilon, d, M})$-SAVI, where $D(\textbf{u}) = ||\textbf{u}||_2 / B$ and 
\begin{align} \label{eq:f-func}
    F_{k, \epsilon, d, M}(c) = \mathrm{Pr}_{x \sim \chi_k^2}\left[x < \frac{1}{c^2} \left( \sqrt{\gamma_{k, \epsilon}} + \frac{3 \sqrt{kd}}{2M} \right)^2\right] + \mathrm{negl}(\kappa). 
\end{align}
\end{theorem}

The proof consists of three parts: showing that the server computes an aggregate $\textbf{U} = \sum_{i \in H'} \textbf{u}_i$ for some $H' \supseteq H$ with probability at least $1 - \mathrm{negl}(\kappa)$; showing that with probability at least $1 - \mathrm{negl}(\kappa)$, nothing else except $\sum_{\mathcal{C}_i \in \mathcal{C}_H}\textbf{u}_i$ is known about $\textbf{u}_i$ for any $i \in H$; and showing that if $\mathcal{C}_i$ is malicious with $||\textbf{u}||_2 / B = c$, then the probability that $i \in H'$ is bounded by $F_{k, \epsilon, d, M}(c)$.

\vspace{1mm}
\noindent 
\textbf{The server aggregates model updates from honest clients and possibly some malicious clients.} 
The aggregation step in Section~\ref{subsec:secagg} ensures that the server computes an aggregate $\textbf{U} = \sum_{\mathcal{C}_i \in \mathcal{C}_H'} \textbf{u}_i$ for a list $\mathcal{C}_H'$ that is marked as honest clients by the server. We need to prove that this list $\mathcal{C}_H'$ must contain $\mathcal{C}_H$. In fact, if $\mathcal{C}_i$ is honest, then $||\textbf{u}_i|| \leq B$. If $\textbf{b}_1, \dots, \textbf{b}_k$ are sampled i.i.d.\ from the normal distribution $\mathcal{N}(\textbf{0}, M \cdot \textbf{I}_{d})$. Then, 
    \ignore{
    \begin{displaymath}
     \frac{1}{||\textbf{u}||_2^2}\sum_{t=1}^k \langle \textbf{a}_t , \textbf{u} \rangle^2   
    \end{displaymath}
    }
    $\frac{1}{||\textbf{u}||_2^2 M^2}\sum_{t=1}^k \langle \textbf{b}_t , \textbf{u} \rangle^2 $ follows the chi-square distribution $\chi_k^2$.
So the probability that 
\begin{align} \label{eq:chisquare-prob}
        \sum_{t=1}^k \langle \textbf{b}_t, \textbf{u}_i \rangle^2 \leq B^2M^2\gamma_{k, \epsilon}
\end{align}
is at least $1 - \epsilon$.
Define $\mathsf{round} : \mathbb{R} \to \mathbb{Z}$ by $\mathsf{round}(n + \alpha) = n$ for $n \in \mathbb{Z}$, $-1/2 \leq \alpha < 1/2$. Define $\textbf{a}_t$ by $a_{tj} = \mathsf{round}(b_{tj})$ for $t \in [1, k]$. After rounding $\textbf{b}_t$ to $\textbf{a}_t$, (\ref{eq:chisquare-prob}) implies the relaxed inequality
\begin{align}  
\label{eq:lem_sum_bound}
    \sum_{t=1}^k \langle \textbf{a}_t, \textbf{u}_i \rangle^2 \leq B_0
\end{align}
If~(\ref{eq:lem_sum_bound}) holds, $\mathcal{C}_i$ can produce a proof which passes the integrity check.
Since $\epsilon = \mathrm{negl}(\kappa)$, the probability that all of the honest clients can produce a valid proof is at least $1 - \mathrm{negl}(\kappa)$.

\vspace{1mm}
\noindent 
\textbf{Security of honest clients.} 
    Cryptographically, the values of $\Psi_{r_i}$, $C(\textbf{u}_i, r_i)$, and $\pi_i$ do not reveal any information about $\textbf{u}_i$ or $r_i$. 
    VSSS ensures that nothing is revealed from the $\leq m$ shares $\{r_{ij}\}_{j \notin \mathcal{C}_H}$ of the secret $r_i$. 
    At the secure aggregation step in Section~\ref{subsec:secagg}, VSSS ensures that only $\sum_{\mathcal{C}_i \in \mathcal{C}_H} r_i$ is revealed. 
    From $\sum_{\mathcal{C}_i \in \mathcal{C}_H} r_i$, the only value that can be computed is $\sum_{\mathcal{C}_i \in \mathcal{C}_H} \textbf{u}_i$.
Therefore, nothing else except $\sum_{\mathcal{C}_i \in \mathcal{C}_H}\textbf{u}_i$ is known about $\textbf{u}_i$ for any $i \in H$.

\vspace{1mm}
\noindent 
\textbf{Pass rate of malicious clients is bounded by $F$.} 
Suppose that $\mathcal{C}_i$ is malicious and $D(\textbf{u}_i) = ||\textbf{u}_i||/B = c > 1$. Suppose that $\textbf{b}_1, \dots, \textbf{b}_k$ are sampled i.i.d.\ from the normal distribution $\mathcal{N}(\textbf{0}, M \cdot \textbf{I}_{d})$ and $\textbf{a}_t$ is rounded from $\textbf{b}_t$ as before. With probability at least $1 - \mathrm{negl}(\kappa)$, if $\mathcal{C}_i$ passes the integrity check, it must happen that $\sum_{t=1}^k\langle\textbf{a}_t, \textbf{u}\rangle^2 \leq B_0$. If $\sum_{t=1}^k\langle\textbf{a}_t, \textbf{u}\rangle^2 \leq B_0$ holds, then the following relaxed inequality on $\textbf{b}_t$ must hold:
\begin{align} \label{eq:chisquare-bound}
         \sum_{t=1}^k \langle \textbf{b}_t , \textbf{u} \rangle^2  \leq \frac{1}{c^2} \left( \sqrt{\gamma_{k,\epsilon}} + \frac{3 \sqrt{kd}}{2M} \right)^2.
    \end{align}
The probability that~(\ref{eq:chisquare-bound}) happens is bounded by $F_{k,\epsilon,d,M}(c)$.

\begin{table*}[t]
\setlength{\tabcolsep}{4pt}
\caption{Cost comparison (g.e. = group exponentiation, f.a. = field arithmetic)}
\centering
\begin{tabular}{| l l | c c | c | c | c c |} 
\hline \multirow{2}{*}{} & \multirow{2}{*}{} & \multicolumn{2}{|c|}{EIFFeL} & \multirow{2}{*}{RoFL} & \multirow{2}{*}{ACORN} & \multicolumn{2}{|c|}{\ourtech{}} \\
 & & g.e. & f.a. & & & g.e. & f.a.   \\
  \hline \multirow{4}{*}{\shortstack[c]{Client \\ comp.}} & commit. & $O(md)$ & $O(nmd)$  & $O(d)$ & $O(d)$& $O(d)$ & small \\
  & proof gen. & 0 & $O(bnmd)$ & $O(db)$   & $O(d)$ & $O(d / \log d)$ & $O(kd)$  \\
  & proof ver. & $ O(nmd / \log (md))$ & $O(bnmd)$ & 0  & 0 & 0 & 0\\
  & total & $O(nmd  / \log (md))$ & $O(bnmd)$ & $O(db)$ & $O(d)$ & $O(d)$ & $O(kd)$  \\
  \hline \multirow{4}{*}{\shortstack[c]{Server \\ comp.}} & prep. & 0 & 0 & 0  & 0 & $ O(k d \log M / \log d \log p)$ & small  \\
  & proof ver. & 0 & small & $O(ndb/\log (db))$  & $O(nd/\log(d))$ & $O(nd / \log d)$ & small \\
  & agg. & 0 & $O(nmd)$ & $O(nd / \log p)$ & $O(nd / \log(p))$ & $O(nd/ \log p)$ & small  \\
  & total & 0 & $O(nmd)$ & $O(ndb/\log (db))$ & $O(nd / \log(d))$ & $O(d(n + k \log M / \log p) / \log d)$ & small  \\
  \hline
  \multicolumn{2}{|c|}{\shortstack[c]{Comm. per client}}& \multicolumn{2}{|c|}{$\approx 2dnb$} &$\approx 12d$ & $\approx (b + \log n)/\log(p)$ & \multicolumn{2}{|c|}{$\approx d$} \\ 
  \hline
\end{tabular}
\label{table:theoretical_cost_compare}
\end{table*}

\vspace{1mm} 
\noindent 
\textbf{Discussion.} 
Now we discuss the effect of $k$ on $F_{k,\epsilon, d, M}$ and the maximum damage, which is illustrated in Figure~\ref{fig:discussion}. 
We set $\epsilon = 2^{-128}$ by default in this paper. 
We set $M = 2^{24}$ so that when $k \leq 10^4$ and $d \leq 10^6$, the term $\frac{3 \sqrt{kd}}{2M}$ is insignificant in Eqn~\ref{eq:f-func}. 
Figure~\ref{subfig:discussion-f} shows the trend of $F_{k,\epsilon, d, M}$ with $d = 10^6$ and different choices of $k$. 
We can see that $F_{k, \epsilon, d, M}(c)$ is very close to 1 when $c$ is slightly larger than 1, and it drops to negligible rapidly as $c$ continues to increase. 
Specifically, it means that when $k = 1000$, a malicious $\textbf{u}_i$ with $||\textbf{u}_i||_2 \leq 1.2 B$ will very likely pass the integrity check, but when $||\textbf{u}_i||_2 \geq 1.4 B$, the check will fail with close-to-1 probability.

This is the downside of the probabilistic $L_2$-norm checking, i.e., slightly out-of-bound vectors may pass the check. 
We can quantify the size of damages caused by such malicious vectors. 
For example, with strict checking, a malicious client can use a malicious $\textbf{u}$ with $||\textbf{u}||_2=B$ which passes the check, so that it can do damage of magnitude $B$ to the aggregate. 
With the probabilistic check, the malicious client can use slightly larger $||\textbf{u}||_2$ at a low failure rate. 
The expected damage it can do to the aggregate is $||\textbf{u}||_2 \cdot F_{k, \epsilon, d, M}(||\textbf{u}||_2 / B)$.
By choosing a suitable $||\textbf{u}||_2$, the maximum expected damage is 
\begin{align}
   B \cdot \max\{c \cdot F_{k, \epsilon, d,M}(c) : c \in (1, +\infty)\}.
\end{align}
The value $\max\{c \cdot F_{k, \epsilon, d,M}(c) : c \in (1, +\infty)\}$ can be calculated easily as $c^* \cdot F_{k,\epsilon, d, M}(c^*)$ because $c \cdot F_{k, \epsilon, d,M}(c)$ is increasing for $c \in (1, c^*]$ and decreasing for $c \in [c^*, +\infty)$ for some $c^*$.
Figure \ref{subfig:discussion-dmage} shows the maximum expected damage with respect to $k$ when $B = 1$.
It turns out that with $\epsilon = 2^{-128}$ and $k \geq 10^3$, the maximum expected damage is close to 1. 
In other words, the magnitude of damage that a malicious client can do to the aggregate is only slightly more than the one under the strict $L_2$-norm check protocol. 
We will experiment with $k = 1\text{K}, 3\text{K}, 9\text{K}$ in Section~\ref{sec:experiment}, corresponding to the ratio of the magnitudes of damages $1.24$, $1.13$, $1.08$, respectively.

\ignore{
\begin{figure}[h]
    \centering
    \includegraphics[width=0.4\textwidth]{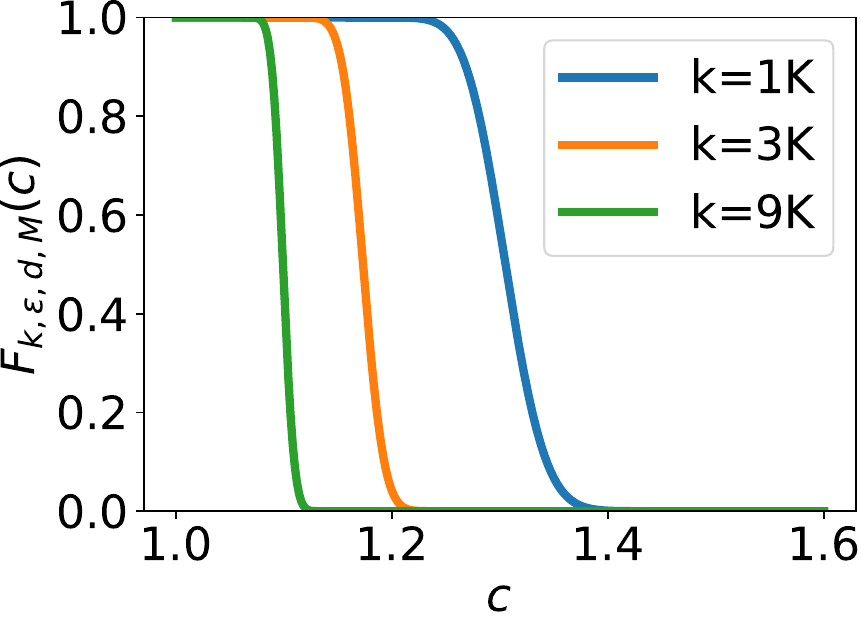}
    \caption{The graph of $F_{k, \epsilon}$ with different numbers of samples, \needcheck{to be updated}}
    \label{fig:security}
\end{figure}
}


\ignore{
\begin{figure}[h]
    \centering
    \includegraphics[width=0.4\textwidth]{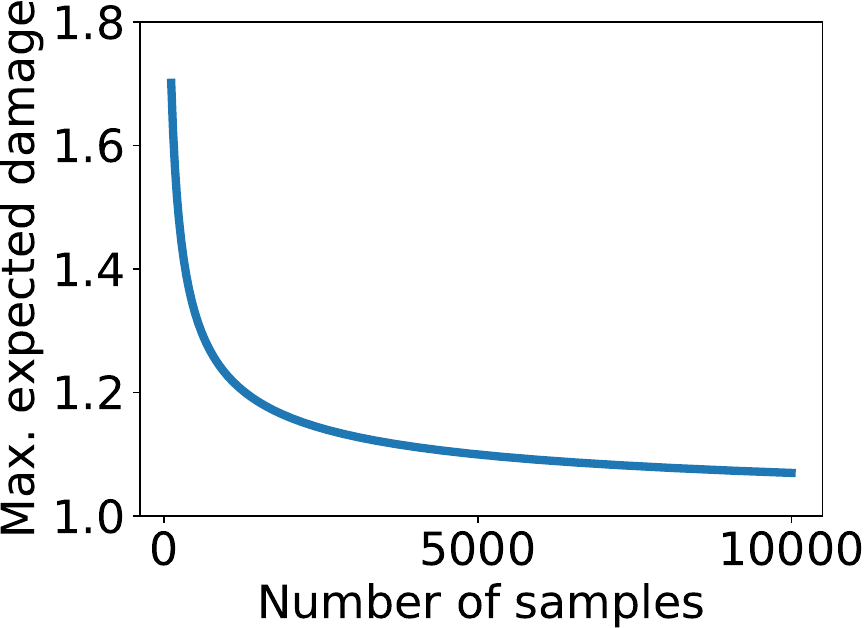}
    \caption{The maximum damage w.r.t.\ number of samples \needcheck{to be updated}}
    \label{fig:damage}
\end{figure}
}

\subsection{Cost Analysis} \label{subsec:cost}

We theoretically analyze the cost of \ourtech{} under the assumption of $d >> k$ by comparing it to \eiffel{}, \rofl{} and ACORN as summarized in Table~\ref{table:theoretical_cost_compare}.
Here we count the number of cryptographic group exponentiations (g.e.) and finite field arithmetic (f.a.) separately for \eiffel{} and \ourtech{}. 
The communication cost per client is measured in the number of group elements. 


\ignore{
\begin{table*}[t]
\small
\caption{Cost comparison (g.e. = group exponentiation, f.a. = field arithmetic)}
\centering
\begin{tabular}{| l l | c c | c c | c c |} 
\hline \multirow{2}{*}{} & \multirow{2}{*}{} & \multicolumn{2}{|c|}{EIFFeL} & \multicolumn{2}{|c|}{RoFL} & \multicolumn{2}{|c|}{\ourtech{}} \\
 & & g.e. & f.a. & g.e. & f.a. & g.e. & f.a. \\
  \hline \multirow{4}{*}{\shortstack[c]{Client \\ comp.}} & Commitment & $d(2m+1)$ & $O(nmd)$   & $2d$ & small & $d$ & small\\
  & Generate proof & 0 & $O(bnmd)$ & $d(8 + 4O(b))$ & small & $d / \log(d)$ & $O(kd)$   \\
  & Verify proof & $ O(nmd / \log md)$ & $O(bnmd)$ & 0 & 0 & 0 & 0 \\
  & Total & $d(2m + O(nm / \log md))$ & $O(bnmd)$ & $d(10 + 4 O(b))$ & small & $d(1 + O(1 / \log(d)))$ & $O(kd)$ \\
  \hline \multirow{4}{*}{\shortstack[c]{Server \\ comp.}} & Preparation & 0 & 0 & 0 & 0 & $ O(k d \log M / \log d \log p)$ & small  \\
  & Verify Proof & 0 & small & $nd O(2b/\log 2bd)$ & small & $n O(d / \log(d))$ & small  \\
  & Aggregation & 0 & $O(nmd)$ & $2 n O(d / \log p)$ & small & $n O(d/ \log p)$ & small \\
  & Total & 0 & $O(nmd)$ & $nd O(2b/\log 2bd)$ & small & $d O((n + k \log M / \log p) / \log d)$ & small \\
  \hline
  \multicolumn{2}{|c|}{\shortstack[c]{Comm. per client}}& \multicolumn{2}{|c|}{$\approx 2dnb$ elements} & \multicolumn{2}{|c|}{$\approx 12d$ elements} & \multicolumn{2}{|c|}{$\approx d$ elements}\\ 
  \hline
\end{tabular}
\label{table:theoretical_cost_compare}
\end{table*}
}

\vspace{1mm}
\noindent
\textbf{\eiffel{}.}
The commitment includes the Shamir secret shares and the check string for each coordinate. 
The Shamir shares incur a cost of $O(nmd)$ f.a. operations. 
The check strings use information-theoretic secure VSSS\footnote{The Feldman check string \cite{feldman1987practical} is not secure because weight updates are small. 
$u_{il}$ can be easily computed from $g^{u_{il}}$ because $u_{il}$ has short bit-length.}~\cite{pedersen2001non}, 
which involves $O(m)$ group exponentiations per coordinate for Byzantine tolerance of $m$ malicious clients.
The proof generation\footnote{In order to prevent overflow in finite field arithmetic, client $\mathcal{C}_i$ has to prove that each $u_{il}$ is bounded, so it has to compute shares of every bit of $u_{il}$.} and verification\footnote{We use the multiplicative homogeneity of Shamir's share to compute shares of the sum of squares at the cost of requiring $m < (n-1)/4$. 
This is discussed in \cite[Section 11.1]{roy2022eiffel}. 
The corresponding cost is $O(d)$ per sum of squares. 
In comparison, the polynomial interpolation approach in \cite[Section 11.1]{roy2022eiffel} is actually $O(d^2)$ per sum of squares
because one needs to compute $d$ values of polynomials of degree $d$, even if the Lagrange coefficients are precomputed.}  (excluding verification of check strings) costs $O(bnmd)$ f.a.\ and about $2dnb$ elements of bandwidth on the client side, where $b$ is the bit length of the update. 
The verification of the check strings of one client takes a mult-exponentiation of length $(m+1)d$, or $O(md/\log md)$ g.e.\ using a Pippenger-like algorithm \cite{pippenger1980evaluation}. 
The server cost is small compared to client cost.
%

\begin{table*}[htb!]
\setlength{\tabcolsep}{4pt}
\caption{Breakdown cost comparison w.r.t. the number of model parameters $d$, where $k = 1000$}
\centering
\begin{tabular}{|c|c|c c c c|c c c c|c|}
\hline
\multirow{2}{*}{$\#$Param.} & \multirow{2}{*}{Approach} & \multicolumn{4}{|c|}{Client Computation (seconds)} & \multicolumn{4}{|c|}{Server Computation (seconds)} & \multirow{2}{*}{\shortstack[c]{Comm. Cost \\ per Client (MB)}} \\
& & \multirow{1}{*}{\shortstack[c]{commit.}} & \multirow{1}{*}{\shortstack[c]{proof gen.}} & \multirow{1}{*}{\shortstack[c]{proof ver.}} & \multirow{1}{*}{\shortstack[c]{total}} & \multirow{1}{*}{prep.} & \multirow{1}{*}{\shortstack[c]{proof ver.}} & \multirow{1}{*}{\shortstack[c]{agg.}} & \multirow{1}{*}{\shortstack[c]{total}} & \\
\hline
\multirow{4}{*}{$d$ = 1K} & \eiffel{} & 0.865 & 3.63 & 11.7 & 16.2 & - & - & 0.182 & \textbf{0.182} & 125 \\
& \rofl{} & 0.051 & 4.43 & - & $\;\;\,$4.5 & - & 91.2 & 0.040 & 91.3 & 0.37 \\
& ACORN & 0.076 & 2.49 & - & $\;\;\,$2.6 & - & 58.9 & 0.040 & 58.9 & \textbf{0.004}\\
& \textbf{\ourtech{} (ours)} & 0.054 & 1.48 & 0.08 & $\;\;\,$\textbf{1.6} & 1.17 & 75.6 & 0.071 & 76.8 & 0.44 \\
\hline
\multirow{4}{*}{$d$ = 10K} & \eiffel{} & 8.38 & 36.8 & 115 & 161 & - & - & 1.81 & \textbf{1.81} & 1250 \\
& \rofl{} & 0.51 & 46.4 & - & 46.9 & - & 860 & 0.41 & 860 & 3.66 \\
& ACORN & 0.75 & 24.5 & - & 25.3 & - & 522 & 0.41 & 523 & \textbf{0.03} \\
& \textbf{\ourtech{} (ours)} & 0.49 & $\;\;\,$1.8 & 0.08 & \textbf{2.3} & 8.61 & 82.5 & 0.71 & 91.8 & {0.71} \\
\hline
\multirow{4}{*}{$d$ = 100K} & \eiffel{} & 84.7 & 382 & 1070 & 1536 & - & - & 18.8 & \textbf{18.8} & 12500 \\
& \rofl{} & $\;\;\,$5.1 & 496 & - & 502 & - & 8559 & $\;\;\,$4.1 & 8563 & 36.6 \\
& ACORN & $\;\;\,$7.6 & 253 & - & 261 & - & 5087 & $\;\;\,$4.1 & 5091 & $\;\;\,$\textbf{0.3} \\
& \textbf{\ourtech{} (ours)} & $\;\;\,$4.8 & 4.5 & 0.08 & \textbf{9.3} & 73.3 & 139 & $\;\;\,$7.2 & 219 & $\;\;\,${3.5} \\
\hline
\multirow{4}{*}{$d$ = 1M} & \eiffel{} & OOM & OOM & OOM & OOM & OOM & OOM & OOM & OOM & OOM \\
& \rofl{} & OOM & OOM & OOM & OOM & OOM & OOM & OOM & OOM & OOM \\
& ACORN & OOM & OOM & OOM & OOM & OOM & OOM & OOM & OOM & OOM \\
& \textbf{\ourtech{} (ours)} & 48.0 & 31.2 & 0.08 & \textbf{79.3} & 653 & 612 & 72.1 & \textbf{1338} & \textbf{30.9} \\
\hline
\end{tabular}
\label{table:breakdown}
\end{table*}

\vspace{1mm} 
\noindent
\textbf{\rofl{}.}
It uses the ElGammal~\cite{Elgamal85} commitment $(g^{u_{il}} h^{r_{il}}, g^{r_{il}})$ for each coordinate $u_{il}$ with a separate blind $r_{il}$, which costs $O(d)$ g.e. in total.
The dominating cost of proof generation includes the generation of a well-formedness proof, which involves $O(d)$ g.e., $4d$ elements, a commitment, a proof of squares ($O(d)$ g.e., $6d$ elements), and a proof of bound of each coordinate ($4\log(bd)$ multi-exponentiations of length $bd$, or $O(bd)$ g.e.).
%
%
The proof verification is executed by the server, where the verification of the bound proof per client takes 1 multi-exponentiation of length about $2bd$. 
The aggregation cost is small.

\vspace{1mm} 
\noindent
\textbf{ACORN.} 
The ZKP part uses a variant of Bulletproofs by Leveraging Lagrange’s four-square theorem. 
Compared to \rofl{}, its computational cost does not depend on $b$. It uses PRG-SecAgg~\cite{BellBGL020} whose communication cost is only $(1+\log(n)/b)$ times more than plaintext.  

\vspace{1mm} 
\noindent
\textbf{\ourtech{} (Ours).}
We use Pederson commitment $g^{u_{il}} w_i^{r_i}$ for each coordinate, which costs $O(d)$ g.e.\ in total. 
On the client side, the main cost of proof generation is to verify the correctness of $\textbf{h}$ received from the server, using $\mathsf{VerCrt}$. 
It costs one multi-exponentiation of length $d$, or $O(d / \log d)$ g.e., plus $O(kd)$ f.a.\ to compute $\textbf{b}\cdot \textbf{A}$. 
On the server side, the main cost can be divided into two parts: 
(1) the computation of multi-exponentiations $\textbf{h}$ at the preparation stage; 
(2) the verification of the correctness of $\textbf{e}$ for each client using $\mathsf{VerCrt}$ at the proof verification stage. 
Each $h_t$, $t \in [1,k]$ is a multi-exponentiation of length $d$ where the powers are discrete normal samples from $\mathcal{N}(0,M^2)$, whose bit-lengths are $\log(M)$. 
The cost of computing such an $h_t$ is  $O(d \log M / \log d \log p)$, a factor of $\log M / \log p$ faster than a multi-exponentiation whose powers have bit-lengths $\log p$.
The verification of $\textbf{e}$ costs $O(d / \log d)$ g.e. 
The aggregation cost is small. 

\ignore{
The benefits of our approach are two folds. 
\begin{itemize}[leftmargin=*,topsep=0pt,itemsep=-0.5ex]
\item Low commitment cost. 
The commitment cost of $\ourtech{}$ shrinks by a factor of $2m+1$ compared to $\eiffel{}$ while still tolerating Byzantine clients, which is not supported by $\rofl{}$. 
%
\item Low cost of computation and verification of Pedersen commitments of inner products. The Pedersen commitment of an inner product can be computed from the Pedersen commitments of individual coordinates:
\begin{align}
    e_{il} = \prod_l y_{il}^{a_{tl}}.
\end{align}
Client $i$ can compute $e_{il}$ in $O(1)$ time once it receives and trusts the server's computation of $h_t = \prod_l w_l^{a_{tl}}$. 

\end{itemize}
}



\section{Experiments} \label{sec:experiment}

We implement the proposed \ourtech{} system in C/C++.
The cryptographic primitives are based on libsodium\footnote{https://doc.libsodium.org/} which implements the Ristretto group\footnote{https://ristretto.group/} on Curve25519 that supports 126-bit security.
The implementation consists of 9K lines of code in C/C++.

\ignore{
\begin{figure*}[htb!]
    \centering
    \subfloat[Client computational time]{{\includegraphics[width=0.33\textwidth]{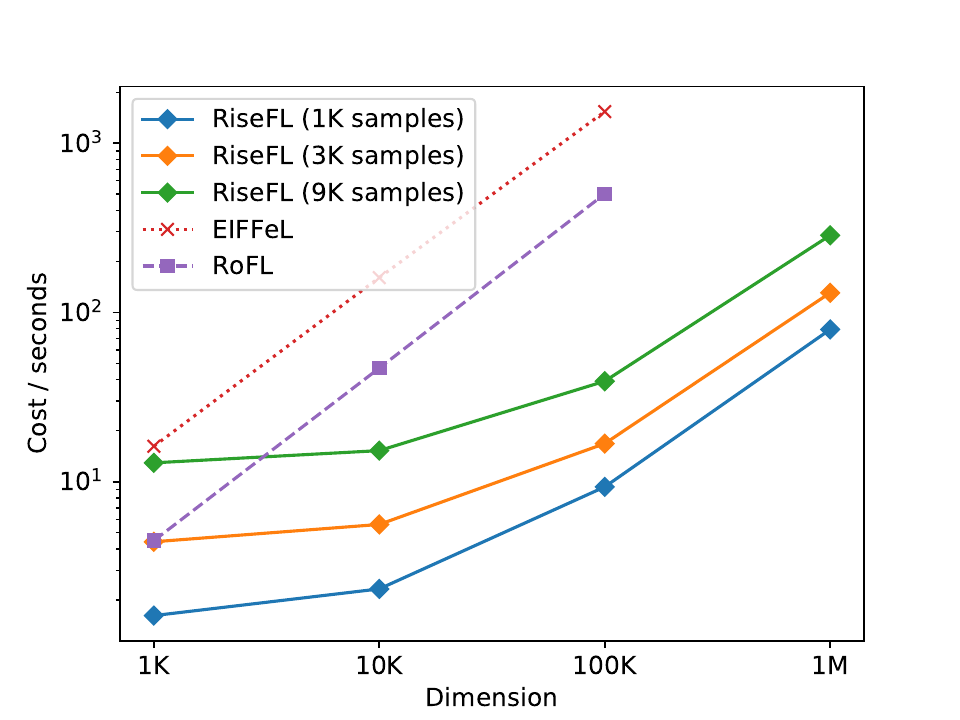} }}%
    \subfloat[Server computational time]{{\includegraphics[width=0.33\textwidth]{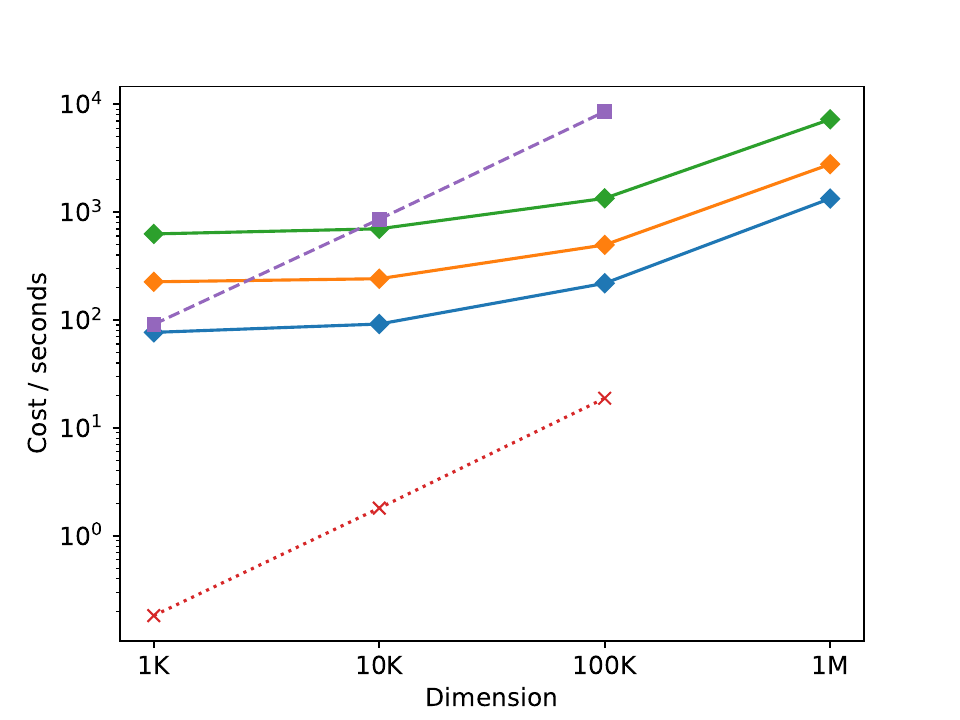} }}%
    \subfloat[Communication cost per client]{{\includegraphics[width=0.33\textwidth]{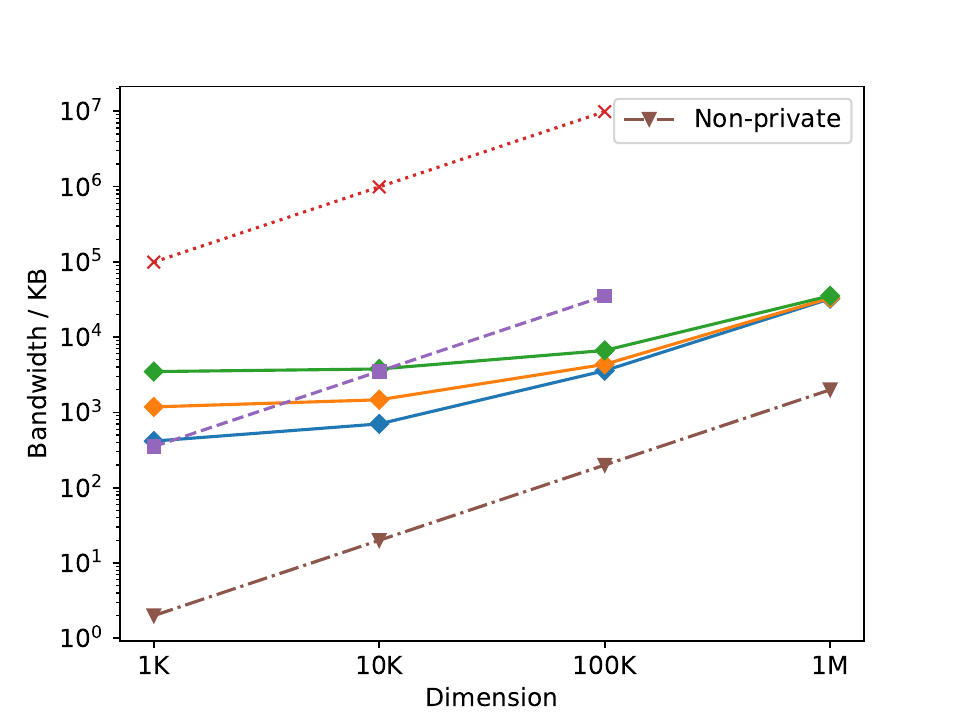} }}%
    \caption{Cost and bandwidth comparison with baselines (vary dimension) \needcheck{(may merge this figure with Tables 2-3 as there are some duplicated values).}}
    \label{fig:baseline_vary_dim}
\end{figure*}
}

\subsection{Methodology} \label{subsec:exp-method}

\textbf{Experimental Setup.} 
We conduct the micro-benchmark experiments on a single server equipped with Intel(R) Core(TM) i7-8550U CPU and 16GB of RAM.
The experiments of the federated learning tasks are simulated on a server with Intel(R) Xeon(R) W-2133 CPU, 64GB of RAM, and GeForce RTX 2080 Ti.
%
The default security parameter is {126} bits. 
We set $\epsilon = 2^{-128}$ to ensure that the level of security of \ourtech{} matches the baselines.
For the fixed-point integer representation, we use 16 bits to encode the floating-point values by default.
As discussed in Section~\ref{sec:analysis}, we set $M = 2^{24}$ to make sure that the rounding error of discrete normal samples is small.


\vspace{1mm}
\noindent
\textbf{Datasets.} 
We use three real-world datasets, namely, 
OrganAMNIST~\cite{medmnistv1,medmnistv2}, OrganSMNIST~\cite{medmnistv1,medmnistv2} and Forest Cover Type~\cite{misc_covertype_31}, 
to run the FL tasks and measure the classification accuracy.
The OrganAMNIST and OrganSMNIST datasets are medical image datasets based on abdominal clinical computed tomography. These datasets comprise 58850 and 25221 images of 784 numerical features (28 $\times$ 28) in 11 classes, respectively.
The Forest Cover Type dataset is a tabular dataset consisting of 581012 rows of 10 numerical features, 44 categorical features and a classification label from 7 label classes.
We also generate a synthetic dataset for micro-benchmarking the computational and communication costs.

\vspace{1mm}
\noindent
\textbf{Models.} We employ a CNN model for the OrganAMNIST dataset,  ResNet-18~\cite{HeZRS16} for the OrganSMNIST dataset, and TabNet~\cite{arik2021tabnet} for the Forect Cover Type dataset. 
The CNN model consists of four layers and 31K parameters. 
The ResNet-18 model consists of 18 layers and 11M parameters. 
The TabNet model consists of 471K parameters.
For the micro-benchmark experiments on synthetic datasets, we use synthetic models with 1K, 10K, 100K, and 1M parameters for the evaluation.

\begin{figure*}[t]
   \centering
   \includegraphics[width=0.85\textwidth]{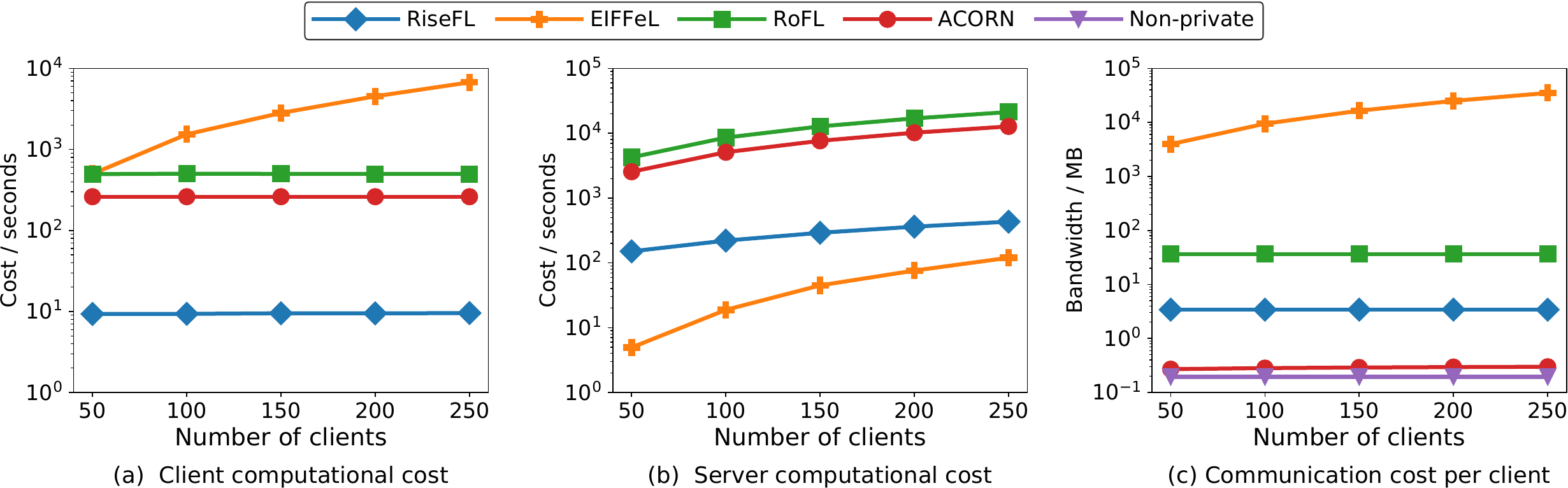}    \caption{Cost comparison w.r.t. the number of clients.}
   \label{fig:baseline_vary_n_m_unify}
\end{figure*}

\vspace{1mm}
\noindent
\textbf{Baselines.} 
We compare \ourtech{} with three secure 
baselines, namely \rofl{}~\cite{burkhalter2021rofl}, \eiffel{}~\cite{roy2022eiffel}, and ACORN~\cite{bell2023acorn}, using the same secure parameter, to evaluate the performance. 
Furthermore, we compare \ourtech{} with two non-private baselines for input integrity checking to evaluate model accuracy. 
The following describes the baselines:

\begin{itemize}[leftmargin=*]
    \item \textit{RoFL}~\cite{burkhalter2021rofl} adopts the ElGammal commitment scheme and uses the strict checking zero-knowledge proof for integrity check. 
    It does not guarantee Byzantine robustness. 
    \item \textit{EIFFeL}~\cite{roy2022eiffel} employs the verifiable Shamir secret sharing (VSSS) scheme and secret-shared non-interactive proofs (SNIP) for SAVI, ensuring Byzantine robustness. 
    We implement \eiffel{} as it is not open-sourced.
    \item \textit{ACORN}~\cite{bell2023acorn} adopts the PRG-SecAgg protocol~\cite{BellBGL020} and uses the strict checking zero-knowledge proof for integrity check. 
    It does not guarantee Byzantine robustness. 
    \item \textit{NP-SC} is a non-private baseline with strict integrity checking, i.e., the server checks each client's update and eliminates the update that is out of the $L_2$-norm bound.
    \item \textit{NP-NC} is a non-private baseline without any checking on clients' updates. 
    In this baseline, malicious clients can poison the aggregated models through malformed updates. 
\end{itemize}

\vspace{1mm}
\noindent
\textbf{Metrics.} We utilize three metrics to evaluate the performance of our \ourtech{} system. 
\begin{itemize}[leftmargin=*]
    \item \textit{Computational Cost} refers to the computation time on each client and on the server. 
    It measures the computational efficiency of the protocol. 
    \item \textit{Communication Cost per Client} refers to the size of messages transmitted between the server and each client.
    It measures the communication efficiency.  
    \item \textit{Model Accuracy} refers to the ratio of correct predictions for the trained FL model.
    It measures the effectiveness of the integrity check method in \ourtech{}. 
\end{itemize}

\subsection{Micro-Benchmark Efficiency Evaluation} \label{subsec:exp-micro}

We first compare the cost of our \ourtech{} with \eiffel{}, \rofl{}, and ACORN, using micro-benchmark experiments. 
Unless otherwise specified, we set the number of clients to 100 and the maximum number of malicious clients to 10 in this set of experiments. 
We use 16 bits\footnote{The effect of bit-length on the computational cost of $\ourtech{}$ is small.} 
for encoding floating-point values and run the experiments on one CPU thread on both the client side and the server side. 

\ignore{
\needcheck{(The following may put to appendix as the implementation details of the \eiffel{} baseline.)}
There are some differences with our \eiffel{} experiments and original ones. 

1. We added bound checks for every coordinate of the weight updates, used the information-secure VSSS, used the multiplicative homogeneity of Shamir's share to compute shares of sum of squares, used batch checking to verify check strings of VSSS, as discussed in Section~\ref{sec:analysis}.




2. We used $3m+1$ shares, instead of $n$ shares, to perform robust reconstruction~\cite{gao2003new} that tolerates $m$ errors. This saves the cost of robust reconstruction by a factor of $n^2 / (3m+1)^2$.

}

\ignore{
\begin{figure*}[htb!]
    \centering
    \subfloat[Client computational time]{{\includegraphics[width=0.33\textwidth]{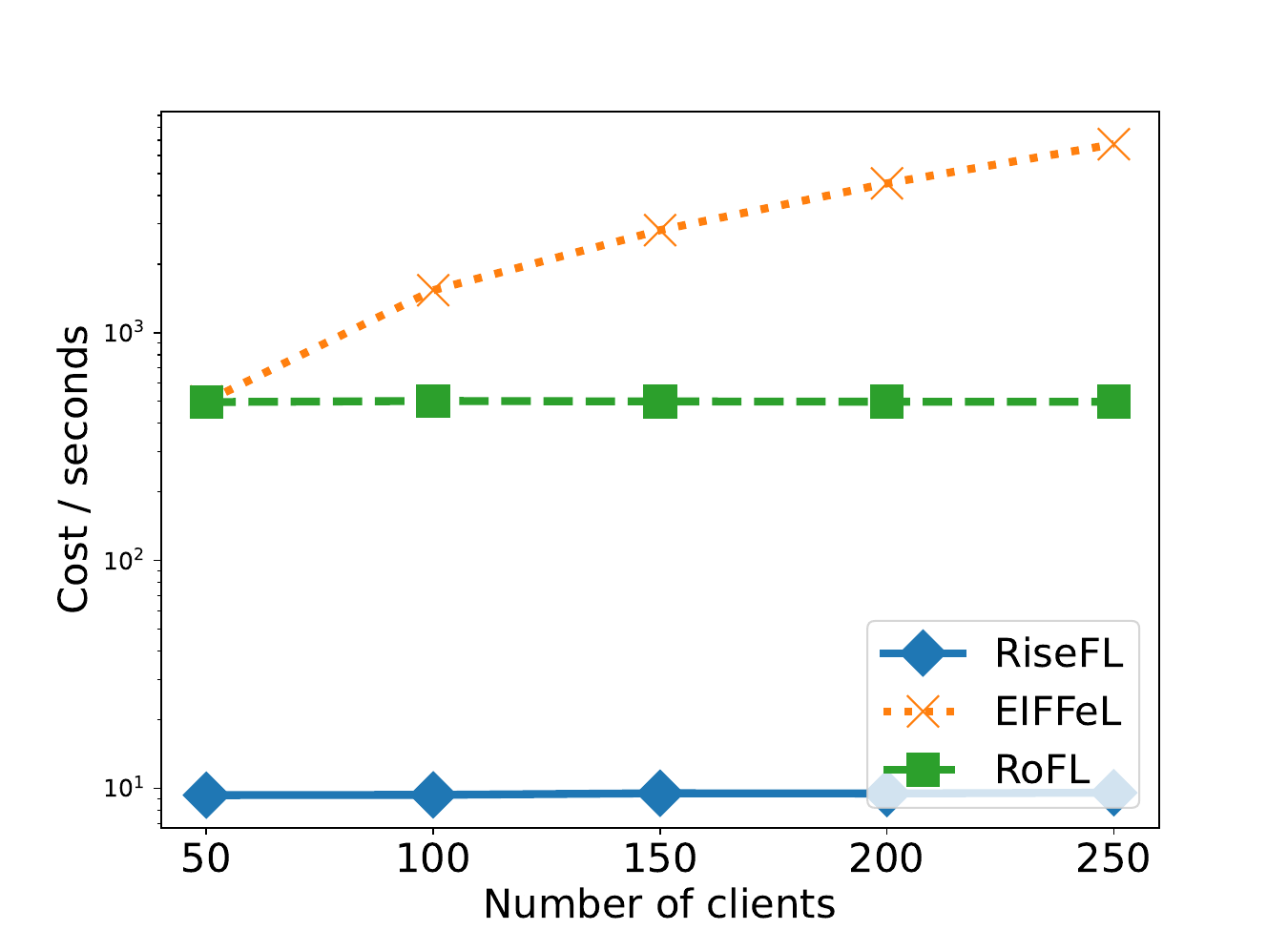} }}%
    \subfloat[Server computational time]{{\includegraphics[width=0.33\textwidth]{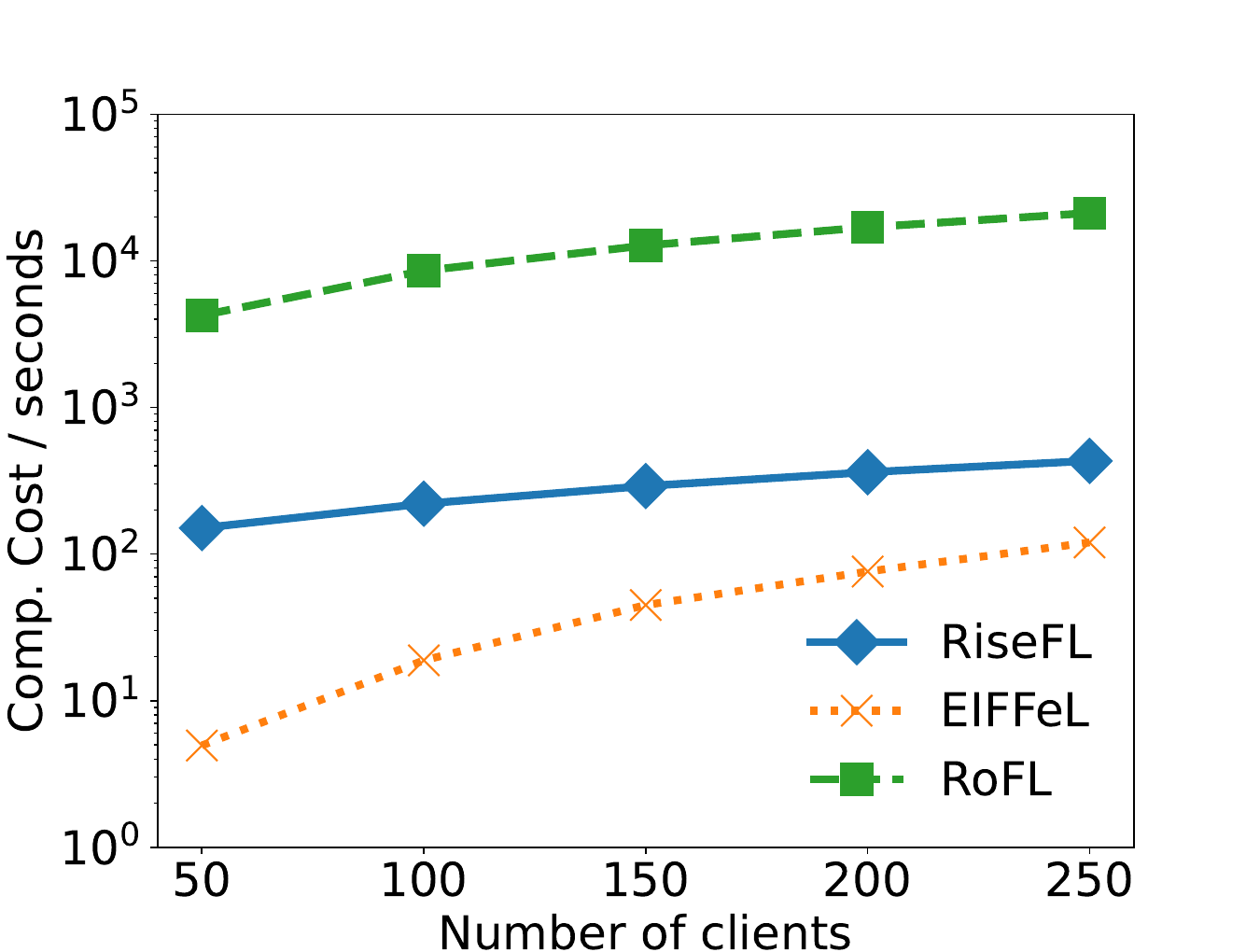} }}%
    \subfloat[Communication cost per client]{{\includegraphics[width=0.33\textwidth]{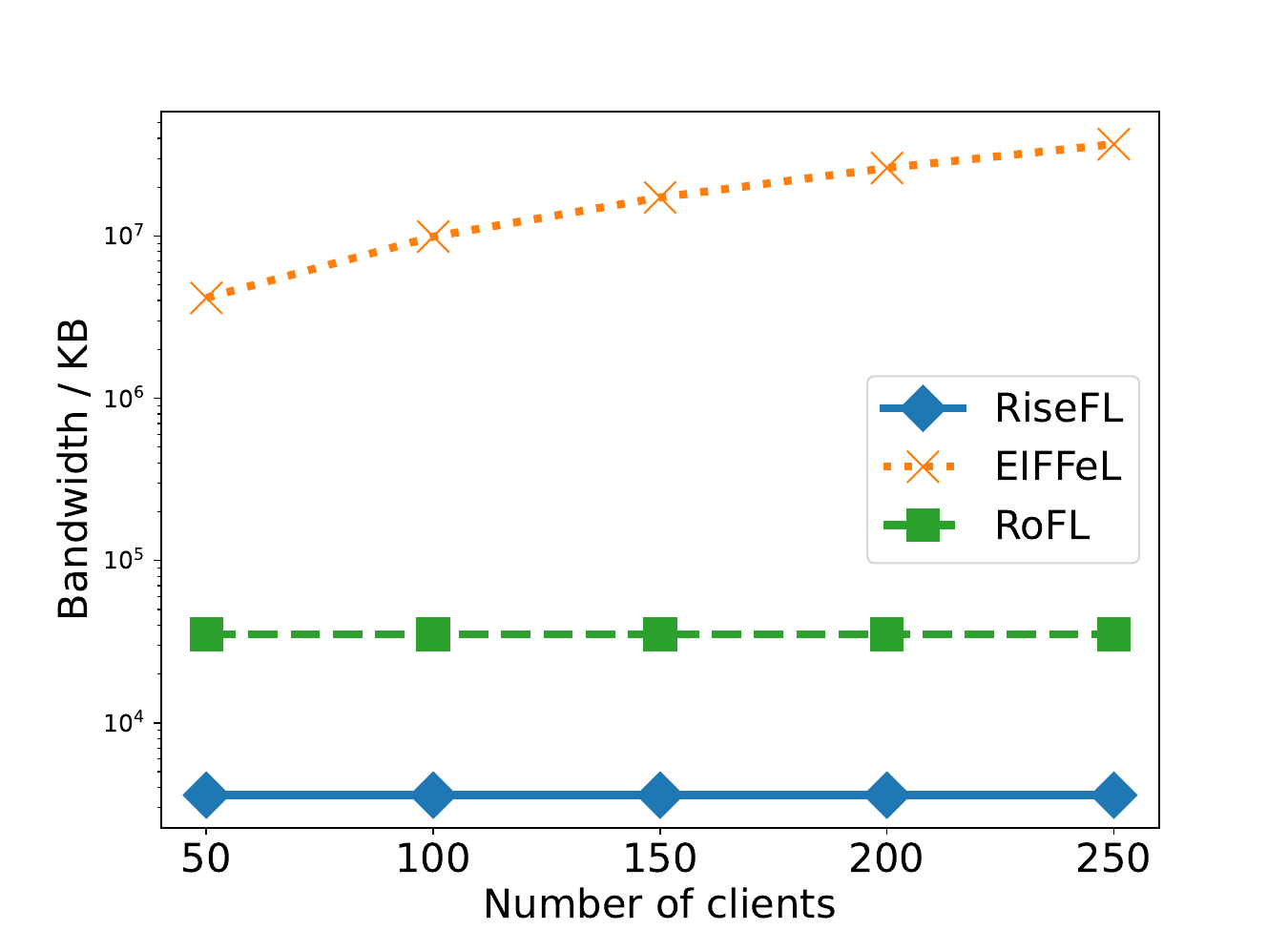} }}%
    \caption{Cost and bandwidth comparison with baselines (vary number of clients and malicious clients)}
    \label{fig:baseline_vary_n_m}
\end{figure*}
}

\ignore{
\begin{figure*}[t]
    \centering
    \subfloat[Client computational cost]{{\includegraphics[width=0.25\textwidth]{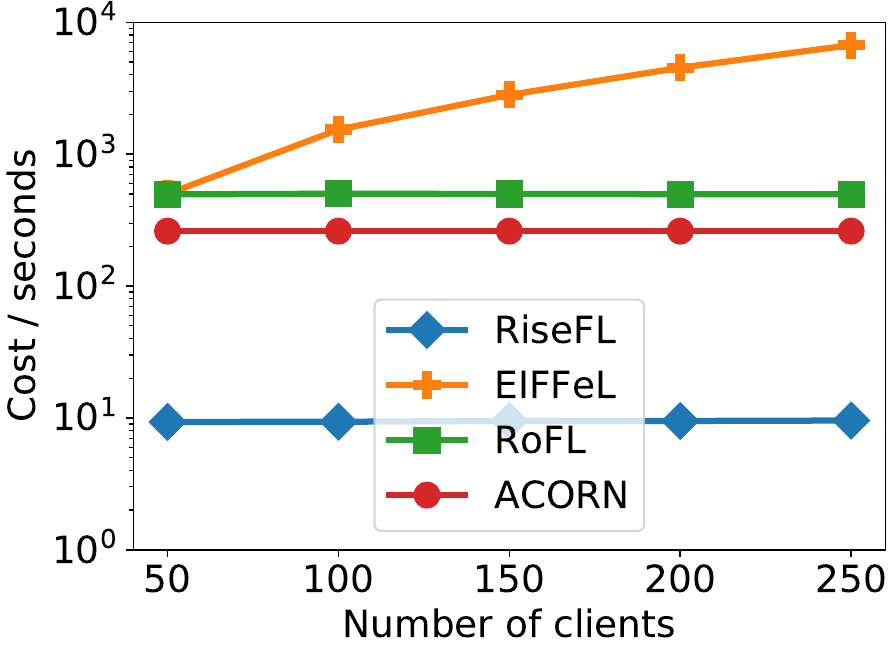} } \label{subfig:vary-n-client-comp}} \hspace{2mm}%
    \subfloat[Server computational cost]{{\includegraphics[width=0.25\textwidth]{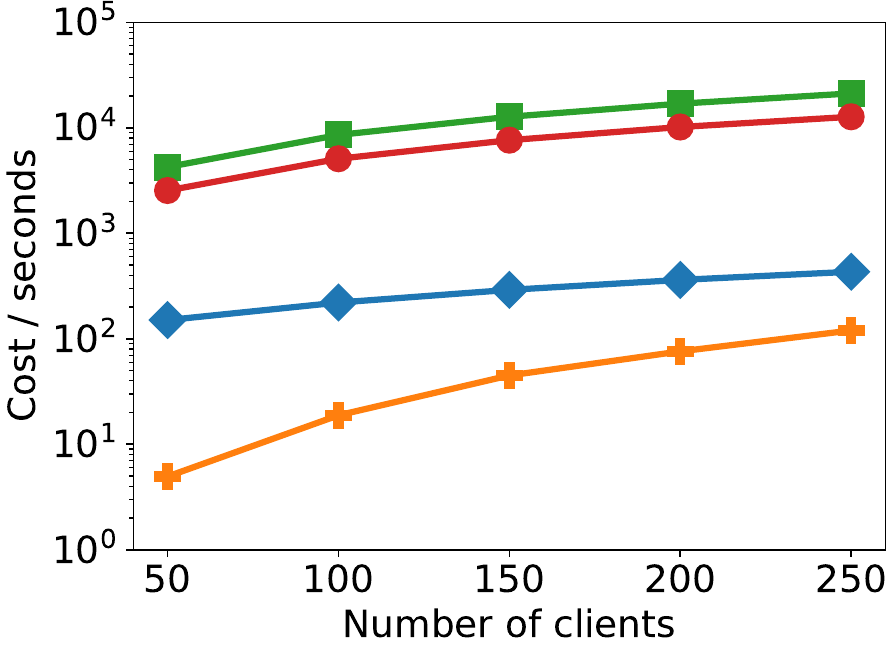} } \label{subfig:vary-n-server-comp}} \hspace{2mm}%
    \subfloat[Communication cost per client]{{\includegraphics[width=0.25\textwidth]{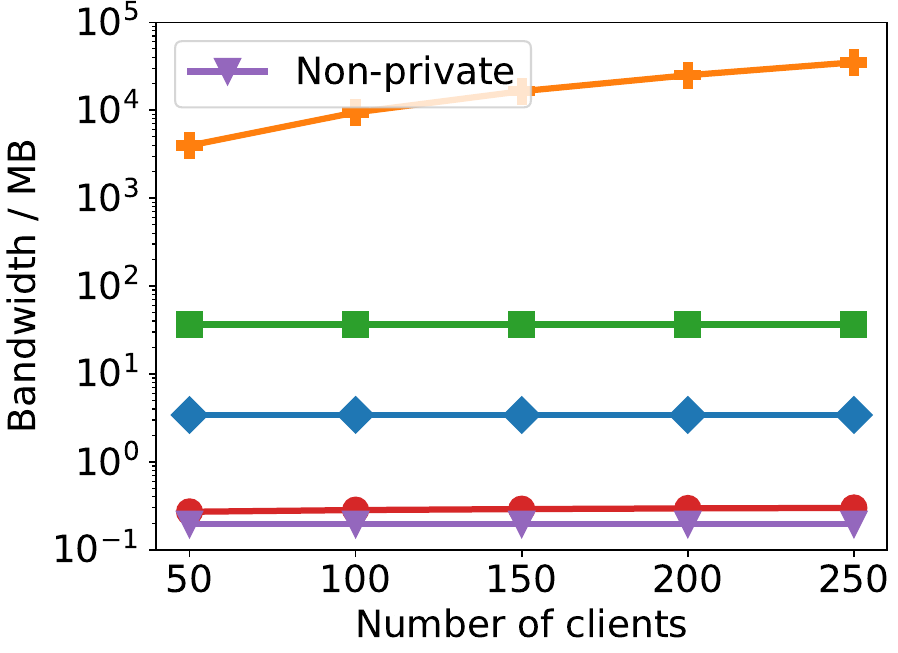} } \label{subfig:vary-n-client-comm}}%
    \caption{Cost comparison w.r.t. the number of clients}
    \label{fig:baseline_vary_n_m}
\end{figure*}
}

\vspace{1mm} 
\noindent
\textbf{Effects of $d$.}
Tables~\ref{table:breakdown} shows the breakdown cost comparison of $\ourtech{}$, $\eiffel{}$, $\rofl{}$, and ACORN w.r.t. the number of parameters $d$ in the FL model, on the client computational cost, the server computational cost, and the communication cost per client. We set the number of samples $k = 1000$. 
We failed to run the experiments with $d = 1$M on $\eiffel{}$, $\rofl{}$, and ACORN due to insufficient RAM, i.e., out of memory (OOM).
%

From the table, we observe that $\ourtech{}$ is superior to the baselines when $d >> k$. 
For example, when $d = 100$K, the client cost of $\ourtech{}$ is 28x smaller than ACORN, $53$x smaller than $\rofl{}$ and $164$x smaller than $\eiffel{}$. 
Compared to $\rofl{}$ and ACORN, the savings occur at the proof generation stage, which is attributed to our probabilistic check technique. 
The cost of $\eiffel{}$ is large as every client is responsible for computing a proof digest of every other client's proof in the proof verification stage, which dominates the cost. 
In comparison, the client-side proof verification cost of $\ourtech{}$ is negligible since every client $\mathcal{C}_i$ only verifies the check string of the Shamir share of one value $r_j$ from every other client $\mathcal{C}_j$. 
This is consistent with the theoretical cost analysis in Table~\ref{table:theoretical_cost_compare}.

With regard to server cost, $\ourtech{}$ is $23$x faster than ACORN\footnote{We did not experiment with server-side batch optimization in ACORN. In ACORN,  batch optimization at the server-side can significantly reduce server-side costs only when few clients are malicious. With the $10\%$ malicious rate in our experiment, the improvement with batch optimization is small.} and $39$x faster than $\rofl{}$ when $d = 100$K, due to our probabilistic check method. 
$\ourtech{}$ incurs a higher server cost than $\eiffel{}$ because in $\eiffel{}$, the load of proof verification is on the client side, and the dominating cost of the server is at the aggregation stage. 
In fact, the total server cost of $\ourtech{}$ is $7$ times smaller than the cost of $\eiffel{}$ of every client.

For communication cost, when $d = 100$K, $\ourtech{}$ results in 16x more than transmitting the weight updates in clear text: the committed value of a 16-bit weight update is 256-bit long. 
The proof size is negligible because $k << d$. 
ACORN transmits only about $2$x more than clear text, due to its use of PRG-SecAgg. 
$\rofl{}$ transmits about $10$x more elements than \ourtech{} in the form of proofs of well-formedness and proofs of squares. 
The communication cost of $\eiffel{}$ is three orders of magnitude larger than $\ourtech{}$ as it transmits the share of every bit of every coordinate of the weight update to every other client.

\vspace{1mm}
\noindent
\textbf{Effects of $n$.}
Figure~\ref{fig:baseline_vary_n_m_unify} shows the comparison of the computational cost and communication cost per client in terms of the number of clients. 
We vary the number of clients $n \in \{50, 100, 150, 200, 250\}$, and set the number of parameters $d = 100$K, and the maximum number of malicious clients $m = 0.1n$. 
We set the number of samples $k = 1$K for \ourtech{} in this experiment.
From Figures~\ref{fig:baseline_vary_n_m_unify}(a) and (c), we can observe that both the client computational cost and communication cost per client of $\ourtech{}$
are at least one order of magnitude lower than $\rofl{}$, $\eiffel{}$, and ACORN, while the server cost of $\ourtech{}$ is linear in $n$. 
In addition, the cost of $\eiffel{}$ both on the client side and the server side, increases quadratically in $n$. 
When $n$ gets larger, the advantage of \ourtech{} is even larger compared to \eiffel{}.

\begin{figure}[t]
    \centering
    \subfloat[Client cost]{{\includegraphics[width=0.23\textwidth]{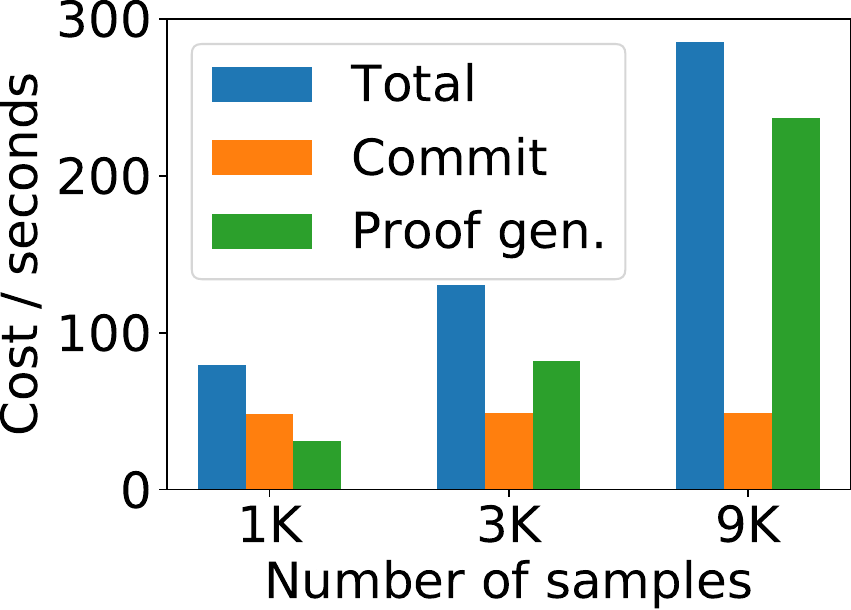} }}%
    \subfloat[Server cost]{{\includegraphics[width=0.23\textwidth]{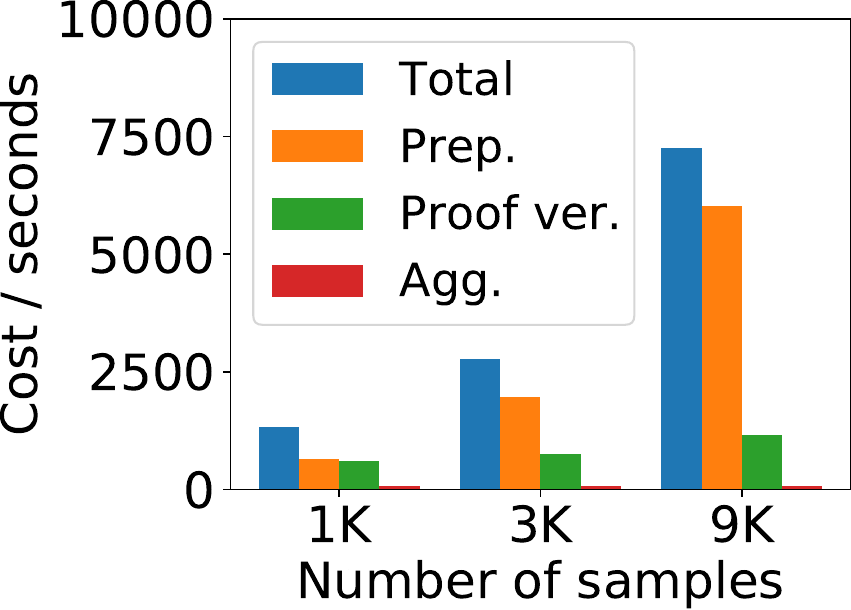} }}%
    \caption{Cost v.s. the number of samples ($d=1$M).}
    \label{fig:vary_samples}
\end{figure}

\vspace{1mm} 
\noindent
\textbf{Effects of $k$.}
Figure~\ref{fig:vary_samples} shows the per-stage breakdown of $\ourtech{}$ with $d = 1$M by varying $k \in \{1\text{K}, 3\text{K}, 9\text{K}\}$. 
On the client side, proof generation is the only stage that scales with $k$. 
On the server side, as $k$ gets larger, the preparation cost of computing $\textbf{h}$ becomes dominant. 
The effects of $k$ on the cost breakdown can be explained by Table~\ref{table:theoretical_cost_compare}.
Specifically, the terms that scale linearly with $k$ are the $O(kd)$ f.a. term of the client's proof generation and the server's preparation costs.
The linear-in-$k$ terms become dominant as $k$ becomes larger.


\ignore{
\begin{table}[t]
\small
\caption{Single-thread client side cost / seconds}
\centering
\begin{tabular}{|c| c | c c c |} 
 \hline
\multicolumn{2}{|c|}{Stage} & $d=1$K & $d=10$K  & $d=100$K \\
\hline
\multirow{3}{*}{\shortstack[c]{Commitment \\ Generation}} & \ourtech{} (Ours) & 0.0535 & 0.487 & 4.78\\
&\eiffel{}& 0.865& 8.38& 84.7\\
&\rofl{}& 0.0512& 0.509& 5.08\\
 \hline
 \multirow{3}{*}{\shortstack[c]{Proof \\ Generation}} & \ourtech{} (Ours) & 1.48 & 1.76 & 4.45 \\
&\eiffel{}& 3.63& 36.8& 382\\
&\rofl{}& 4.43& 46.4 & 496 \\
 \hline
 \multirow{3}{*}{\shortstack[c]{Proof \\ Verification}}& \ourtech{} (Ours) & 0.0822 & 0.0810 & 0.0774 \\
&\eiffel{}&  11.7& 115 & 1070 \\
&\rofl{}& - & - &  - \\
\hline
 \multirow{3}{*}{Total} & \ourtech{} (Ours) & 1.62& 2.32 & 9.31\\
&\eiffel{}& 16.2 & 161 & 1536\\
&\rofl{}& 4.48 & 46.9 & 502\\
  \hline
\end{tabular}
\label{table:microbench_client}
\end{table}

\begin{table}[t]
\small
\caption{Single-thread server side cost / seconds}
\centering
\begin{tabular}{|c| c | c c c |} 
 \hline
\multicolumn{2}{|c|}{Stage} & $d=1$K & $d=10$K  & $d=100$K \\
\hline
\multirow{3}{*}{Preparation} & \ourtech{} (Ours) & 1.17 & 8.61 & 73.3\\
&\eiffel{}& - &  - &  -\\
&\rofl{}& - & - & -\\
\hline
\multirow{3}{*}{\shortstack[c]{Proof \\ Verification}} & \ourtech{} (Ours) & 75.6 & 82.5 & 139\\
&\eiffel{}& - &  - &  -\\
&\rofl{}& 91.2 & 860 & 8559\\
 \hline
 \multirow{3}{*}{\shortstack[c]{Update \\ Aggregation}} & \ourtech{} (Ours) & 0.0711 & 0.712 & 7.24 \\
&\eiffel{}& 0.182 & 1.81 & 18.8 \\
&\rofl{}& 0.0399& 0.405& 4.13\\
 \hline
 \multirow{3}{*}{Total} & \ourtech{} (Ours) & 76.8 & 91.8 & 219\\
&\eiffel{}& 0.182 & 1.81 & 18.8\\
&\rofl{}& 91.3 & 860 & 8563\\
  \hline
\end{tabular}
\label{table:microbench_server}
\end{table}
}

\begin{figure*}[t]
    \centering
    \subfloat[OrganAMNIST sign flip attack with \\ $L_2$-norm bound check] {{\includegraphics[width=0.225\textwidth]{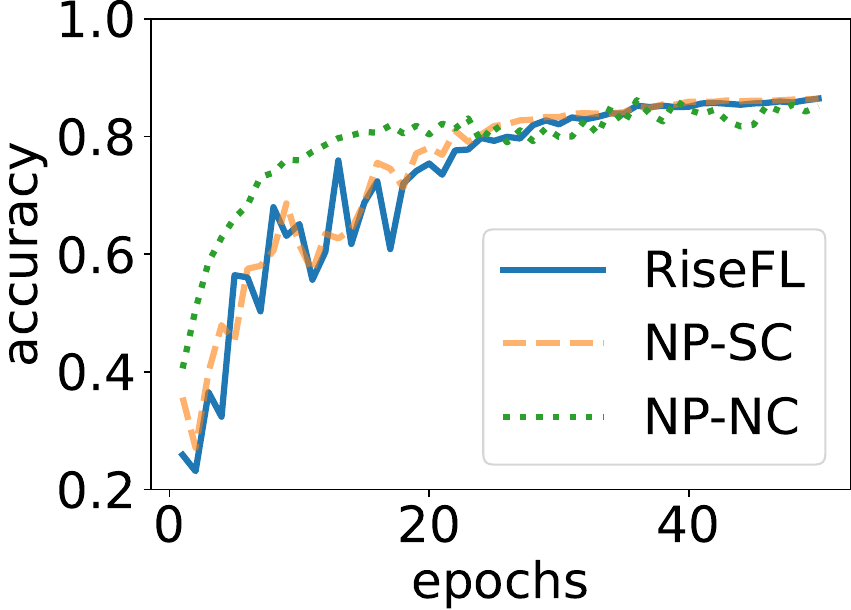} } \label{subfig:mnist-flip}}%
    \hspace{0.1mm}
    \subfloat[OrganAMNIST scaling attack with \\ $L_2$-norm bound check] {{\includegraphics[width=0.225\textwidth]{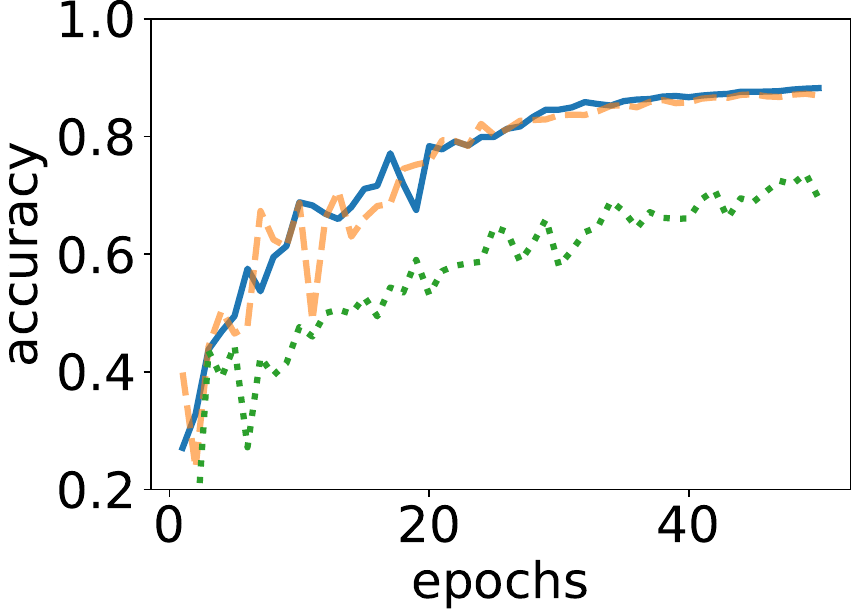} } \label{subfig:mnist-scale}}%
    \hspace{0.1mm}
    \subfloat[OrganAMNIST label flip attack with \\ cosine similarity check] {{\includegraphics[width=0.225\textwidth]{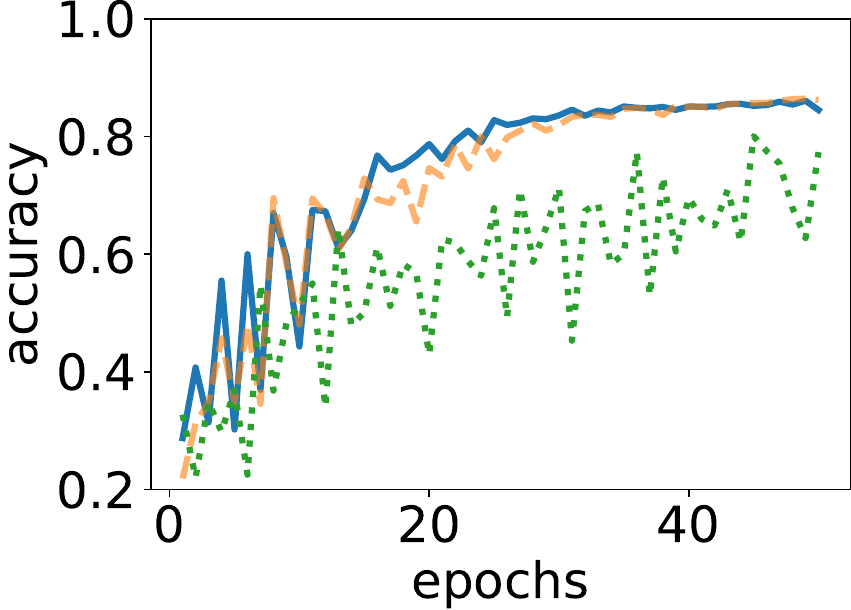} } \label{subfig:mnist-label}}%
    \hspace{0.1mm}
    \subfloat[OrganAMNIST noise attack with \\ sphere defense check] {{\includegraphics[width=0.225\textwidth]{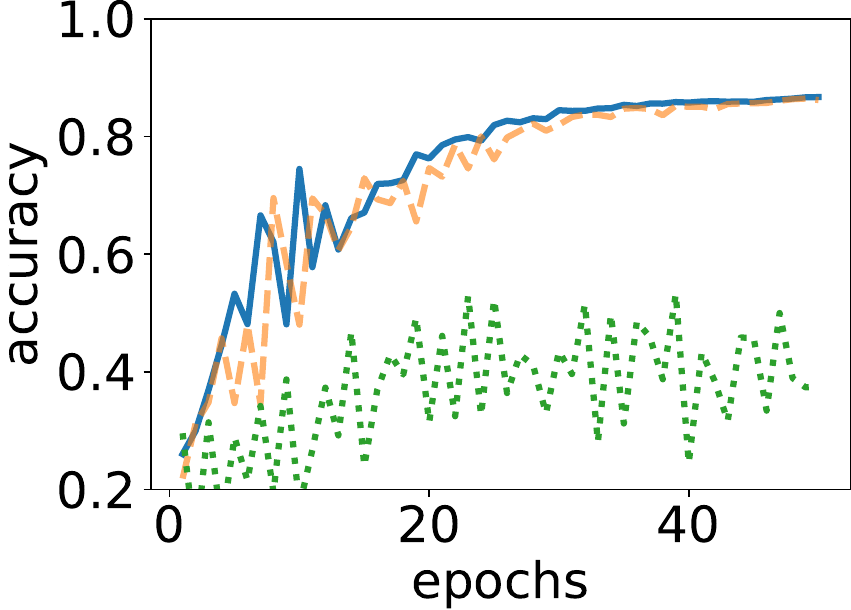} } \label{subfig:mnist-noise}}\\%
    \subfloat[OrganSMNIST sign flip attack with \\ $L_2$-norm bound check]{{\includegraphics[width=0.225\textwidth]{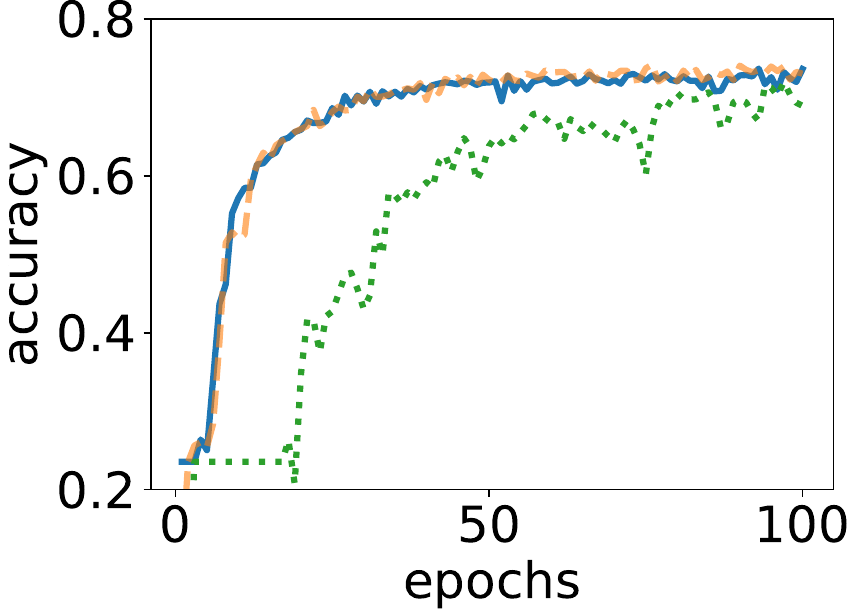} } \label{subfig:cifar-flip}}%
    \hspace{0.1mm}
    \subfloat[OrganSMNIST scaling attack with \\ $L_2$-norm bound check]{{\includegraphics[width=0.225\textwidth]{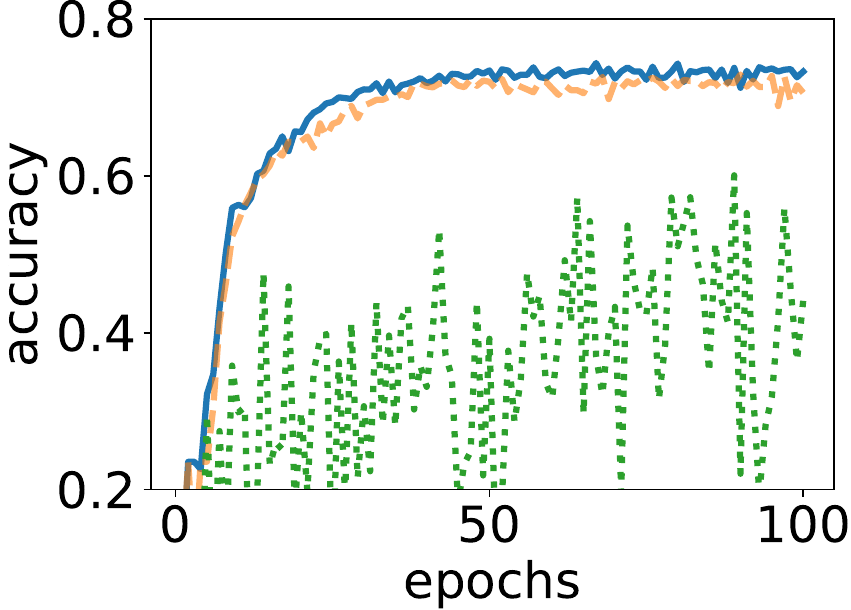} } \label{subfig:cifar-scale}}%
    \hspace{0.1mm}
    \subfloat[OrganSMNIST label flip attack with \\ cosine similarity check]{{\includegraphics[width=0.225\textwidth]{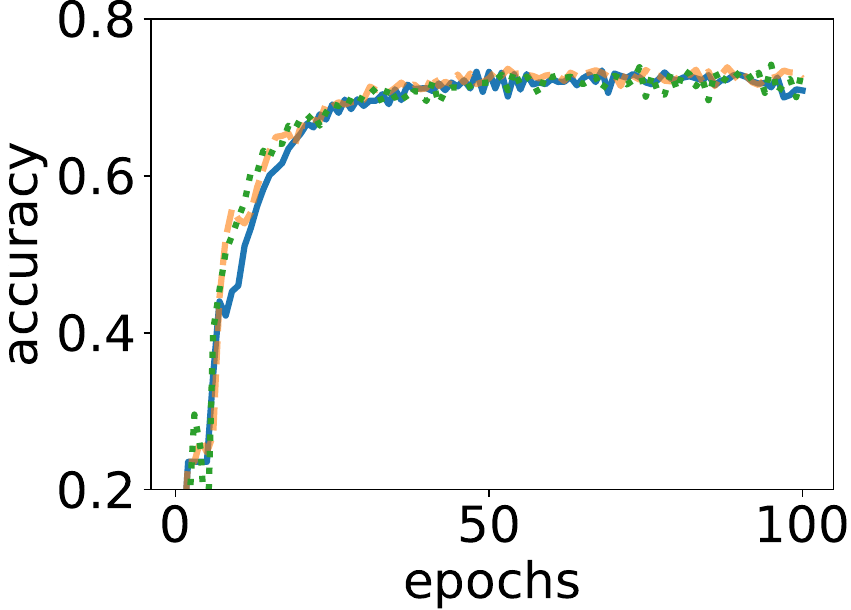} } \label{subfig:cifar-label}}%
    \hspace{0.1mm}
    \subfloat[OrganSMNIST noise attack with \\ sphere defense check]{{\includegraphics[width=0.225\textwidth]{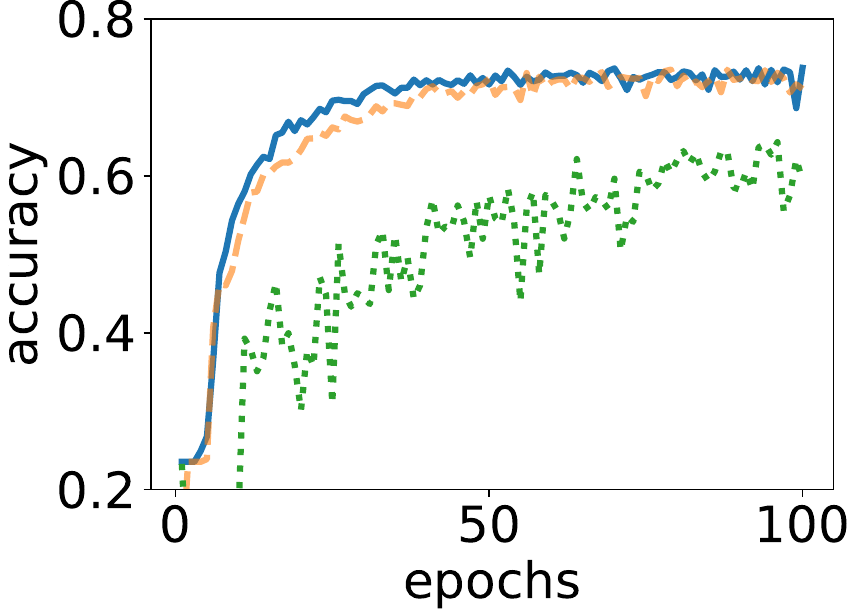} } \label{subfig:cifar-noise}}\\%
    \subfloat[Forest Cover Type sign flip attack with $L_2$-norm bound check]{{\includegraphics[width=0.225\textwidth]{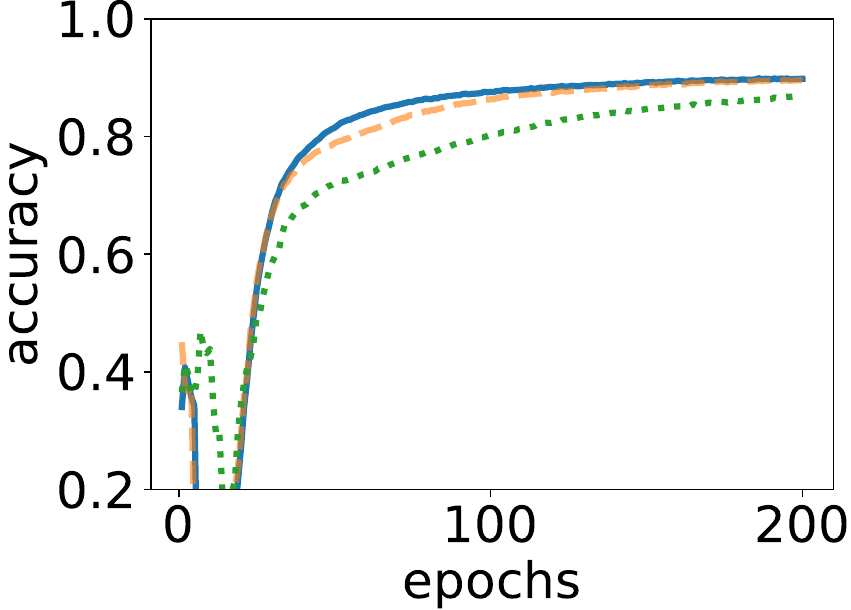} } \label{subfig:cifar-flip}}%
    \hspace{0.1mm}
    \subfloat[Forest Cover Type scaling attack with $L_2$-norm bound check]{{\includegraphics[width=0.225\textwidth]{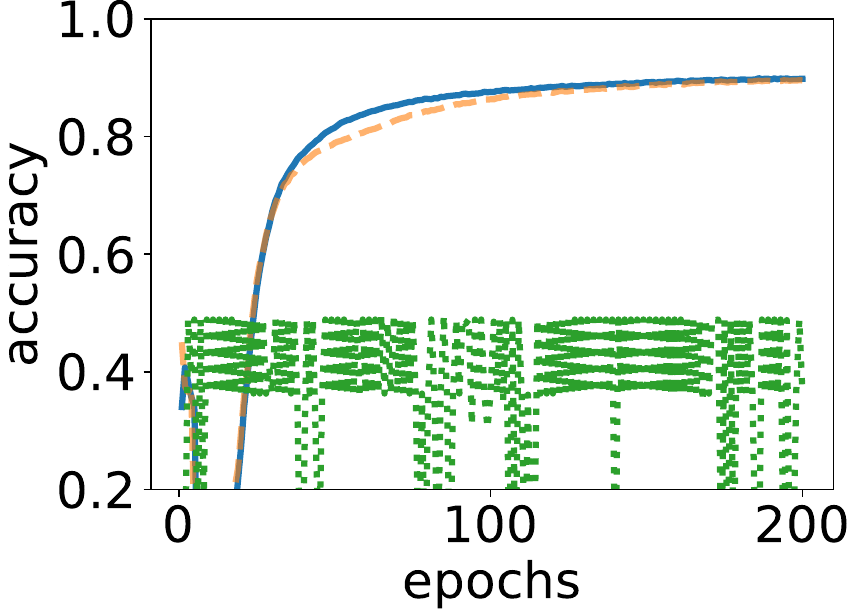} } \label{subfig:cifar-scale}}%
    \hspace{0.1mm}
    \subfloat[Forest Cover Type label flip attack with cosine similarity check]{{\includegraphics[width=0.225\textwidth]{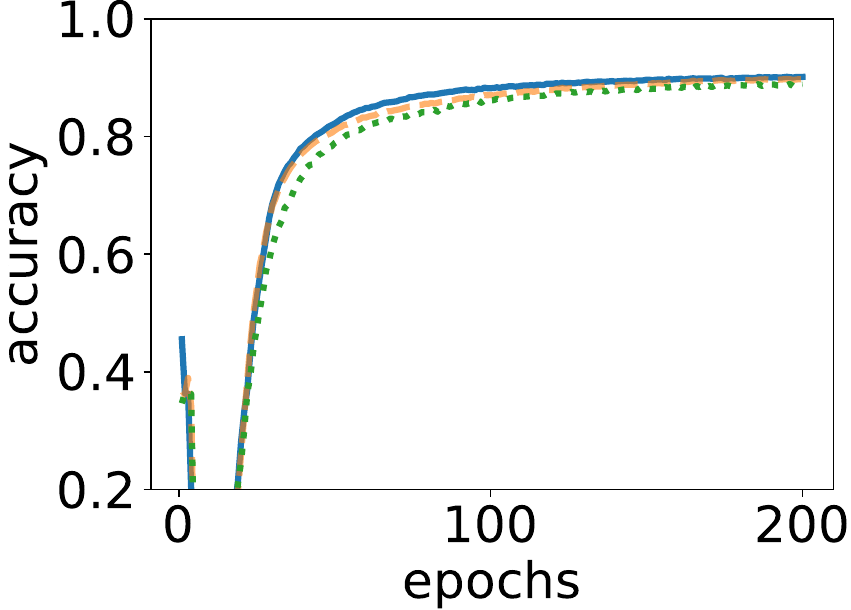}}  \label{subfig:cifar-label}}%
    \hspace{0.1mm}
    \subfloat[Forest Cover Type noise attack with sphere defense check]
     {\includegraphics[width=0.225\textwidth]{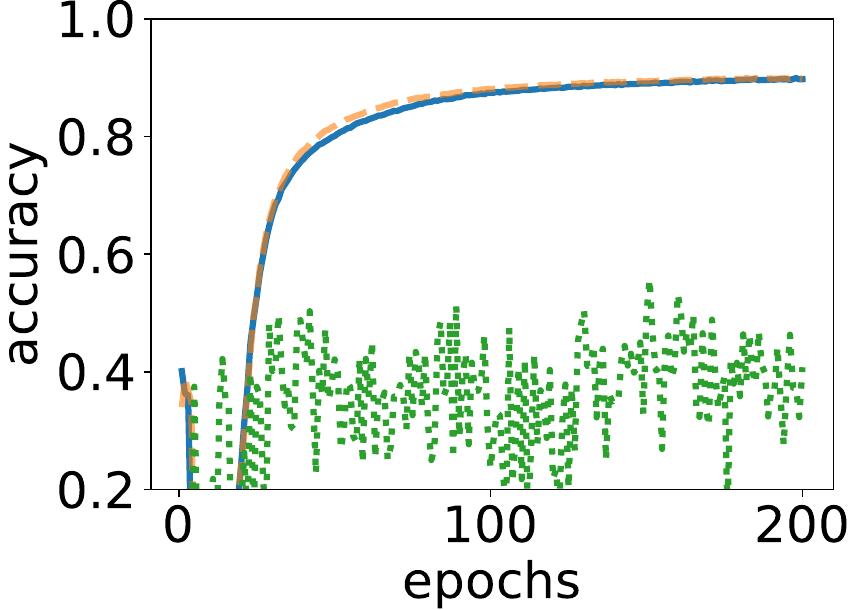} } 
    \caption{Comparison of training curves, where NP-SC and NP-NC denote the non-private methods with strict check and with no check, respectively.}
    \vspace{-1mm}
    \label{fig:training_curves}
\end{figure*}

\subsection{Robustness Evaluation} \label{subsec:exp-macro}


To evaluate the effectiveness of our probabilistic input integrity check method, 
we show the FL model accuracy of \ourtech{} against four commonly used attacks on OrganAMNIST, OrganSMNIST and Forest Cover Type datasets, respectively. 
The first is the sign flip attack~\cite{damaskinos2018asynchronous}, 
where each malicious client submits $- c \cdot \textbf{u}$ as its model update with $c > 1$. 
The second is the scaling attack~\cite{BhagojiCMC19}, 
where each malicious client submits $c \cdot \textbf{u}$ as its model update with $c > 1$.
The third is the label flip attack~\cite{sun2019can}, which labels one category of instances as another. 
The fourth is the additive noise attack~\cite{li2019rsa}, which adds Gaussian noises to the model update. 

We experiment with three different defenses: 
the $L_2$ norm defense, 
the sphere defense~\cite{SteinhardtKL17}, and the cosine similarity defense~\cite{BagdasaryanVHES20, CaoF0G21}. 
%
\ignore{
\begin{itemize}[leftmargin=*]
    \item Sign flip attack~\cite{damaskinos2018asynchronous}: each malicious client submits $- c \cdot \textbf{u}$ as its model update, where $c > 1$. 
    \item Scaling attack~\cite{BhagojiCMC19}: each malicious client submits $c \cdot \textbf{u}$ as its model update, where $c > 1$. 
\end{itemize}
}
In this set of experiments, 
We use FLSim\footnote{https://github.com/facebookresearch/FLSim} to simulate federated learning with 100 clients and 10 malicious clients.
Figure~\ref{fig:training_curves} shows the training curves of $\ourtech{}$ with two non-private baselines, NP-SC and NP-NC (see Section~\ref{subsec:exp-method}). 
There are two main observations. 
First, $\ourtech{}$ achieves better accuracy than the no-checking baseline NP-NC. 
This is expected as the malicious clients can poison the aggregated models by invaliding model updates if the server does not check the input integrity, 
leading to lower accuracy 
or non-converging curves.
Second, the training curves of $\ourtech{}$ and the strict $L_2$ norm check baseline NP-SC are very close. 
This confirms the effectiveness of \ourtech{} in identifying malformed updates and robust aggregation under a variety of attacks. 

\section{Related Works} \label{sec:related}


\noindent
\textbf{Secure Aggregation.} To protect the client's input privacy in federated learning (FL), a number of studies have explored the secure aggregation~\cite{bonawitz2017practical, ZhengLLYYW23, KairouzL021}, which enables the server to compute the aggregation of clients' model updates without knowing individual updates. 
A widely adopted approach~\cite{bonawitz2017practical} is to let each client use pairwise random values to mask the local update before uploading it to the server. 
The server can then securely cancel out the masks for correct aggregation. 
Nonetheless, these solutions do not guarantee input integrity, as malicious clients can submit arbitrary masked updates. 

\vspace{1mm}
\noindent
\textbf{Robust Learning.} Several works have been proposed for robust machine learning, including \cite{0017DG18, ShenS19, SteinhardtKL17, YinCRB18, YinCRB19, PanZWXJY20, CaoF0G21, MaSWLCD22, XuJZZJS22, Xiang0L023}.
However, some of these approaches, such as \cite{0017DG18, ShenS19, SteinhardtKL17}, are designed for centralized training and require access to the training data, which makes them unsuitable for FL. 
On the other hand, solutions like \cite{YinCRB18, YinCRB19, PanZWXJY20, CaoF0G21, MaSWLCD22, XuJZZJS22} specifically address the FL setting and focus on ensuring Byzantine resilient gradient aggregation.
These solutions operate by identifying and eliminating client updates that deviate significantly from the majority of clients' updates, as they are likely to be malformed updates. 
However, it is worth noting that these approaches require the server to access the plaintext model updates, which compromises the client's input privacy.

\vspace{1mm}
\noindent
\textbf{Input Integrity Check with Secure Aggregation.} 
In this research direction, Prio~\cite{corrigan2017prio} and Elsa~\cite{rathee2023elsa} ensure both input privacy and input integrity 
with multiple non-colluding servers. Under the stronger assumption of at least two non-colluding servers, secure aggregation can be achieved with a lower computation cost. 
In contrast, we focus on a single-server setting. 
Under this setting, \cite{burkhalter2021rofl, roy2022eiffel, bell2023acorn} ensure both input privacy and integrity. \rofl{}~\cite{burkhalter2021rofl} utilizes homomorphic commitments~\cite{Elgamal85} that are compatible with existing mask-based secure aggregation methods and adopt the zero-knowledge proof~\cite{bunz2018bulletproofs} to validate the clients' inputs. ACORN~\cite{bell2023acorn} utilizes PRG-SecAgg~\cite{BellBGL020} which is bandwidth-efficient.
However, neither \rofl{} nor ACORN supports Byzantine-robust aggregation. \eiffel{}~\cite{roy2022eiffel} designs an approach based on verifiable secret sharing~\cite{shamir1979share} and secret-shared non-interactive proofs (SNIP)~\cite{corrigan2017prio} techniques, which can tolerate Byzantine attacks. 
Nevertheless, the efficiency of these three solutions is quite low, especially when the number of model parameters is large. 
In contrast, in \ourtech{}, we propose a hybrid commitment scheme and design a probabilistic input integrity check method, 
providing support for Byzantine-robust aggregation and achieving significant efficiency improvements.

\section{Conclusions} \label{sec:conclusion}

In this paper, we propose \ourtech{}, a robust and secure federated learning system that guarantees both input privacy and input integrity of the participating clients. 
We design a hybrid commitment scheme based on Pedersen commitment and verifiable Shamir secret sharing, and present a probabilistic $L_2$-norm integrity check method, which achieves a comparable security guarantee to state-of-the-art solutions while significantly reducing the computation and communication costs. 
The experimental results confirm 
the efficiency and effectiveness of our solution.
%


\bibliographystyle{ACM-Reference-Format}
\bibliography{ref}

\end{document}